\documentclass[12pt]{article}
\pdfoutput=1

\usepackage{amsmath,amssymb,graphicx,multirow,xspace}
\usepackage[colorlinks=true,urlcolor=blue,anchorcolor=blue,citecolor=blue,filecolor=blue,linkcolor=blue,menucolor=blue,pagecolor=blue]{hyperref}
\usepackage[compress,numbers]{natbib}
\long\def\symbolfootnote[#1]#2{\begingroup%
\def\thefootnote{\fnsymbol{footnote}}\footnote[#1]{#2}\endgroup}

\input prepictex
\input pictex
\input postpictex
\newdimen\tdim
\tdim=\unitlength
\def\stpltsmbl{\setplotsymbol ({\small .})}

\newcommand{\makeblue}[1]{\color[rgb]{0.4,0.4,1}#1\color[rgb]{0,0,0}}
\newcommand{\makered}[1]{\color[rgb]{0.7,0,0}#1\color[rgb]{0,0,0}}

\renewcommand{\slash}[1]{#1\!\!\!/}
\newcommand{\MET}{$\slash E_T$\xspace}

\newcommand{\go}{{\tilde g}}
\newcommand{\Ho}{{\tilde H}}

\newcommand{\sq}{{\tilde q}}
\newcommand{\st}{{\tilde t}}
\newcommand{\sbo}{{\tilde b}}

\newcommand{\qq}{\qquad}
\newcommand{\comment}[1]{}

\newcommand{\beq}{\begin{equation}}
\newcommand{\eeq}{\end{equation}}
\newcommand{\beqa}{\begin{eqnarray}}
\newcommand{\eeqa}{\end{eqnarray}}

\begin{document}

\begin{titlepage}

\begin{flushright}
\small{RUNHETC-2013-19}\\
\end{flushright}

\vspace{0.5cm}
\begin{center}
\Large\bf
Toward Full LHC Coverage of Natural Supersymmetry
\end{center}

\vspace{0.2cm}
\begin{center}
{\sc Jared A. Evans,$^1$\symbolfootnote[1]{jaevans@physics.rutgers.edu}
Yevgeny Kats,$^{1,2}$\symbolfootnote[4]{yevgeny.kats@weizmann.ac.il}
David Shih$^1$\symbolfootnote[2]{dshih@physics.rutgers.edu}
and Matthew J. Strassler$^3$\symbolfootnote[3]{strassler@physics.harvard.edu}}\\
\vspace{0.6cm}
\textit{$\,^1$New High Energy Theory Center\\
Department of Physics and Astronomy\\
Rutgers University, Piscataway, NJ 08854, USA}\\
\vspace{0.5cm}
\textit{$\,^2$Department of Particle Physics and Astrophysics\\
Weizmann Institute of Science, Rehovot 76100, Israel}\\
\vspace{0.5cm}
\textit{$\,^3$Department of  Physics\\
Harvard University, Cambridge, MA 02138, USA}
\end{center}

\vspace{0.4cm}

\begin{abstract}
We argue that combining just a handful of searches for new physics at Run I of the LHC is sufficient to exclude most supersymmetric extensions of the Standard Model in which the gluino is kinematically accessible and the spectrum is natural. Such models typically give rise to significant \MET, top quarks and/or high object multiplicity, and we show that having even one of these signatures generally results in stringent limits. We also identify, among models that lack these signatures, the few gaps in coverage remaining, and propose search strategies to close these gaps. Our results are general and independent of the details of the spectrum, assumptions about minimality, $R$-parity, etc. Our analysis strategy should remain applicable when the LHC moves to higher energy.  Central to our argument are ATLAS and CMS searches for many jets and low \MET, a proposed lepton + many jets search, an ATLAS search for 6-7 high-$p_T$ jets, and a reexamination of the control and signal regions of the CMS black hole search.
\end{abstract}

\end{titlepage}

\vspace{0.2cm}
\noindent

\tableofcontents
\newpage

\section{Introduction}
\label{sec:introduction}

As we prepare for the restart of the LHC in 2015, it is an ideal time to take stock of what we have learned from Run I at 7-8~TeV. Since most LHC searches have been updated to the full dataset (20~fb$^{-1}$ at 8~TeV), we can now endeavor to glean as much insight from them as possible. The discovery of a Standard Model-like Higgs at 125~GeV~\cite{Aad:2012tfa,Chatrchyan:2012ufa} reinforces the urgency of the hierarchy problem, for which supersymmetry (SUSY) remains one of the best-motivated candidates for a natural solution. For reviews encompassing both the theory of natural SUSY and recent LHC results, see~\cite{Feng:2013pwa,Craig:2013cxa}.   Numerous SUSY searches at the LHC have so far only provided null results, which have been used to place stringent limits on a diverse array of models and ``simplified models''. However, it is far from clear how to use this collection of limits to reach general conclusions about all possible forms of natural SUSY.

In this paper, we aim to clarify the status of natural SUSY by focusing on the question: {\it to what extent do existing 8~TeV searches exclude natural models with a kinematically accessible gluino?} Large classes of models have clearly been excluded, but it is plausible that gaps in coverage remain.   Our objective in this paper is to seek these gaps and to understand their origins.   We will accomplish this by viewing the space of natural supersymmetric models in a very general way, and then exploring the constraints placed on this space by relevant LHC searches.

We are motivated to focus on gluino pair production in this paper, because the gluino has the highest production rate (for a given mass) of any particle required by naturalness to be accessible at the LHC (though possibly not until 14~TeV). The kinematic limit at Run I of the LHC ($\sim$~20~fb$^{-1}$ at  8~TeV) corresponds to  $m_{\go} \sim 1.5$~TeV. Clearly there will be some models with gluinos between 1 and 1.5~TeV that are kinematically accessible but difficult to detect, or that require a statistical combination of multiple searches for sensitivity (see, e.g., the recent study of~\cite{Buchmueller:2013exa}), which we will not pursue here.  Gaps where the gluino can still be lighter than 1~TeV, where more than $\sim 500$ gluino pairs would have been produced, would be much more interesting.  Here, we will argue that there are very few classes of models that are not yet convincingly ruled out for $m_\go < 1$~TeV, and we will identify how such models evade current searches.

In order for our conclusions about natural SUSY to be as general as possible, we will not restrict ourselves to any specific theoretical framework. In particular, we will not concern ourselves with quantifying fine-tuning in specific weak-scale models (which requires making assumptions about the UV theory), preferring instead to focus on the more experimentally driven question of what the  current limits are on general spectra, with our only assumption being the presence of light higgsinos, $m_\Ho \lesssim 400$~GeV. The upper limit on the higgsino mass is motivated by the fact that this mass typically contributes to the electroweak vev at tree level. If one assumes that the electroweak scale is protected {\it purely by supersymmetry} (not combined with any other symmetry or dynamical effect), then the lightest higgsino mass and the Higgs boson mass cannot be too far separated without fine tuning.
(The models we study will typically contain light stops, $m_\st \lesssim 1$~TeV, which is also demanded by naturalness, but this assumption is not crucial for our conclusions.  Examples of naturalness discussions which push the higgsinos and stops to near these boundaries are found in~\cite{Hall:2011aa,Gherghetta:2012gb}.)
We will be agnostic about both the rest of the superpartner spectrum (including particles beyond the MSSM) and how it arises from a high-energy theory, allowing the low-energy phenomenology to be completely general in principle.  In practice, we will attempt to address this very wide set of scenarios by studying a limited number of models that we believe are sufficient for us to draw conservative lessons about natural SUSY.

To understand our approach, whose overall logic we will explain more carefully in section~\ref{sec:approach}, it is very useful to keep in mind a benchmark scenario with a minimal natural spectrum: a gluino ($\go$), top squark ($\st$) and higgsinos ($\Ho_{1,2}^0$, $\Ho^\pm$), where the neutralino $\Ho_1^0$ is a stable LSP and $m_\go > m_\st > m_\Ho$.  (For some recent studies of models of this type in the context of the LHC, see~\cite{Buchmueller:2013exa,Papucci:2011wy,Essig:2011qg,Brust:2011tb,Kowalska:2013ica,Arvanitaki:2013yja}.) Here, decays of pair-produced gluinos such as
\begin{equation}
\go \,\to\, \st + \bar t,\,\,\,\st^* + t\,,\qquad \st \,\to\, t + \Ho^0,\,\,\, b + \Ho^+
\label{eqn:gluinodecays}
\end{equation}
result in three characteristic properties that can be used to distinguish them from Standard Model (SM) backgrounds:
\begin{itemize}
\item Substantial missing transverse momentum (\MET)
\item Top quarks (resulting in a $b$ jet, plus either $jj$ or $\ell\nu$)
\item High multiplicity of objects  ($N\gtrsim 8$)
\end{itemize}

Importantly, {\it these three features are common far beyond this simplified model.}  A large \MET signal is typical in simple $R$-parity
preserving SUSY models and even many $R$-parity violating ones.
Gluino decays into tops are a generic feature of a vast range of natural
models due to the presence of top squarks. Even when $\go\to\st\bar t$
is kinematically forbidden, off-shell processes such as $\go\to t\bar
b\Ho^-$ can dominate the gluino decays.  Finally, high multiplicity is
very common, especially in models with little or no \MET.  Whereas in many classic SUSY models gluinos commonly decay to two jets plus an invisible SM LSP (the lightest superpartner of any Standard Model state), the
visible or mostly visible decay of the SM LSP typically provides at least two
additional objects.  Any additional steps in the cascade (e.g., from
$W$, $Z$ or $h$ emission or from decays via sleptons) also add to the
multiplicity, as do top quarks.  Thus, in many models, eight or more
objects are common in gluino pair events.  In short, the value of
focusing on limits from \MET, tops and high multiplicity is that few
natural models lack all of these signatures. It is easy to eliminate
one or even two, but far from trivial to eliminate all three.

In this paper, our strategy will be to argue that {\it any one} of the three signatures listed above is typically enough to rule out the model for $m_\go\lesssim 1$~TeV.  To build our case, we will study a number of modifications to the minimal benchmark scenario which allow us to interpolate between different combinations of these three signatures.  To construct such models, we will be employing one or more of the following features:
\begin{itemize}
\item {\it Baryonic $R$-parity violation (RPV)}. By allowing the LSP to decay to jets through a $UDD$ operator (for a review, see~\cite{Barbier:2004ez}), we can completely eliminate the \MET\,--- but generally at the cost of high multiplicity.
\item {\it Hidden valley (HV) scenarios}. Here we allow the LSP to decay to additional ``hidden-valley'' particles~\cite{Strassler:2006im,Strassler:2006qa}. By tuning the masses of these particles, the \MET can be smoothly interpolated between typical $R$-parity conserving MSSM values and zero, again with a higher, but easily adjustable, multiplicity.
\item {\it Light second-generation squarks.} Although the first-generation squarks are highly constrained by current searches due to their large production rate through valence quarks, the second-generation squarks are free to be much lighter~\cite{Mahbubani:2012qq}. By lowering the masses of second-generation squarks, we can reduce the branching ratios of final states with tops by providing more decay modes $\tilde g\to q\tilde q$.
\end{itemize}
By reinterpreting a comprehensive set of ATLAS and CMS searches, we have studied the limits on the gluino mass as a function of the parameters of these scenarios.  Since the limits we present come from searches that are signature-driven and are not optimized for particular models, we argue that our results are largely independent of the specific models we chose to study, and thus apply broadly to all natural models that are dominated by similar signatures.

\begin{figure}[t]
\begin{center}
\includegraphics[width=0.75\textwidth]{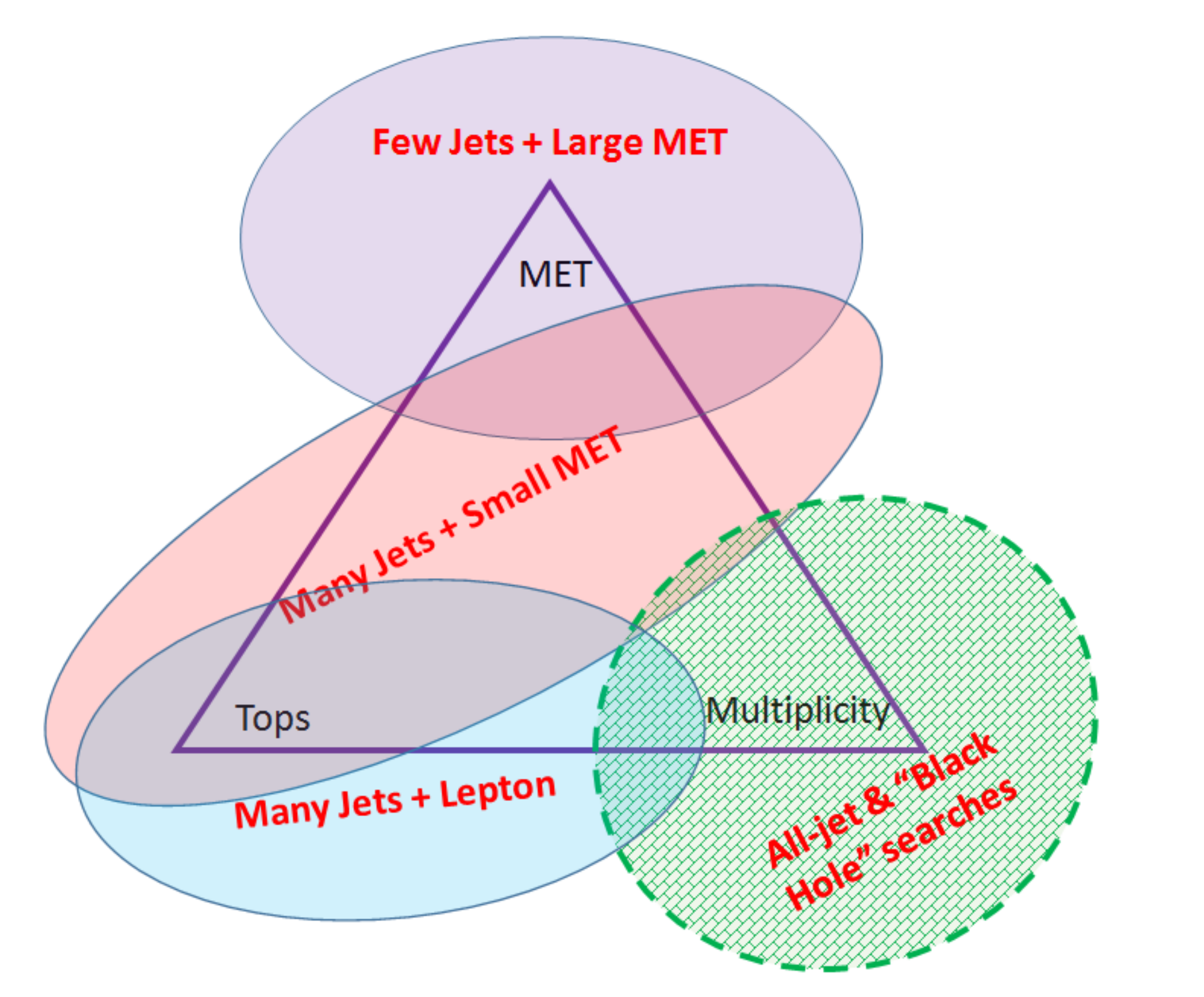}
\caption{Schematic representation of the common gluino decay features (corners of the triangle) and the searches that are sensitive to them.  The brickwork shading indicates that the coverage in the high multiplicity region is incomplete, as will be discussed in section~\ref{subsec:jetpthierarchy}. }
\label{fig:money}
\end{center}
\end{figure}

One of our main findings is that out of the multitude of existing LHC searches, only a handful of ATLAS and CMS searches are truly essential for constraining most natural SUSY models. These are listed in table~\ref{tab:searches}. Of course, many other searches (e.g., multi-lepton searches) constrain some part of natural SUSY parameter space. But the ones we highlight here form a minimal inclusive set which we believe cover the greatest range of natural SUSY models with an accessible gluino, not including models that are easily excluded by the presence of many leptons or photons.\footnote{We will not address models in which exotic objects, such as long-lived particles that decay in flight, particles with non-standard tracks, or lepton-jets, are common.  Specialized searches are often required for such models and a separate theoretical study is required.}  In particular,
\begin{itemize}
\item To constrain models with \MET requires the standard low-multiplicity high-\MET searches (such as~\cite{ATLAS-CONF-2013-047} and part of~\cite{CMS-PAS-SUS-13-012}) combined with high-multiplicity low-\MET searches (such as~\cite{Aad:2013wta} and another part of~\cite{CMS-PAS-SUS-13-012}).
\item To constrain models with tops as effectively as possible requires the same high-multiplicity/low-\MET searches combined with a search, previously proposed in~\cite{Lisanti:2011tm}, for a lepton, a $b$ and many jets, with little or no \MET requirement.  We will frequently refer to this as the LSST search, after the authors of~\cite{Lisanti:2011tm}.  See appendix~\ref{app:LSST} for details.
\item To constrain models with high multiplicity requires yet again the high-multiplicity low-\MET searches, combined with the constraints from the ATLAS search for 6-7 high-$p_T$ jets (without a \MET requirement)~\cite{ATLAS-CONF-2013-091} and the control and signal regions of the CMS black hole (BH) search~\cite{Chatrchyan:2013xva}, which we re-analyzed in a very conservative manner for this particular purpose.
\end{itemize}
Few natural models with a gluino below 1~TeV survive this lethal net of searches.\footnote{It is noteworthy that either ATLAS searches alone or CMS searches alone appear sufficient to rule out nearly all models in this class, though with somewhat different methods and coverage at very low \MET.  Said another way, most models that are ruled out at all are covered by two or more LHC searches, which provides confidence that the exclusion is robust.  We will see some exceptions later.}  This is illustrated schematically in figure~\ref{fig:money}. Among models that can survive are ones that lack \MET and tops, and whose jets have a large hierarchy in $p_T$, such that they fail the uniformly hard $p_T$ cuts of the ATLAS 6-7 jets search.

\begin{table}[t]
\begin{center}
\begin{tabular}{|c|c|c|}\hline
Search                       & Data (fb$^{-1}$) & Reference \\\hline
ATLAS 2-6 jets + large \MET  &       20.3       & \cite{ATLAS-CONF-2013-047} \\\hline
ATLAS 7-10 jets + low \MET   &       20.3       & \cite{Aad:2013wta} \\\hline
CMS jets + \MET              &       19.5       & \cite{CMS-PAS-SUS-13-012} \\\hline
ATLAS 6-7 high-$p_T$ jets    &       20.3       & \cite{ATLAS-CONF-2013-091} \\\hline
\multirow{2}{*}{CMS black holes (BH)} & \multirow{2}{*}{12.1} & our re-analysis of~\cite{Chatrchyan:2013xva} \\
& & (see appendix~\ref{app:BH})\\\hline
\multirow{2}{*}{LSST lepton + many jets w/$b$} & 20 & our implementation of~\cite{Lisanti:2011tm} \\
& (expected) & (see appendix~\ref{app:LSST}) \\\hline
\end{tabular}
\end{center}
\caption{Searches most important to our study.  All use the 8~TeV LHC data.  See appendix~\ref{app:methodology} for the details concerning our implementation of these searches, and appendix~\ref{app:simulation} for the description of our detector simulator and its validation.}
\label{tab:searches}
\end{table}

In short, we can summarize the results of this paper with the following simple rule of thumb:
\begin{itemize}
\item[] {\it Using the short list of searches in table \ref{tab:searches}, almost any natural SUSY model where gluino decays frequently produce top quarks, or significant \MET, or a high multiplicity of high-$p_T$ objects is excluded for gluino mass at least up to $\sim$~1~TeV.}
\end{itemize}
Our presentation in the rest of the paper is organized as follows. We start by discussing our conceptual approach in section~\ref{sec:approach}.   Next we discuss the limits on classic SUSY models in section~\ref{sec:settingthestage}.   Then in section~\ref{sec:MET} we study models with few or no top quarks, and a parameter that allows us to exchange \MET for jet multiplicity.  Next in section~\ref{sec:tops}, we explore models with variable numbers of top quarks but no other sources of \MET.  Finally in section~\ref{sec:all-hadronic}, we explore models with no \MET and no top quarks but with various parton multiplicities.  The technical details concerning our simulation method and implementation of specific searches relevant for  sections~\ref{sec:settingthestage}-\ref{sec:all-hadronic} are discussed in appendices~\ref{app:methodology} and~\ref{app:simulation}.   In section~\ref{sec:loopholes} we turn to addressing questions we have left unanswered and closing the   loopholes in our argument; we believe we have left very few open. Our summary in section~\ref{sec:conclusions} contains a few general lessons that we have learned from our study. In particular, we will make some recommendations as to how ATLAS and CMS experimentalists might close remaining gaps and strengthen the limits, via further analysis of the 2011-2012 data set.  Our suggestions should also be relevant for the coming higher-energy run.

\section{The Logic of Our Approach}
\label{sec:approach}

In this work, we aim to build a very general picture of the current constraints on a wide class of natural SUSY models.  Our approach involves neither searching exhaustively through model space nor selecting a few benchmarks motivated by theory.  Instead, our examples are carefully chosen to be maximally illuminating of the origin and nature of the experimental constraints.  Since our logic is somewhat non-traditional, it is important that it be transparent.

In this paper, we will focus solely on signatures arising from gluino pair production.\footnote{Note that we will assume the first-generation squarks to be much heavier than the gluino.  Decoupling these states does not affect naturalness.  The potential impact of the first-generation squark masses on the gluino pair-production cross section is not important for our conclusions, as will be discussed in section~\ref{sec:1st-gen-sq}.} As discussed in section~\ref{sec:introduction}, our analysis relies upon the fact that the gluino has an enormous production cross section for its mass, so that model-independent searches can often detect it.   We also noted there that gluino pair-production in natural SUSY models generally leads to abundant \MET, top quarks and/or high object multiplicity.  To examine the extent to which natural SUSY is ruled out for $m_\go\lesssim 1$~TeV, we study carefully selected models that isolate each one of these three signatures. To be conservative, we also arrange for these models to have no other low-background objects (photons, extra $b$ jets, or extra leptons) in their final states. We will show that these models are excluded typically by two or more very general searches whose event selections are not strongly dictated by the models they aim to constrain.  The model-independent nature of these searches assures the results we obtain apply broadly to many models with the same gluino mass and the same basic signature. While constraints vary somewhat from model to model, limits on $m_\go$ from \MET and/or top quarks signatures extend well above 1~TeV, allowing confidence in a general claim of a 1~TeV limit. For models without \MET or top quarks, the limits are considerably less robust.  Generally they still fall in the 1~TeV range, but there are exceptions we will discuss in detail.

Obviously, for scenarios with large \MET, as is commonly the case in simple $R$-parity conserving models, multiple \MET-based ATLAS and CMS searches provide powerful limits.  We would like to ask to what extent the limits persist in cases where the \MET is smaller.  There are several ways in which a \MET signal can be reduced. The possibility of a ``compressed spectrum'', where the gluino is not much heavier than a stable neutral LSP, is no longer compatible with naturalness given current limits from monojet and jets+MET searches.\footnote{In models with small $m_\go-m_{\tilde\chi_1^0}$, limits from jets+\MET searches~\cite{ATLAS-CONF-2013-047} rule out gluinos decaying to $q\bar q\tilde\chi_1^0$ well above $m_{\tilde \chi_1^0}=400$~GeV, beyond which the higgsinos (which must be at least as heavy as the LSP, by definition) are not really natural anymore. Furthermore, since the gluino decay products in this regime are soft, their details are unimportant, so we can view this result as largely independent of the precise gluino decay mode. Small gaps in~\cite{ATLAS-CONF-2013-047} at $m_\go-m_{\tilde\chi_1^0}<25$~GeV are apparently ruled out by the 7~TeV CMS monojet search~\cite{CMS:2011hpa} up to $\sim 440$~GeV~\cite{Dreiner:2012gx}.  The 8~TeV version of the same search~\cite{CMS:rwa}, which has not yet been reinterpreted in this way, surely extends the exclusion even further.   Note also that a gluino LSP decaying to three jets via $\go\to qqq$ is not natural either; it has already been constrained by one search to near 1~TeV~\cite{ATLAS-CONF-2013-091} (and by a second search to near 800~GeV, with a gap near 700~GeV~\cite{CMS-PAS-EXO-12-049}). \label{footnote:compressed}} Intrinsic \MET  can also be reduced in models with complex cascades, and can be eliminated entirely in models without a stable neutral LSP, such as models with baryonic $R$-parity violation. To smoothly interpolate between the classic large-\MET cases and the RPV-like zero-\MET cases we will study certain ``Hidden Valley''-type models where the amount of \MET can be dialed, by adjusting parameters.  (Models of this type include Stealth Supersymmetry~\cite{Fan:2011yu,Fan:2012jf} as a special case.) We will present results only for the simplest example, in which the \MET is replaced by the smallest possible number of jets, but similar results are found for more complex examples with higher multiplicities.  What will be shown is that limits are powerful in both high-\MET low-multiplicity models and low-\MET high-multiplicity models. Only when the intrinsic \MET has nearly disappeared do the limits weaken significantly.

Next, we turn to top quarks, which are common in gluino decays. By gauge invariance, $\go\to \st t$ cannot be forbidden except by kinematics; and even if $\go\to \st t$ is forbidden kinematically, the decay $\go\to tb\Ho^-$ is allowed as long as $m_\go > m_\Ho+m_t+m_b$.  Once produced, the decay of the top provides many handles for searches: a $b$ jet along with either two extra jets or \MET plus a lepton or hadronic tau.   We will see by studying a variety of examples that in the presence of tops, with or without additional sources of \MET, limits on such models are already very strong, and can be improved further.

We will also investigate how the limits weaken as top quark production is ``diluted'' by the presence of other colored particles to which the gluino may decay.  If $\go\to\st t$ is allowed, it may be diluted to a rather small branching fraction if there are {\it many} other colored particle-sparticle pairs to which the gluino can decay on-shell. If only $\go\to t b\Ho^-$ is allowed,
the branching fraction can be reduced to essentially zero by allowing even just a {\it single} two-body, on-shell decay into a colored particle-sparticle pair.  In some cases the loop decay $\go\to g\Ho^0$ (occasionally left out of simplified models used in the literature) can also dilute the top signal.

\begin{figure}[t]
\begin{center}
\includegraphics[width=0.65\textwidth]{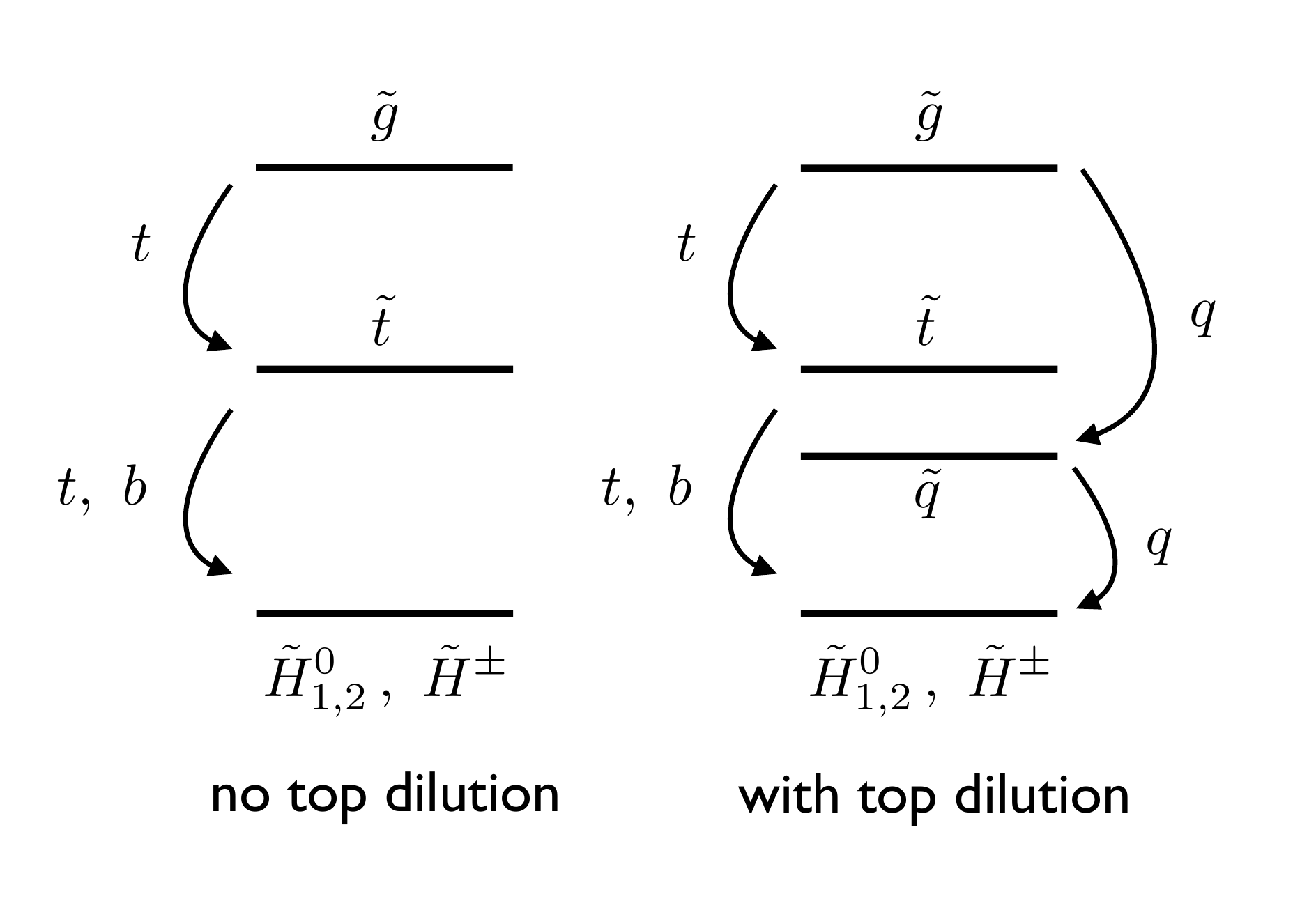}
\caption{Allowing the gluino to decay to other colored states besides the stops (e.g., second-generation squarks) can effectively ``dilute'' the amount of tops generated in gluino events.}
\label{fig:topdil}
\end{center}
\end{figure}

To study the effect of partial or complete dilution of the top quark signal, we will allow the charm and strange squarks to be light. This allows bypassing decays via tops, as shown in figure~\ref{fig:topdil}. First-generation squarks' huge cross sections make models easier to exclude, so it is conservative to take them heavy. At the same time, the second-generation squarks can be much lighter without violation of either flavor constraints or direct search bounds, as has been emphasized recently in~\cite{Mahbubani:2012qq}.  Direct searches are even less constraining in theories with reduced \MET, such as RPV or HV models. We will not assume dilution by bottom squarks because it leads to tops via $\tilde b\to t\Ho^-$ or bottom quarks via $\tilde b\to b\Ho^0$, giving signals which are again easier to exclude than those from second-generation squarks.\footnote{Dilution of $\go\to t\st$ may also occur if there are light non-MSSM colored particle-sparticle multiplets $C$, $\tilde C$, so that $\go\to C\tilde C$ is common.  However, this possibility does not introduce anything qualitatively new. If $C$ and $\tilde C$ both decay,  the final states that arise are captured by the broad range of squark dilution scenarios that we study below. If either is stable, then the resulting $R$-hadrons, produced in gluino decays, would have been observed (as can probably be inferred from the results of~\cite{Chatrchyan:2013oca}, for example).}

Finally, we come to our third signature, high object multiplicity. The rate for high multiplicity events from gluino pair production can rival high-multiplicity QCD rates.  As we will see, gluino decays yielding high multiplicity, such as $\go\to q\bar q\Ho$ followed by $\Ho\to qqq$, are typical in scenarios without \MET.  More complex cascades are not unusual, and top quarks can further increase this multiplicity.  After including initial and final state radiation, final states with at least 8 jets (and often 12 or more) are typical.  To obtain signals with low multiplicity and little or no \MET requires models with non-generic  mass spectra.

We will explore several models with varying jet multiplicity and verify that they are ruled out to $m_\go\sim 1$~TeV except in very special circumstances, where narrow gaps in coverage can extend down to much lower gluino masses.   Existing search strategies are least developed here, and backgrounds are largest, so limits are weaker than for \MET or top signatures.  Still, the searches used are highly model-independent, so we believe that the lessons learned about the overall coverage, the gaps we find, and our suggested approach for filling the gaps apply to this whole class of natural SUSY models, and even beyond.

In summary, it is very difficult  to  create natural SUSY models that substantially avoid the three main signatures of \MET, tops, and high multiplicity (leaving out situations that create other easy signatures, such as $\geq 3$ leptons).  One can reduce \MET by allowing the LSP to decay, as in models with $R$-parity violation, or by elaborating the gluino decay chain, but in either case a high multiplicity of hard objects generally results.   Tops in the gluino decay chain can be suppressed or removed by kinematics, but then more energy is typically available for a \MET signal from the LSP, if the latter is stable, or for hard jets, if it is not.  Finally, in the absence of top quarks and \MET, one can avoid high multiplicity without reintroducing \MET only through some adjustment of the sparticle spectrum that forces some objects to be much softer than others, or leads to merging of two or more partons into a single jet, or via some other special situation.  In these special cases a low jet multiplicity and/or a large hierarchy of jet $p_T$'s can result, causing constraints to weaken sharply, as we will see.

Our results rely on the existence of model-independent searches that cover each of the three natural SUSY signatures. Through application of these searches, we can draw very broad, though certainly not airtight, conclusions about natural SUSY models with an accessible gluino. We will argue that the overwhelming majority of such models with a gluino below 1~TeV are now excluded by LHC data, with small identifiable holes in coverage that we believe can all be addressed, using 2011-2012 data, via improvements in the search program.

\section{Classic SUSY Models}
\label{sec:settingthestage}

In this section we briefly consider ``classic'' SUSY simplified models with an invisible LSP below our naturalness bound of 400~GeV.  Here, gluino pair production leads to many events with substantial \MET, except in regions with special kinematics.\footnote{Note our assumption of a light LSP is important in this section, more so than in the sections that follow.  In ``classic'' SUSY models with a heavy LSP, \MET is strongly reduced, because gluinos produced near threshold provide little  visible energy, while for boosted gluinos the LSP momenta tend to be roughly back-to-back.  Limits on such models often disappear for $m_{\rm LSP} \gtrsim 600$~GeV.}
We will first consider simplified models with no tops (which typically requires heavy stops), in which the gluino decays to the LSP via light quarks and possibly additional particles.  Then we will turn to ``minimal natural SUSY models'' motivated by the original works of~\cite{Dine:1993np, Dimopoulos:1995mi, Cohen:1996vb}, containing only a gluino, at least one third-generation squark, and the higgsinos. In these models, tops are abundant in gluino decays.

The point of our discussion in this section is mainly to remind the reader that models of this type are well excluded by existing searches.  Thereafter, we will modify these models, to study how limits weaken when we reduce the amount of \MET in models without tops (section~\ref{sec:MET}), or reduce the amount of tops in models where tops are the only source of \MET (section~\ref{sec:tops}).  For the details concerning our simulations (based on tools developed previously in~\cite{Kats:2011qh,Evans:2012bf}), see appendices~\ref{app:methodology} and~\ref{app:simulation}. 

\subsection{CMSSM-like models}

We begin by reviewing some of the LHC constraints on gluino pair production in simplified SUSY models in which the gluino predominantly decays to light-flavor quarks, giving final states that typically contain several high $p_T$ objects and a large amount of \MET.

Since in a natural theory we expect that the higgsinos should be lighter than $\sim 400$~GeV, the same upper bound applies to  the LSP, whether or not it is a higgsino. If the LSP is neutral and stable, and therefore invisible, gluino pair production must lead (by color conservation) to at least two and generally at least four jets + \MET, possibly along with other
objects.  If the number of jets is two or four, and the gluino lies well
above the LSP, a large \MET signal results. In this case, limits on
the gluino mass approach the kinematic limit. Specifically, the ATLAS
search for 2-6 jets + \MET (figure~7 of~\cite{ATLAS-CONF-2013-047})
almost completely covers the simplified model in which $\go\to q\bar q
\tilde\chi_{\rm LSP}$, where $q$ is not a $t$ quark, with reach up to
$m_{\go} \sim 1400$~GeV for $m_{\rm LSP}<350$~GeV, degrading to $\sim 1000$~GeV
up to $m_{\rm LSP}<450$~GeV.  A model where, due to a special arrangement of the spectrum, most gluino pair events would contain only one hard jet per gluino is similar to the classic simplified model for light-flavor squarks production with $\sq\to q\chi_{\rm LSP}$, but with a factor of $\sim 7$ larger cross section for the same mass. The limits in this case are similarly strong, as can be inferred from figure~19 of~\cite{ATLAS-CONF-2013-047}.

If the gluino cascade decay involves more steps, it will produce more visible particles during the transition to the LSP.  In this case, the \MET signal is generally reduced, but this loss is compensated by increasing the number of objects in the final state. If these extra transitions yield non-jet visible objects --- leptons, photons, hadronic taus --- the signal is generally easier to observe.  The worst case scenario is to have only jets produced in these decays.  (This is virtually impossible in the MSSM, but need not be so in non-minimal SUSY.)  By reinterpreting an existing search, we can see that, for fixed $m_\go$ and $m_{\rm LSP}$, limits do not degrade significantly for simple cascades.  An 8 jet + \MET state may result if $\go\to q\bar q\tilde\chi_2$ and $\tilde\chi_2\to q\bar q\tilde\chi_1$ (where $\tilde\chi$ refers to either a chargino or neutralino) via on- or off-shell $W$, $Z$ or $h$ emission. For the case where a $W$ is emitted, the limits from jets + \MET searches can be seen in figure~8 of~\cite{ATLAS-CONF-2013-047} and figure~10 of~\cite{Aad:2013wta}. Accounting for the branching fractions of the $W$ (about half of the events contain no leptons) and using figure~21 of~\cite{ATLAS-CONF-2013-047} to rescale the cross section up by a factor of $\sim 2$, we can see the constraints on $m_\go$ and $m_{\rm LSP}$ are comparable to the case without a cascade. Thus we see that even for many-jet + \MET final states (without top quarks), existing searches are successful at constraining gluino production (assuming a naturally light LSP) up to 1~TeV and beyond.

\subsection{Minimal natural SUSY models}
\label{sec:minimal}

Now let us turn to a simplified model that produces tops and \MET.  For a long time, a number of experimental and theoretical considerations have motivated the study of ``minimal'' natural SUSY spectra~\cite{Dine:1993np, Dimopoulos:1995mi, Cohen:1996vb}, where one decouples all states except those that are absolutely necessary for naturalness -- the gluino, stop (and possibly sbottom) and higgsinos.  We will see that (for a stable higgsino) the limits on the gluino are very constraining,  restricting $m_\go$ to 1~TeV or above.

\begin{figure}[t]
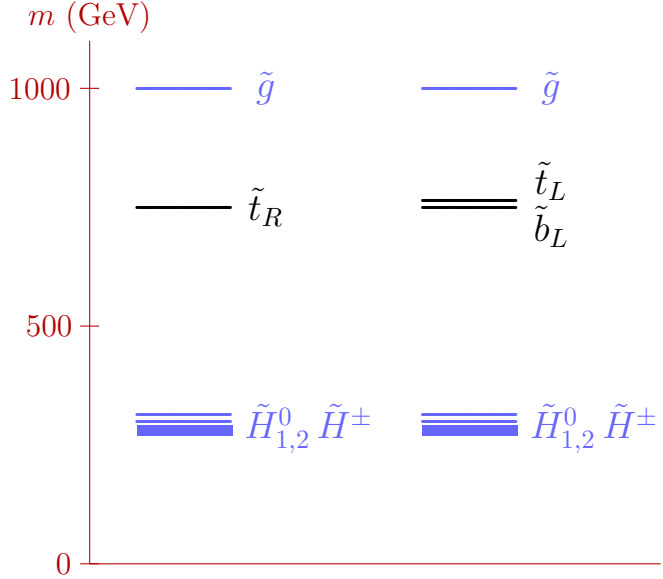

\begin{center}
$$\beginpicture
\setcoordinatesystem units <0.18\tdim,0.18\tdim>
\stpltsmbl
\makered{
\putrule from -100 0 to -100 1100
\putrule from -100 0 to 1100 0
\put {$m$ (GeV)} at -100 1150
\putrule from -120 0 to -80 0
\putrule from -120 500 to -80 500
\putrule from -120 1000 to -80 1000
\put {0} at -155 0
\put {500} at -185 500
\put {1000} at -205 1000
}
\setsolid
\makeblue{\plot 0 1000 200 1000 /
\put {\large$\go$} at 270 1000 }
\plot 0 750 200 750 /
\put {\large$\st_R$} at 270 750
\setsolid
\makeblue{
\plot 0 315 200 315 /
\plot 0 300 200 300 /}
\linethickness=4pt
\makeblue{
\putrule from 0 280 to 200 280
\put {\large$\Ho^0_{1,2}\,\Ho^\pm$} at 360 285 }
\setsolid
\makeblue{\plot 600 1000 800 1000 /
\put {\large$\go$} at 870 1000 }
\plot 600 750 800 750 /
\plot 600 765 800 765 /
\put {\large$\sbo_L$} at 870 710
\put {\large$\st_L$} at 870 805
\setsolid
\makeblue{
\plot 600 315 800 315 /
\plot 600 300 800 300 /}
\linethickness=4pt
\makeblue{
\putrule from 600 280 to 800 280
\put {\large$\Ho^0_{1,2}\,\Ho^\pm$} at 960 285 }
\color[rgb]{0,0,0}
\endpicture$$
\caption{Mass spectra of the models studied in this section.  A thick line indicates that the lighter $\Ho^0$ is stable.  Blue lines indicate masses that we will vary.}
\label{fig:spectraCh4}
\end{center}
\end{figure}

We will consider two benchmark simplified models containing a gluino, the higgsinos, and either the $\st_R$ or the $\st_L,\sbo_L$ multiplet; see figure~\ref{fig:spectraCh4}.  In each case there are three mass parameters,
\begin{itemize}
\item $m_\go$, $m_{\st_R}$, $m_{\Ho}$  (we will refer to this as the $\go-\st_R-\Ho$ model)
\item $m_\go$, $m_{\tilde Q_3}$, $m_{\Ho}$
\end{itemize}
and we set $\tan\beta = 10$. Naturalness leads us to expect both the right-handed stop and the left-handed stop-sbottom multiplet to be relatively light, but it is possible that one of them dominates the gluino decays because it happens to be much lighter than the other, so we will focus on one and then the other to isolate the different types of decays that can occur.\footnote{If both multiplets take part in gluino decays, and there is significant mixing between the stops, then transitions between these states, with $W$, $Z$ or $h$ emission, can be important. Except for the occasional leptons that may make these signals easier to detect, we do not expect the presence of such events to greatly affect our conclusions.}
The masses of the two neutral and the charged higgsino are set by a single parameter, $m_{\Ho^0_1},\, m_{\Ho^0_2},\, m_{\Ho^\pm} \approx \mu$ (to which we will refer as $m_\Ho$); they are split only by mixing with gauginos, and for our choice of parameters the splitting is always less than 5~GeV. This near-degeneracy is sufficient to ensure that particles emitted in transitions between the different higgsino states are essentially unobservable.  The lightest higgsino is stable and provides a source of large \MET, as long as $m_\go-m_\Ho$ is not too small.

The main gluino decay chain in the first model is that of eq.~(\ref{eqn:gluinodecays}), namely $\tilde g \to \tilde t t$, followed   by $\st \to b \Ho^+$ or $t\Ho^0$.  Note that in many experimental searches, limits are placed on a slightly different scenario (gluino-stop-bino) so gluino cascades with $\tilde t\to b\Ho^+$ are not probed.  In the second model, the lightest sbottom  is also present, so $\tilde g\to \tilde b b$ is present too, with predominantly $\tilde b\to t\Ho^-$ (because the lightest sbottom is assumed to be mostly left-handed), if allowed by phase space.

\begin{figure}[t]
\begin{center}
\includegraphics[width=0.47\textwidth]{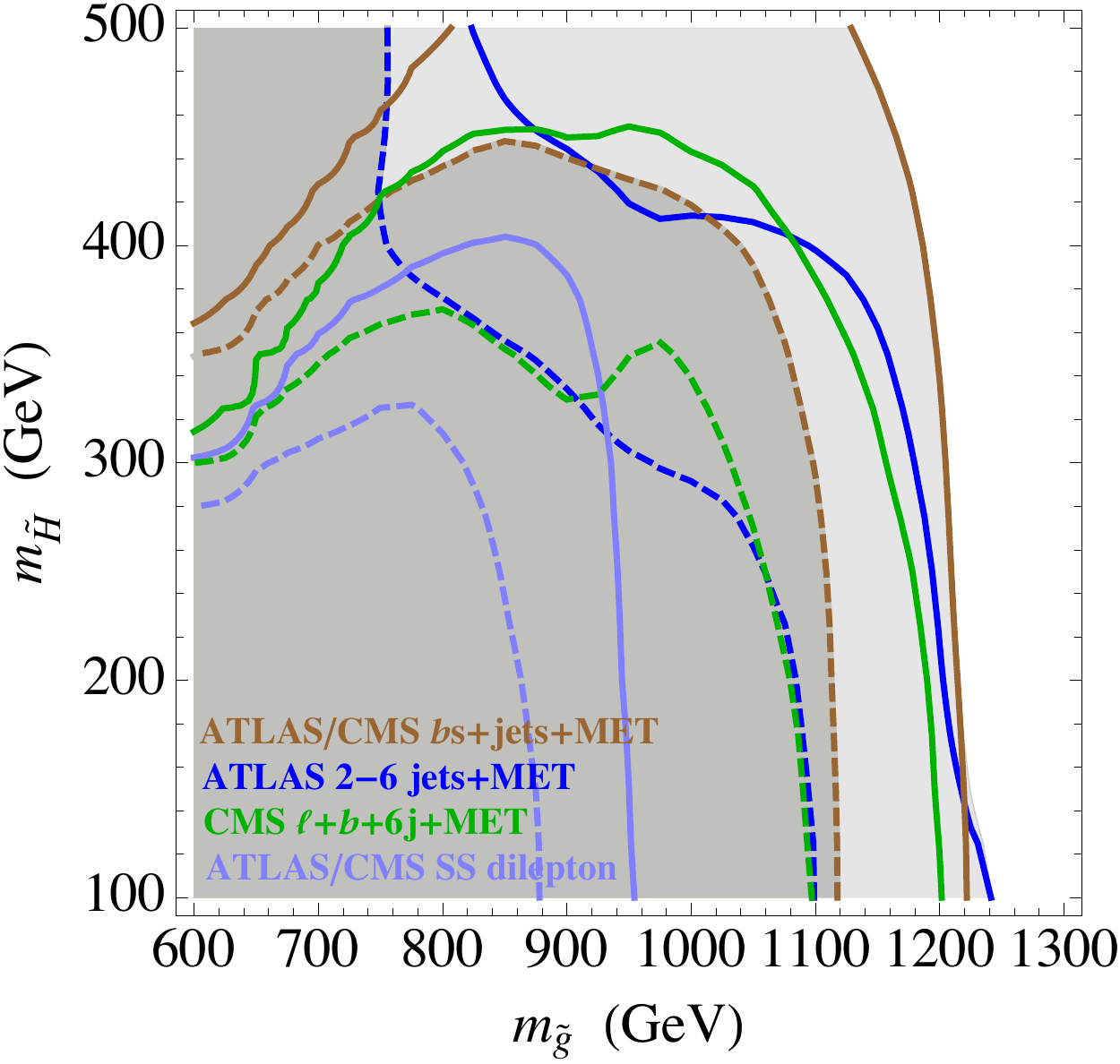}\qq
\includegraphics[width=0.47\textwidth]{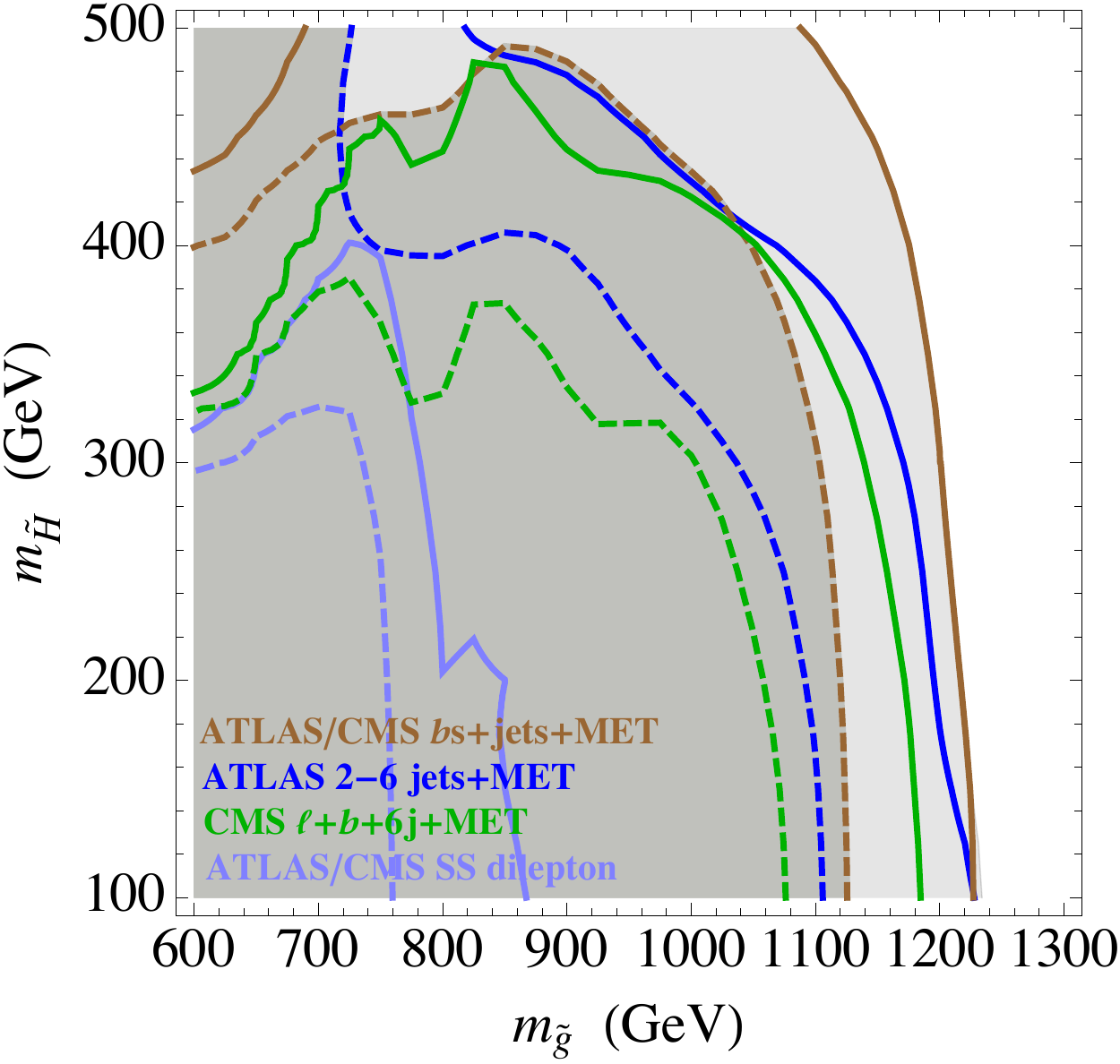}
\caption{Limits on scenarios with a stable higgsino LSP, in the gluino-higgsino mass plane, with a right-handed stop (left plot) or a left-handed stop and sbottom (right plot) at 750~GeV, as described in figure~\ref{fig:spectraCh4}. The searches shown are ATLAS 2-6 jets+\MET~\cite{ATLAS-CONF-2013-047}, ATLAS and CMS $b$+jets+\MET searches~\cite{ATLAS:2012ksq,Chatrchyan:2013wxa} (combined into one contour), CMS $\ell$+$b$+$6j$+\MET~\cite{CMS:wwa}, and the ATLAS and CMS same-sign dilepton searches~\cite{ATLAS:2013tma,CMS-PAS-SUS-13-013,Chatrchyan:2012paa,CMS-PAS-B2G-12-015} (combined into one contour). Dashed lines indicate regions in which the exclusion is stronger than a factor of~2.}
\label{fig:go_vs_ho_min}
\end{center}
\end{figure}

Figure~\ref{fig:go_vs_ho_min} presents our limits from various searches as a function of the higgsino and gluino masses. The lightest stop mass ($m_{\st_R}$ on the left and $m_{\tilde Q_3}$ on the right) is fixed at 750~GeV, which is within the range of naturalness we consider, and lies slightly out of reach of current limits on direct production of stops and sbottoms with simple decays, even for a massless LSP.  The searches shown in the plots are representative of the most constraining ones among those we analyzed (see appendices~\ref{app:methodology} and~\ref{app:simulation} for details of how these limits are derived). {\it Nominal} limits are indicated in solid lines, with {\it robust} limits (exclusion by a factor of 2) shown in dashed lines.\footnote{Given the success of our simulations in reproducing experimental exclusion limits, we believe our results should be accurate to within a factor of two, so that the true exclusion limits obtained by the experiments would almost certainly be stronger than  our robust exclusion limits.  Note also that other  models that are similar to these but have slightly different branching fractions may easily populate a final state by amounts that differ by factors of order 2 or so; thus robust exclusion of a specific model assures that most similar models will also be excluded.}

As expected, since the events contain copious top quarks, high-$p_T$ ($b$-)jets, and \MET, the constraints on these scenarios, for generic mass splittings, are quite stringent (generally around 1.2~TeV, which is quite close to the kinematic limit). We display the limits from the ATLAS and CMS searches for multiple $b$s+jets+\MET~\cite{ATLAS:2012ksq,Chatrchyan:2013wxa}, the ATLAS search for 2-6 jets+\MET~\cite{ATLAS-CONF-2013-047} and the CMS search for $\ell$+$b$+$6j$+\MET~\cite{CMS:wwa}. (The other searches from table~\ref{tab:searches} are not shown for clarity.) 
Across nearly all of the natural region with $m_\Ho<400$~GeV, at least one of these searches is providing \emph{robust} exclusion (indicated by the region of darker shading) for gluino masses nearly up to the kinematic limit.  Notice that limits from same-sign (SS) dilepton searches~\cite{ATLAS:2013tma,CMS-PAS-SUS-13-013,Chatrchyan:2012paa,CMS-PAS-B2G-12-015} -- sometimes regarded as a good path for discovering natural SUSY spectra -- are not as strong, due to the low probability for obtaining a SS pair of leptons from top decays. Although $b$-jet based searches set the strongest limits for these simplified models,  the ATLAS 2-6 jets+\MET search with no $b$-tag requirement from table~\ref{tab:searches}  nominally excludes the entire region of $m_\go<1000$~GeV, $m_\Ho<400$~GeV (and beyond). This illustrates both the usefulness and the limitations of our short list of searches in table~\ref{tab:searches}. While for a given model the searches in table~\ref{tab:searches} may not set the best possible limit, in general they will cover natural SUSY up to gluino masses of $\sim 1$~TeV.

\section{Trading \MET for Extra Jets}
\label{sec:MET}

We argued in the previous section that models with a lot of \MET from a stable LSP (such as the higgsino) receive very strong constraints on the gluino mass, approaching the kinematic limit ($\sim 1.2-1.4$~TeV). Now we will ask by how much the \MET must be reduced for the limit on the gluino mass to weaken appreciably.  We will show that only a sharp reduction of the \MET can degrade the limits from \MET-based searches.  However, when this happens, the extra jets that replace the \MET give rise to limits from searches that do not require \MET at all.

\subsection{Benchmark scenario}

\begin{figure}[t]
\begin{center}
\includegraphics[width=0.5\textwidth]{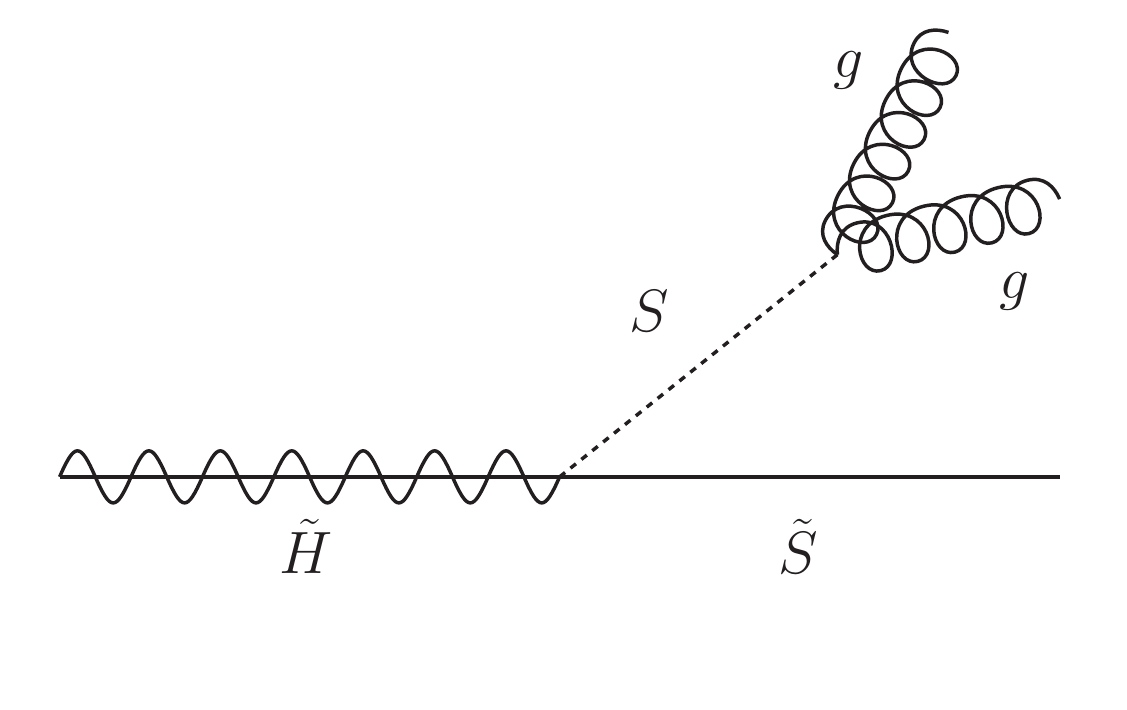}
\qq\qq
$\beginpicture
\setcoordinatesystem units <0.18\tdim,0.18\tdim>
\stpltsmbl
\makered{
\putrule from -100 0 to -100 1100
\putrule from -100 0 to 530 0
\put {$m$ (GeV)} at -100 1150
\putrule from -120 0 to -80 0
\putrule from -120 500 to -80 500
\putrule from -120 1000 to -80 1000
\put {0} at -155 0
\put {500} at -185 500
\put {1000} at -205 1000 }
\setsolid
\makeblue{\plot 0 1000 200 1000 /
\put {\large$\go$} at 270 1000 }
\plot 0 600 200 600 /
\put {\large$\st_R$} at 270 600
\plot 0 500 200 500 /
\put {\large$\sq$} at 270 500
\plot 0 215 200 215 /
\plot 0 200 200 200 /
\plot 0 185 200 185 /
\put {\large$\Ho^0_{1,2}\,\Ho^\pm$} at 360 220
\makeblue{\plot 0 130 200 130 /
\put {\large$S$} at 260 132 }
\linethickness=4pt
\makeblue{\putrule from 0 60 to 200 60
\put {\large$\tilde S$} at 260 50 }
\color[rgb]{0,0,0}
\endpicture$\qq\qq\qq
\caption{{\bf Left:} Higgsino decay in the ``minimal Hidden Valley'' model used in this section to interpolate between large and zero \MET.   {\bf Right:} Mass spectrum of the model studied in this section.  The label $\sq$ indicates the four second-generation squarks ($\tilde c_R$, $\tilde s_R$, $\tilde c_L$, $\tilde s_L$).  A thick line indicates that $\tilde S$ is stable.  Blue lines indicate masses that we will vary.}
\label{fig:spectraCh5}
\end{center}
\end{figure}
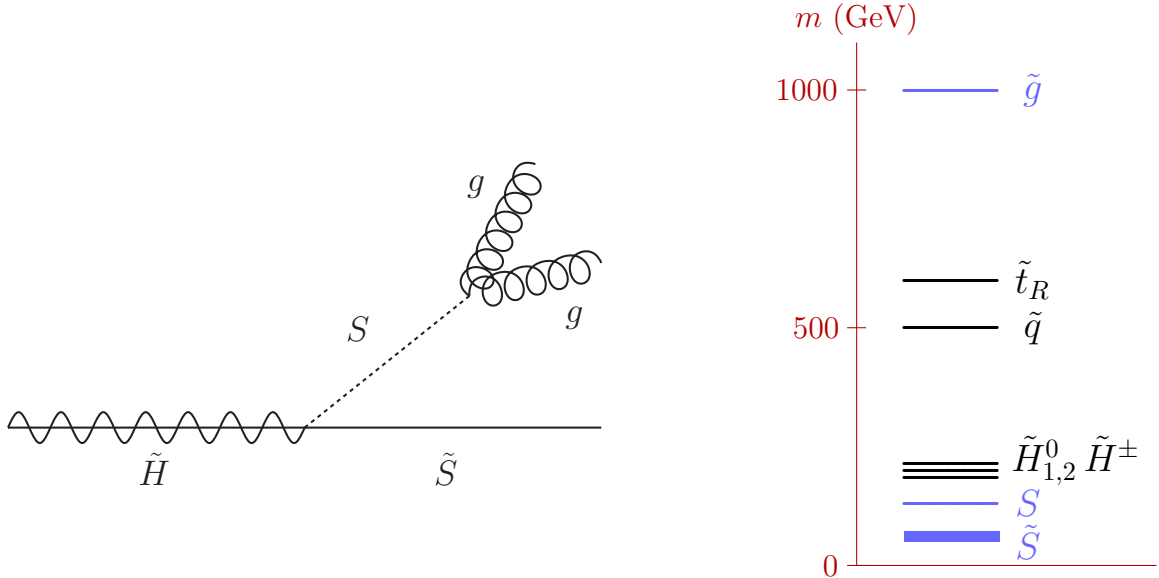

In this section, we will utilize a model where the amount of
\MET may be smoothly adjusted between that of a classic MSSM scenario
with large \MET and one which has virtually no intrinsic \MET.
Since we are looking for gaps in the LHC search strategy, we want to examine the
limits on the most conservative cases.  We expect these to be found when the
reduction of \MET only increases the number of light-flavor and gluon jets.
Any similar reduction in which leptons, photons, taus, $b$'s, or $W/Z/h$ are commonly produced at the expense of the \MET should lead to more easily excluded models, since such signals will have smaller backgrounds.\footnote{Of course, if searches for such signals have not been done or updated yet, the {\it current} limits will not be as strong as they potentially could be.  A deeper assessment of possible loopholes invoking these particles will be presented in section~\ref{sec:loopholes}.} This motivates us to consider a Hidden Valley (HV) type model in which the SM LSP decays promptly, {\it preserving} $R$-parity, into low-mass SM-singlets, some of which decay visibly (and promptly) to jets. An example of this, which we will make use of in this section, is shown in figure~\ref{fig:spectraCh5} (left). Also, since we are interested in isolating the effect of \MET on exclusion limits, we would like to minimize MSSM sources of tops and $b$'s.  (We will return top quarks to the final states in section~\ref{sec:tops}, to explore the effect of their presence.) We will modify the minimal $\go-\st_R-\Ho$  benchmark model of section~\ref{sec:minimal} to achieve these goals.

The full mass spectrum of the model is summarized in figure~\ref{fig:spectraCh5} (right). As in section~\ref{sec:minimal}, the spectrum contains a gluino $\go$ (of variable mass), a right-handed stop $\st$ (which we will take at 600~GeV), and the higgsinos $\Ho$ (which we will fix at 200~GeV).  We have already assumed the other third-generation squarks are somewhat heavier, but to reduce top quarks in the gluino decays we have to do something about the decay $\go\to\st\bar t$, as well as its version with an off-shell stop $\go\to t\bar b\Ho^-$.  We ``dilute'' these decay modes by adding light second-generation squarks (recall figure~\ref{fig:topdil} from section~\ref{sec:approach}).  With the two charm and the two strange squarks comparable in mass to $\st_R$, the dilution is roughly 80\% when $\go\to \st \bar t$ is kinematically allowed.  This number can be smaller or larger depending on the squark masses and certain mixing angles.  If $\go\to\st\bar t$ is kinematically forbidden, then having just one second-generation squark light is enough for dilution (of the 3-body decay $\go\to t\bar b\Ho^-$) to reach $\sim100\%$. For our immediate purposes, we will place second-generation squarks $\sq$ at 500~GeV.  This greatly reduces the number of tops, eliminating them altogether for $m_\go < 775$~GeV.

Next, to reduce the \MET, we expand the MSSM by adding a minimal HV sector (referred to in the following as the mHV model) containing a singlet scalar $S$ and its fermionic superpartner $\tilde S$, with masses $m_S$ and $m_{\tilde S}$. As shown in figure~\ref{fig:spectraCh5}, we require the neutral LSP of the SM, $\Ho$, to decay as $\Ho\to S \tilde S$, and $S$ to decay as $S\to gg$. Meanwhile, $\tilde S$ is the true LSP, and is stable and invisible. The limit $m_S\to 0, m_{\tilde S}\to m_{\Ho}$ gives $\Ho$ an essentially invisible decay, in which case the model retains its MSSM-like large \MET signal.  The opposite limit, $m_S\to m_{\Ho}$, $m_{\tilde S}\to 0$, gives a fully visible decay and very little if any intrinsic \MET.

\subsection{Results}

\begin{figure}[t]
\begin{center}
\includegraphics[width=0.47\textwidth]{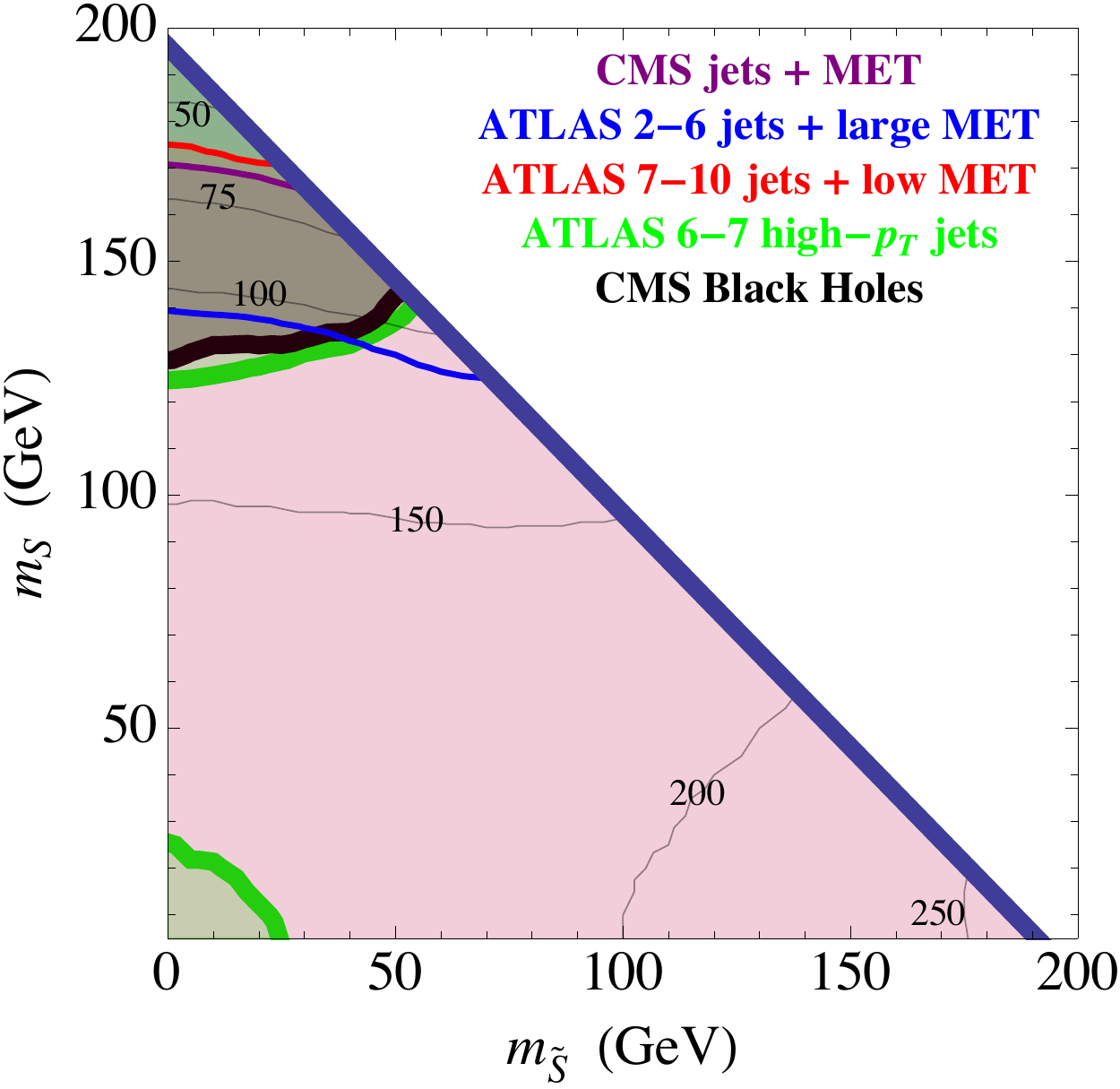}\qq
\includegraphics[width=0.47\textwidth]{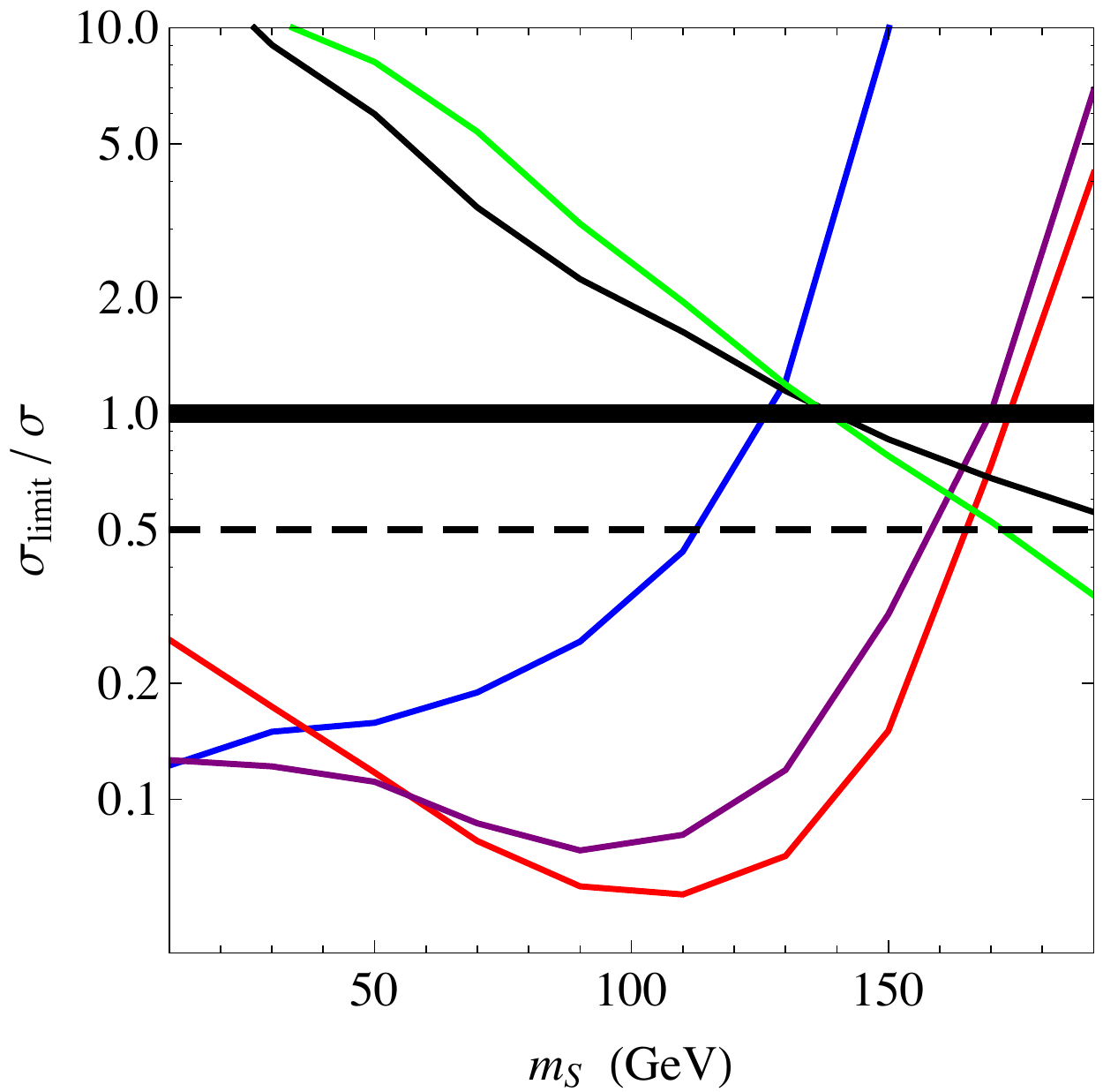}
\caption{Limits on the mHV model with $m_\go = 750$~GeV, $m_\st = 600$~GeV, $m_\sq = 500$~GeV, $m_\Ho = 200$~GeV (see figure~\ref{fig:spectraCh5}). The left plot shows the limits in the $(m_S,m_{\tilde S})$ plane, with light gray contours showing the average \MET (in GeV, including jet energy mismeasurement).  The \MET-based searches~\cite{ATLAS-CONF-2013-047,CMS-PAS-SUS-13-012,Aad:2013wta} (thin colored lines) exclude the lower region (nearly the entire plot), while the CMS black hole search~\cite{Chatrchyan:2013xva} and ATLAS 6-7 high-$p_T$ jets search~\cite{ATLAS-CONF-2013-091} (thick colored lines) exclude the upper region and the lower-left corner. The right plot focuses on the diagonal line $m_S + m_{\tilde S} = 190$~GeV and shows the exclusion power of the various searches. Nominal (robust) exclusion occurs where the line corresponding to the search drops below 1 (0.5).}
\label{fig:go_750_ho_200_HV1}
\end{center}
\end{figure}

In figure~\ref{fig:go_750_ho_200_HV1} (left), we show a plot that explores the mHV model as a function of the masses $m_S$ and $m_{\tilde S}$, at fixed gluino mass of 750~GeV.  At this value of $m_\go$, all gluinos decay as $\go\to j \sq \to jj \Ho \to jj S \tilde S \to jjjj \tilde S$; dilution of decays that produce top quarks is complete.  The triangle is the kinematically allowed region, where $m_S+m_{\tilde S}<m_\Ho$.  The contours labeled by numbers are contours of constant average \MET, after accounting for jet-energy mismeasurement.  They show that the lower-right corner represents the MSSM-like region, where there is no loss of \MET, while the upper-left is the RPV-like region where the \MET has been entirely replaced by two jets per gluino.  Note that even without any intrinsic \MET, jet energy mismeasurement in high multiplicity events leads to measured ``fake'' \MET of order 50~GeV for gluinos of this mass.

\begin{figure}[t]
\begin{center}
\includegraphics[width=0.47\textwidth]{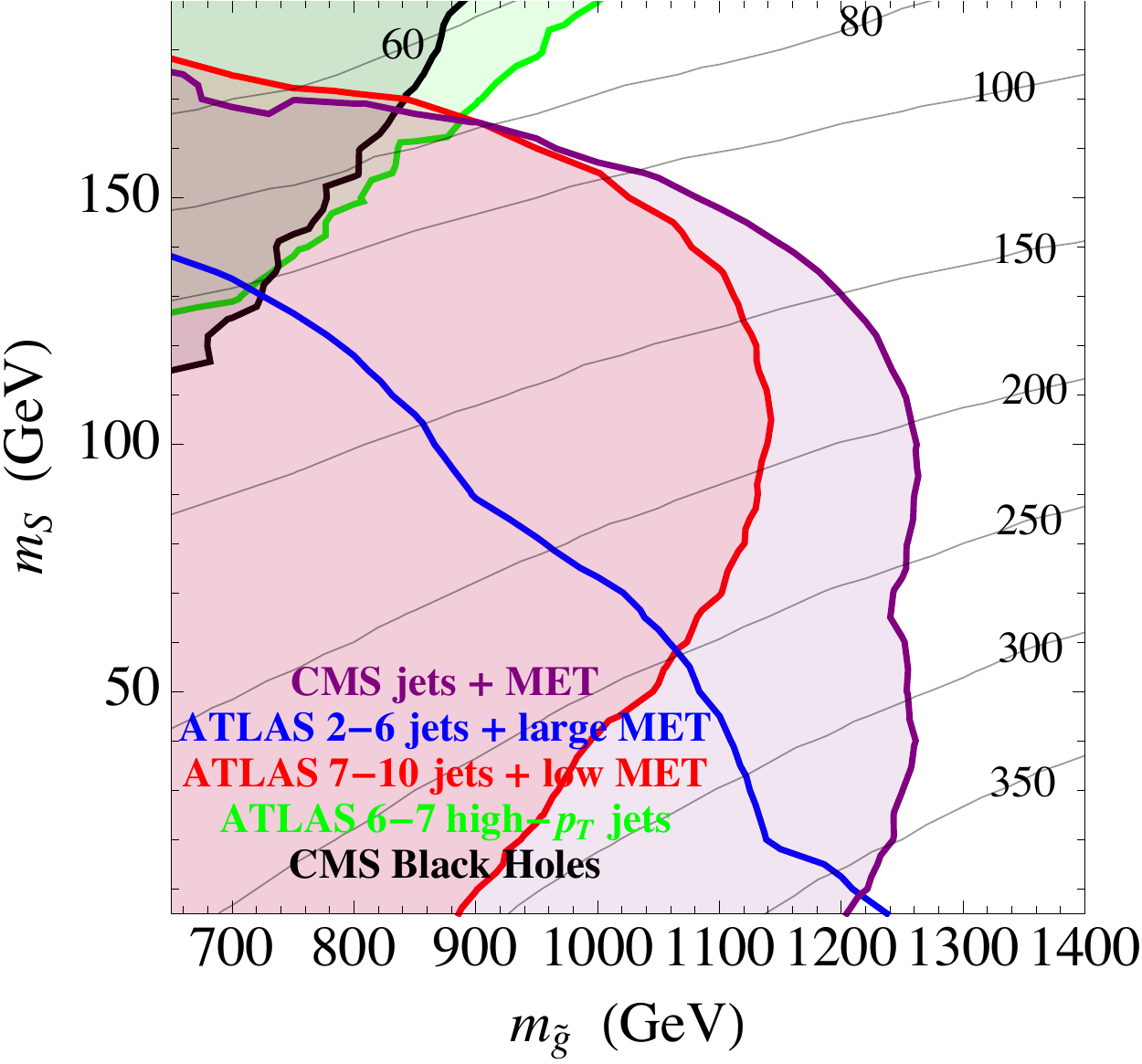}\qq
\includegraphics[width=0.47\textwidth]{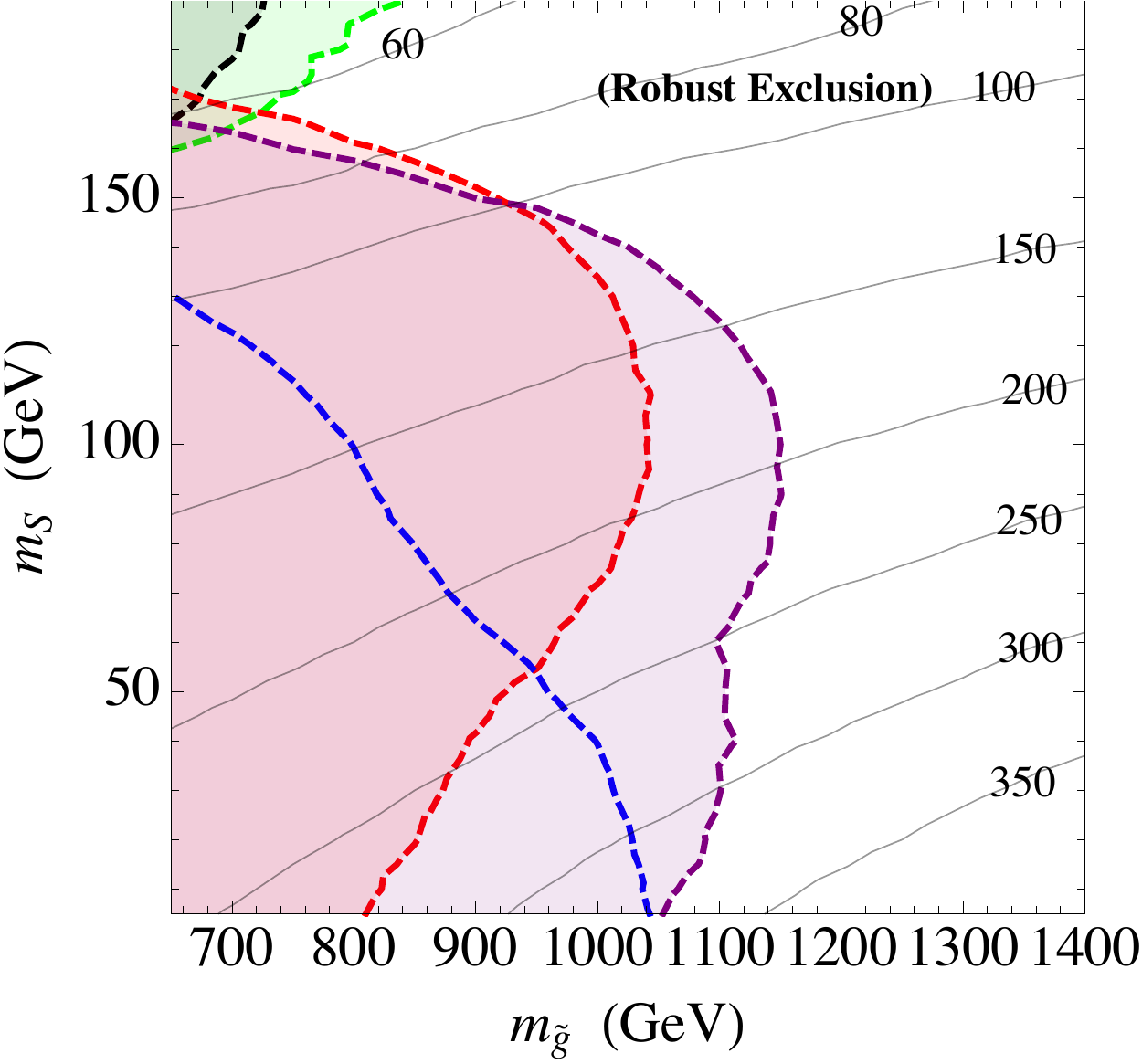}
\caption{Limits on the mHV model as a function of the gluino and the HV scalar masses, for $m_\st = 600$~GeV, $m_\sq = 500$~GeV, $m_\Ho = 200$~GeV, $m_S + m_{\tilde S} = 190$~GeV (see figure~\ref{fig:spectraCh5}). The curves at left are for nominal exclusion; those on the right are for robust exclusion. The light gray numbered contours inside the plots show the average \MET (in GeV).}
\label{fig:go_ho_200_HV1}
\end{center}
\end{figure}

The colored lines on the plot represent our estimates for limits from various experimental searches. The \MET-based searches (thin red, purple, and blue lines) exclude downwards in the plot, while the non-\MET-based searches (thick black and green lines) have coverage in the upper-left and lower-left regions. So we see in this plot that these two classes of searches are complementary:
\begin{itemize}
\item Standard low-multiplicity high-\MET searches, such as the ATLAS one~\cite{ATLAS-CONF-2013-047} we show here in blue, become ineffective in the upper-left (RPV-like) region, though at this low gluino mass they still do rather well across most of the kinematic triangle.
\item The gap in the upper left is mostly filled by the ATLAS and CMS high-multiplicity low-\MET searches (in red and purple, respectively)~\cite{Aad:2013wta,CMS-PAS-SUS-13-012}, except for the extreme corner.\footnote{In the region with very low \MET, the ATLAS and CMS high-multiplicity \MET-based searches (red and purple curves) must rely on the tail of the \MET distribution, and become sensitive to our modeling of jet energy mismeasurement.  To avoid these tails, we allow exclusion only by bins in which the signal efficiency is above $\sim 10^{-4}$ (for more details, see appendix~\ref{app:simulation}). This affects the \MET-based searches in the very low-\MET regions of figures~\ref{fig:go_750_ho_200_HV1}, \ref{fig:go_ho_200_HV1} and~\ref{fig:topdilution-grid}; but in these regions, the searches that do not require \MET, and are not dependent on jet energy mismeasurement, provide stronger limits anyway. Thus our overall results do not depend upon this issue.}
\item The upper left corner is comfortably covered by the ATLAS 6-7 high-$p_T$ jet search~\cite{ATLAS-CONF-2013-091} (green) and by our conservative application of the CMS BH search~\cite{Chatrchyan:2013xva} (black), see appendix~\ref{app:BH} for details. (The ATLAS search is also somewhat effective in the lower-left corner, where the two gluons from a very light and energetic $S$ often form a single energetic jet.)
\end{itemize}

Figure~\ref{fig:go_750_ho_200_HV1} (right) illustrates these same lessons in more detail by taking a 1D slice of the 2D plot.  It shows how effective the searches are as a function of $m_S$, for $m_S+m_{\tilde S}$ held fixed at $190\mbox{ GeV}=m_\Ho-10$~GeV (i.e., along a diagonal line from near the lower-right corner of the triangle to near the upper-left corner), so that the left side of the plot is the MSSM-like high-\MET region and the right side is the RPV-like low-\MET region.  Each line represents the effectiveness of a particular search; if a curve lies below the solid black line at 1 (0.5), that means the search nominally (robustly) excludes the signal. Note the powerful exclusion (by a factor of $\sim 10$) by the \MET-based searches across most of the $m_S$ range; and where the high-multiplicity low-\MET searches weaken in the very low-\MET region, the ATLAS 6-7 jet search and the CMS BH search come into play and deliver a robust limit.

In figure~\ref{fig:go_ho_200_HV1}, we look at the limits as a function of the gluino mass, while again varying the scalar mass $m_S$ along the line $m_S+m_{\tilde S}=190$~GeV. The numbers and color-coded curves are as before. The curves in the left plot indicate that gluinos are excluded up to $\sim 1.25$~TeV at moderate to high \MET, but this limit drops slightly below 1~TeV at very low \MET, for $m_S \gtrsim 160$~GeV.  If we demand robust exclusion (by a factor of 2), the right plot shows this limit dropping to near 700~GeV; TeV-scale robust limits survive up to $m_S\sim 145$~GeV.

A larger choice of higgsino mass does not dramatically change the situation.  It  weakens limits a bit in the high-\MET region (analogous to increasing $m_\Ho$ in figure~\ref{fig:go_vs_ho_min}), but strengthens limits elsewhere, because the heavier higgsino produces higher $p_T$ jets in its decays.

To summarize, we have argued in this section, using the mHV model as a benchmark, that only a very significant reduction of \MET will degrade the limits on $m_\go$ from \MET-based searches down below  1~TeV. And when this happens, complementary limits are  provided by searches that do not require \MET, though these are not entirely robust in the 1~TeV range.  Although the evidence we have presented here is based on a single benchmark model,  we have studied more complex models, and found similar results: the strong limits from the \MET-based searches are lost at low enough \MET, but there the ATLAS 6-7 high-$p_T$ jet search and the CMS BH search provide significant limits. Nevertheless, these limits can sometimes fall well below 1~TeV, an issue we will return to in section~\ref{sec:all-hadronic}.

\section{Constraining Top Quarks}
\label{sec:tops}

As we discussed in section~\ref{sec:approach}, gluino decays in natural SUSY models commonly produce top quarks. In scenarios with tops and large \MET from the LSP, the standard \MET-based searches work well, as we have already seen in the context of the $\go-\st-\Ho$ models of section~\ref{sec:minimal}. On the other hand, in scenarios with little or no \MET from the LSP, the object multiplicity in gluino decays is typically large.  This fact, together with the $b$-jets from top decays and the leptons and/or small \MET from the leptonic decays of the tops, can be used to constrain such scenarios, as we will see.  Moreover, varying the branching fraction for gluinos to produce top quarks  provides an axis in the space of signatures that is roughly orthogonal to the axis in which \MET is varied.  We will explore this axis later in this section.

While there exist several searches (for heavy exotic quarks) that target events with tops and/or $W$'s without additional sources of \MET, these searches are not sufficiently generic for our purposes, as is discussed in appendix~\ref{app:LSST}. Therefore, our goal in this section will be twofold. As in the previous sections, we will examine which of the existing general-purpose searches happen to be effective, and to what extent. In addition, we will discuss what kind of a search \emph{would} be, according to our understanding, most relevant to these scenarios and estimate its potential reach.

Our focus will be on searches aimed at 1-lepton events (where the lepton is $e$ or $\mu$), or 0-lepton events with a small amount of \MET (which arise in the case of a hadronically decaying $\tau$, or a lost $e$ or $\mu$). Searches for the less abundant dilepton events are not competitive for these types of signals.  Opposite-sign dileptons suffer from the same $t\bar t$ backgrounds as do the single-lepton events, and from lower statistics. Same-sign dileptons have lower backgrounds, but the limits from existing SS dilepton searches do not match those of other searches, perhaps partly because existing searches do not require the high jet multiplicity characteristic of low-\MET models.  If there are additional sources of leptons, such as cascades involving sleptons, then the models are easily ruled out by multi-lepton searches~\cite{Chatrchyan:2013xsw,CMS-PAS-SUS-13-008,CMS-PAS-SUS-13-010,ATLAS:2012epx,ATLAS:2013rla,ATLAS:2013qla}; we will consider models where this is not the case.

\subsection{Relevant searches}

In gluino pair events that produce top quarks (and not much \MET), it is common for the signal of a lepton, a $b$ quark, and multiple jets (typically 8 or more) to reach or exceed the dominant $t\bar t$+jets background.  A search sensitive to such events, as suggested by LSST~\cite{Lisanti:2011tm}, is therefore highly motivated. Since such a search has not yet been carried out by ATLAS or CMS, we will present expected limits from our own implementation of such a search. The details of our proposed search are discussed in appendix~\ref{app:LSST}.

The two most relevant existing studies (established by recasting all of the searches shown in table~\ref{tab:all-searches} plus many more)  are the searches for many jets plus low \MET by ATLAS~\cite{Aad:2013wta} and CMS~\cite{CMS-PAS-SUS-13-012}. These searches gain their sensitivity by using the low-to-moderate \MET from the neutrino produced in top decays. As we will see, they are comparable in sensitivity to the LSST search as we have estimated it. This is at first glance surprising, since these searches veto on leptons, and neutrinos from top decays are always produced in association with a lepton.  However, hadronically decaying taus are not being vetoed, and electrons and muons sometimes fail identification requirements or are lost inside a jet in this high multiplicity environment, so a substantial number of events pass the event selection for these searches. (Similar observations about these kinds of searches have been made before, see~\cite{Asano:2012gj}.)

Is it possible that systematic uncertainties regarding lepton isolation and identification requirements could be enhanced for signals where the lepton must be lost? A partial cross-check is given by ATLAS's interpretation of its search~\cite{Aad:2013wta} in terms of an RPV scenario where $\go\to\st t$ and $\st\to bj$. ATLAS finds a limit on the gluino mass that nearly reaches 1~TeV, which agrees quite closely with our own estimate for this signal.  We view this as providing evidence that we may apply this search for constraining tops with some confidence.

We should note though, that reinterpretation of these jets + low \MET searches~\cite{Aad:2013wta,CMS-PAS-SUS-13-012} for scenarios where the only source of \MET is $W$s from top decays has a caveat. These searches use lepton + many jets samples as  control regions for estimating the contribution of the $t\bar t$ and $W$+jets backgrounds to their signal regions.  A $\tilde g\to t+X$ signal can contaminate these control regions and potentially limit the reach of the searches, especially if the relative contributions to the control and signal regions are similar to those of the SM backgrounds. Both searches apply an upper bound on $m_T$ in the control regions in order to decrease signal contamination, but this is ineffective for signals where all the \MET is coming from a $W$. This might be avoidable only with a larger set of control regions and use of both 0- and 1-lepton bins. Note that we have not included the effects of control region contamination in deriving limits from the ATLAS and CMS many jets + low-\MET searches~\cite{Aad:2013wta,CMS-PAS-SUS-13-012}, so we might be overestimating their exclusion range in some of the cases that we will study in this section.

Altogether, we see both complementarity and valuable redundancy in having both leptonic and low-\MET non-leptonic high-multiplicity searches.  First, the two are usually comparable in reach, but have very different sources of systematic uncertainty.   We will see cases (including one similar to the RPV scenario just mentioned) in which the lepton-based search appears somewhat more powerful. We also expect that there are models (perhaps ones with partial dilution of the top signal and/or small \MET from the LSP) where the reverse is true.   Moreover, since they rely on non-overlapping event samples, adding a one-lepton search to a zero-lepton search may significantly enhance the overall reach.

\begin{figure}[t]
\begin{center}
$$\beginpicture
\setcoordinatesystem units <0.18\tdim,0.18\tdim>
\stpltsmbl
\makered{
\putrule from -100 0 to -100 1100
\putrule from -100 0 to 1920 0
\put {$m$ (GeV)} at -100 1150
\putrule from -120 0 to -80 0
\putrule from -120 500 to -80 500
\putrule from -120 1000 to -80 1000
\put {0} at -155 0
\put {500} at -185 500
\put {1000} at -205 1000 }
\put {\large Figure~\ref{fig:go_st_350_ho_200}} at 450 1000
\put {\large Figure~\ref{fig:topdilution-grid}} at 1550 1000
\setsolid
\makeblue{\plot 0 800 200 800 /
\put {\large$\go$} at 270 800 }
\plot 0 350 200 350 /
\put {\large$\st_R$} at 270 350
\setsolid
\plot 0 215 200 215 /
\plot 0 200 200 200 /
\plot 0 185 200 185 /
\put {\large$\Ho^0_{1,2}\,\Ho^\pm$} at 360 195
\setsolid
\makeblue{\plot 600 800 800 800 /
\put {\large$\go$} at 870 800 }
\setdashes
\plot 600 315 800 315 /
\plot 600 300 800 300 /
\plot 600 285 800 285 /
\put {\large$\Ho^0_{1,2}\,\Ho^\pm$} at 960 295
\setsolid
\plot 600 100 800 100 /
\put {\large$\st_R$} at 870 100
\setsolid
\makeblue{\plot 1400 800 1600 800 /
\put {\large$\go$} at 1670 800 }
\makeblue{\plot 1400 620 1600 620 /
\put {\large$\sq$} at 1670 620 }
\plot 1400 500 1600 500 /
\put {\large$\st_R$} at 1670 500
\setsolid
\plot 1400 215 1600 215 /
\plot 1400 200 1600 200 /
\plot 1400 185 1600 185 /
\put {\large$\Ho^0_{1,2}\,\Ho^\pm$} at 1760 195
\color[rgb]{0,0,0}
\endpicture$$
\caption{Mass spectra of the models studied in this section.  The label $\sq$ indicates the four second-generation squarks ($\tilde c_R$, $\tilde s_R$, $\tilde c_L$, $\tilde s_L$).  The LSP decays to jets via RPV in all cases.  Dashed lines indicate particles that do not participate in the dominant gluino decay processes.  Blue lines indicate masses that we will vary.}
\label{fig:spectraCh6}
\end{center}
\end{figure}

\subsection{Scenarios with tops and no additional \MET}
\label{subsec:topsundiluted}

\begin{figure}[t]
\begin{center}
\includegraphics[width=0.47\textwidth]{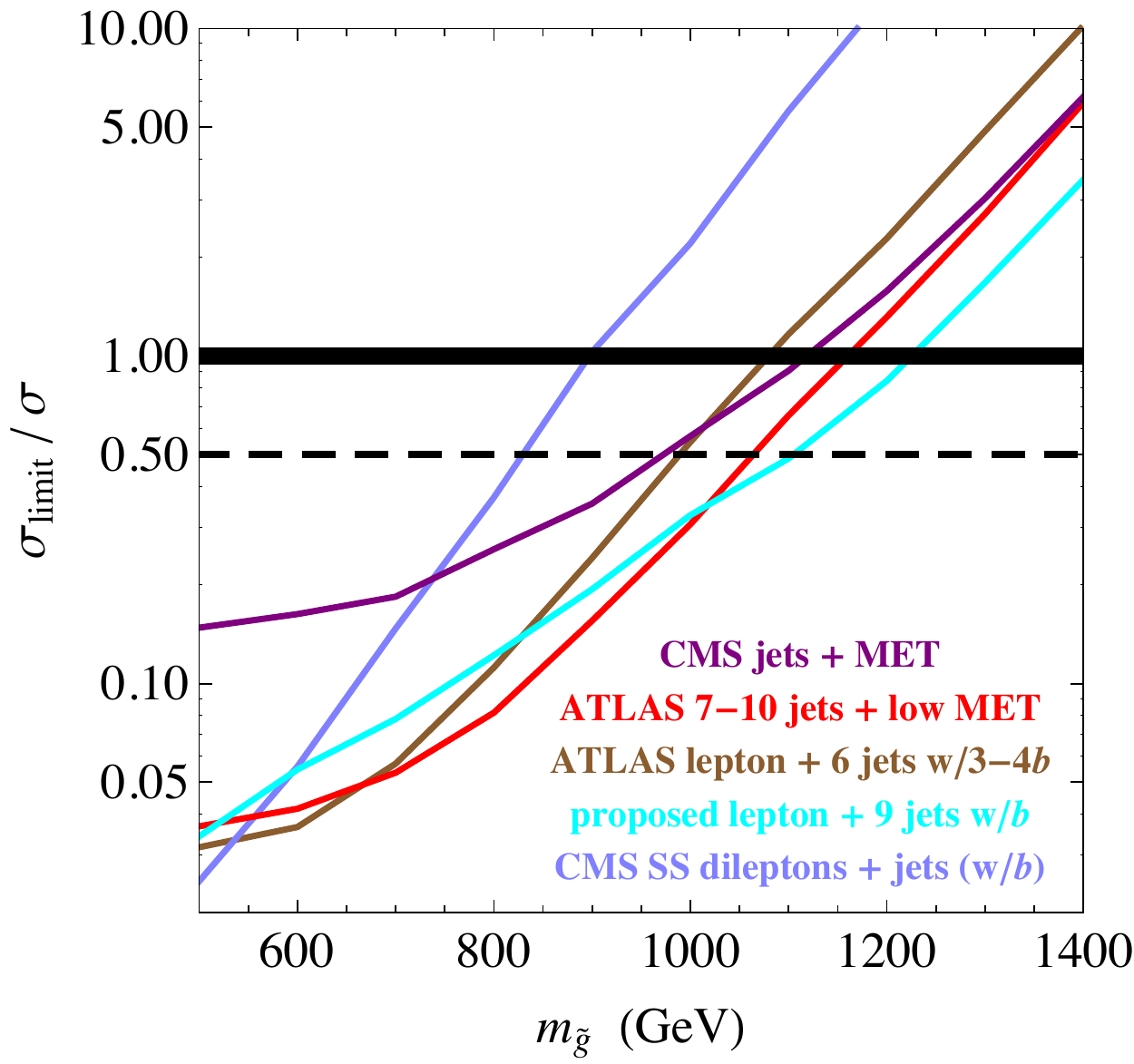}\qq
\includegraphics[width=0.47\textwidth]{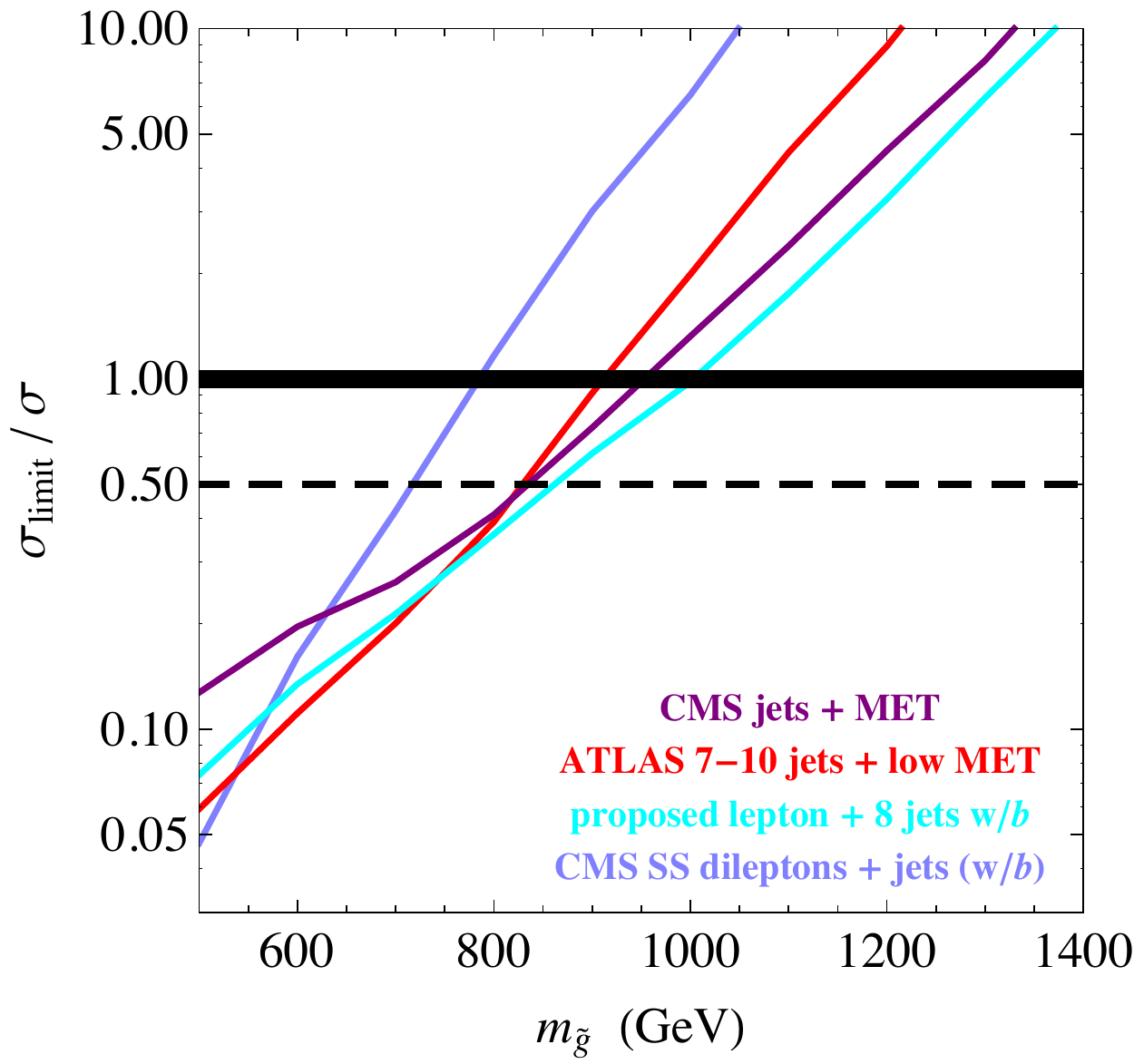}
\caption{As a function of the gluino mass, limits (left) on $\go\to\st t$, $\st\to b\Ho$, $\Ho\to jjj$, for $m_\st = 350$~GeV, $m_\Ho = 200$~GeV, and (right) on $\go\to\st t$, $\st\to jj$, for $m_\st = 100$~GeV (see figure~\ref{fig:spectraCh6}) from the searches~\cite{CMS-PAS-SUS-13-012,Aad:2013wta,ATLAS:2013ima,CMS-PAS-SUS-13-013}, and the expected limit from the LSST-proposed lepton + many jets search.  The CMS BH and ATLAS 6-7 jets searches are somewhat less powerful than the jets+\MET and lepton+jets searches and were left off the graphs for clarity.}
\label{fig:go_st_350_ho_200}
\end{center}
\end{figure}

Now let us start examining the effectiveness of the different searches in several example scenarios. We will again take the minimal $\go-\st_R-\Ho$ benchmark model of section~\ref{sec:minimal} (see
figure~\ref{fig:spectraCh4}) and modify it in various ways. First, we will allow the LSP to decay to jets via baryonic RPV, so that the final states have no intrinsic \MET except possibly from tops. In the leftmost spectrum in figure~\ref{fig:spectraCh6}, the gluino decays to a stop, $\go\to\st\bar t$, and the stop decays to a chargino $\st\to b\Ho^+$,  with the chargino decaying to unobservable, soft particles (due to a small splitting) and a neutralino LSP $\Ho^0_1$. This in turns decays as $\Ho^0_1 \to jjj$ via the RPV coupling $\lambda''_{212}$ (through a diagram involving an off-shell squark). In this example, we assumed the stop to be at 350~GeV and the higgsinos (chargino and neutralinos) near 200~GeV, giving 100\% branching ratio for $\go\to t b jjj$.

Figure~\ref{fig:go_st_350_ho_200} (left) presents cross section limits as a function of the gluino mass, for this scenario. The strongest limits come from the ATLAS~\cite{Aad:2013wta} and CMS~\cite{CMS-PAS-SUS-13-012} no-lepton high-multiplicity low-\MET searches, and are comparable to our estimate of the expected limit from a lepton + many jets (LSST) search. Due to the large number of $b$-jets (four in each event), the ATLAS $t'$ search~\cite{ATLAS:2013ima} (``lepton + 6 jets w/3-4$b$'') is also relevant. Its limits could likely be improved by requiring larger jet multiplicities; the single-lepton events in the signal contain 12 colored partons, but the search demands only $\geq 6$ jets. Said another way, one would expect that adding bins with 3 $b$-tags to the proposed LSST search (as was suggested in~\cite{Lisanti:2011tm}) could make that search quite a bit more powerful, since background is dominantly $t\bar t$ plus jets.  However, it is quite satisfying that simply requiring 1 $b$ tag is enough to put limits well above 1~TeV.

In the second model shown in figure~\ref{fig:spectraCh6}, the gluino decays as $\go\to\st\bar t$, with the top squark decaying to a pair of jets $\st\to jj$ via the RPV coupling $\lambda''_{312}$ (while the higgsinos, assumed to be heavier than the stop, do not participate in gluino decays). This model was studied in~\cite{Han:2012cu}, where an LSST-type search was considered.  In our example, we fix $m_\st=100$~GeV. (At present there is no sensitivity to direct production of such light top squarks, due to trigger limitations.) Our limits are shown in the right-hand plot of figure~\ref{fig:go_st_350_ho_200}. For high $m_\go$ and very low $m_\st$, the jets from $\st \to jj$ often merge, reducing the multiplicity.  This weakens the limits from the CMS and ATLAS high-multiplicity low-\MET searches as well as the proposed LSST search (although jet substructure methods could be useful in this scenario, as explored in \cite{Han:2012cu}).  For heavier stops, the limits (not shown) on the gluino mass become stronger.

Both the ATLAS 6-7 jet search~\cite{ATLAS-CONF-2013-091} and the CMS BH search~\cite{Chatrchyan:2013xva} (neither is shown, for clarity of the figure) give limits about 200~GeV weaker than the jets + low \MET searches, about 1~TeV for the plot at left and about 750~GeV for the plot at right of figure~\ref{fig:go_st_350_ho_200}. Same-sign dileptons (our plots show the limits from~\cite{CMS-PAS-SUS-13-013}; comparable limits are obtained from~\cite{CMS-PAS-B2G-12-015,ATLAS:2013tma,ATLAS-CONF-2013-051}) also are not competitive for models of this type, except at the lowest masses where many searches already exclude the signal.

In both plots, the proposed LSST search gives the strongest limit on $m_\go$. We must caution that the systematic uncertainties on our simulations, and the guesswork involved in our implementation of the LSST search, prevent us from making any strong claims regarding its power relative to the other searches.  However, our evidence shows that it should at least be comparable to the CMS and ATLAS high-multiplicity low-\MET searches, and, as we have discussed above, this makes it complementary and supplementary to those searches.

\subsection{Scenarios with reduced number of tops}
\label{subsec:topsdiluted}

We have seen that with abundant top quarks in gluino final states, limits on $m_\go$ extend above 1~TeV even without \MET arising from the LSP. Now we are going to study what happens to the limits as we dilute decays that produce tops, via the mechanism described in figure~\ref{fig:topdil}. In particular, we will consider the rightmost spectrum of figure~\ref{fig:spectraCh6}, which contains second-generation squarks to which the gluino can decay. The stop mass is fixed at 500~GeV, and the higgsino mass is fixed at 200~GeV.

\begin{figure}[t]
\begin{center}
\includegraphics[width=0.47\textwidth]{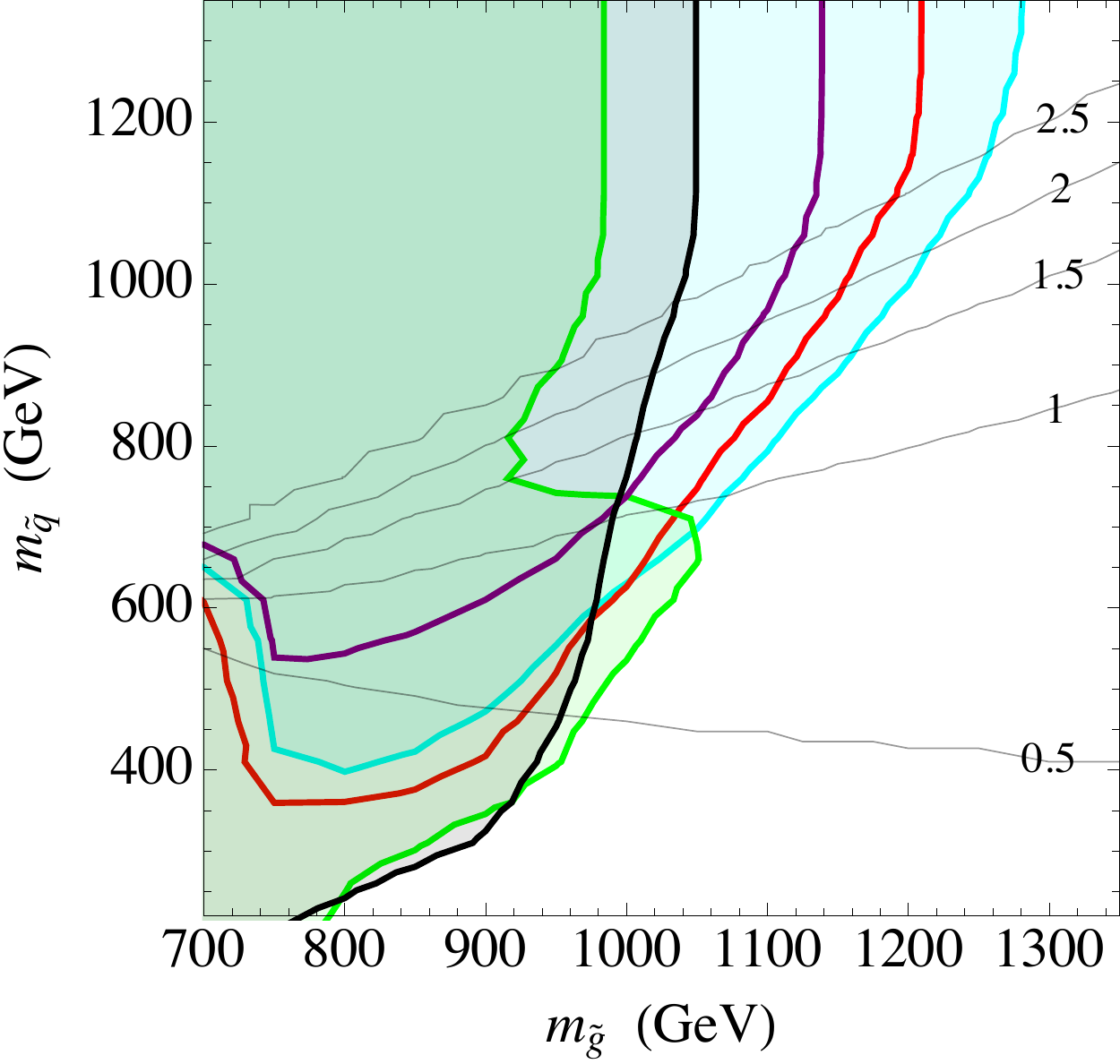}\qq
\includegraphics[width=0.47\textwidth]{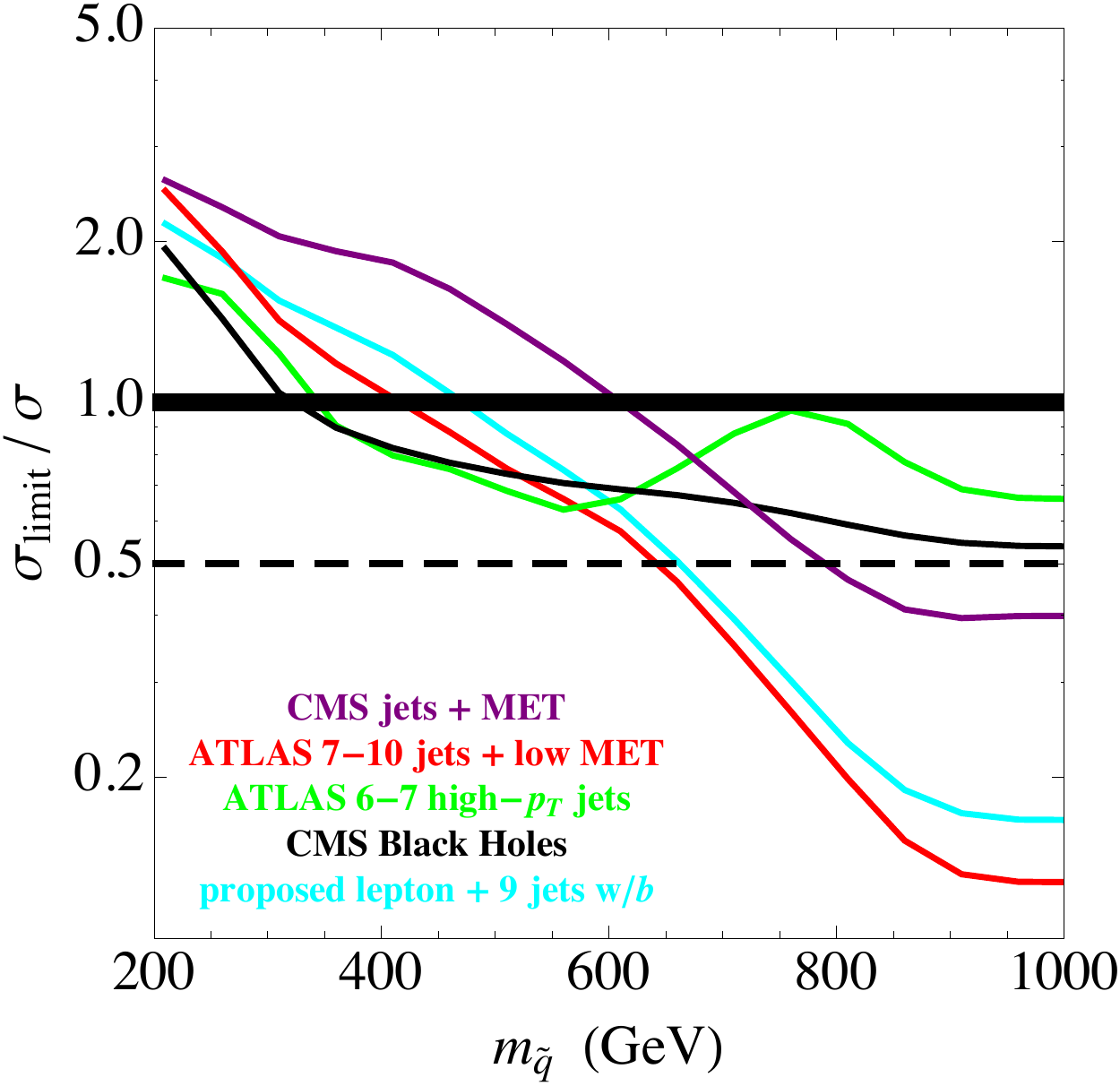}
\caption{{\bf Left:} Exclusion contours in gluino vs.\ second-generation squark masses with $m_\st=500$~GeV, $m_\Ho=200$~GeV and $\Ho\to jjj$ via RPV (see figure~\ref{fig:spectraCh6}).   The competition between the two decays, $\go\to t \st$ and $\go \to q\sq$, is controlled by $m_\sq$, with large branching ratio for the former (latter) at high (low) $m_\sq$.  Light gray contours display the average number of tops in an event.  These lines converge near $m_\go = m_\sq = m_\st + m_t = 675$~GeV because for gluinos below that mass, there is $\sim$ 100\% dilution (0 tops) for $m_\sq < m_\go$ and almost no dilution ($\sim3$ tops per event) for $m_\sq > m_\go$. {\bf Right:}  The sensitivity of each search is shown for a slice through the $(m_\go,m_\sq)$ plane at left, with fixed $m_\go=900$~GeV. At the right edge of the plot, there is no top dilution; moving left, dilution grows toward 100\%.}
\label{fig:topdilution-grid}
\end{center}
\end{figure}

In figure~\ref{fig:topdilution-grid} (left), we present the limits as a function of $m_\go$ and the common mass $m_\sq$ of the four second-generation squarks.  The common squark mass parameterizes the degree of top dilution: smaller $m_\sq$ means more dilution. This is illustrated by the light gray contours of constant average number of tops per event; this number is $\approx 3$ for large $m_\sq$, where the gluino decays entirely through the stop, and it goes to zero for small $m_\sq$, where the gluino decays predominantly through the second-generation squarks.   We see that the strongest limit on the gluino mass of $\approx 1.3$~TeV comes from the proposed LSST search (the $n\geq 9$-jet bin dominates) in the upper-right corner of the plot, with slightly weaker limits from the ATLAS and CMS high-multiplicity low-\MET searches. Here there is no dilution of the top signal, and the leptons and neutrinos from the tops lead to powerful exclusion. As we decrease $m_\sq$, decays with tops become increasingly diluted, and the limits on $m_\go$ from these three searches become progressively weaker.  However, before these searches become ineffective, the ATLAS 6-7 high-$p_T$ jet search and the CMS BH search, which require neither leptons nor \MET, begin to play a major role, maintaining limits at or above 1~TeV until $m_\sq\sim m_\st \sim 500$~GeV.  Yet in the lower-left corner of the plot, where top quarks are virtually absent, the limit on $m_\go$ drops below 800~GeV.\footnote{For the ATLAS 6-7 jets search, the rapid change in the exclusion limit near $m_\sq\sim 750$~GeV is due to the fact that one search bin dominates the exclusion limit above that point and a different one dominates below.} We will explore this regime in more detail in section~\ref{sec:all-hadronic}.

In the right-hand plot of figure~\ref{fig:topdilution-grid}, we have taken a 1D slice through the left-hand plot, with $m_\go=900$~GeV. As before, tops are abundant for large $m_\sq$ and scarce for low $m_\sq$.  This plot illustrates that while the non-\MET based searches are nominally effective in ruling out most of the slice of parameter space, their limits are not robust (in the sense that they may be within the uncertainty of our simulation or not apply to slightly modified scenarios).  This is in contrast to the other three searches, which impose robust exclusion at higher values of $m_\sq$, and whose limits are all strongly correlated with the number of tops in the events.

To summarize, an abundance of tops in gluino decays allows exclusion by the existing high-multiplicity low-\MET searches, and may allow for an even stronger exclusion if the proposed LSST 1-lepton high-multiplicity low-\MET + $b$ search is implemented.  Note that since we did not utilize same-sign dileptons (we found the existing SS dilepton searches to be less sensitive than the 0 and 1-lepton searches), there is no loss of sensitivity for Dirac gluinos, which need not produce same-sign pairs in their decays.  In fact there is a gain, due to the increased cross section.\footnote{Typically, models with Dirac gluinos are thought to lower SUSY cross sections because they eliminate the important $t$-channel production mode of  first-generation squarks. But if these are decoupled, then employing Dirac gluinos only serves to {\it increase} the overall SUSY cross section, because a Dirac gluino has twice as many degrees of freedom as a Majorana gluino.}  Meanwhile, we have also seen, both in this section and in section~\ref{sec:MET}, that in the complete absence of tops and \MET, exclusion limits can fail to reach 1~TeV, sometimes by a substantial margin.

\section{All-Hadronic Final States}
\label{sec:all-hadronic}

We have presented evidence, based on models that interpolate between one signature and another, that gluinos whose decays possess at least a moderate amount of \MET or frequently contain top quarks are excluded through 1~TeV.  Weaker limits appear when both \MET and tops are absent (see figures~\ref{fig:go_ho_200_HV1} and~\ref{fig:topdilution-grid}).  These weaknesses would be severe were it not for two searches that do not require \MET, leptons or photons --- the ATLAS 6-7 high-$p_T$ jets search~\cite{ATLAS-CONF-2013-091}, which constrains low-to-moderate multiplicity final states, and the CMS BH search~\cite{Chatrchyan:2013xva}, which constrains moderate-to-high-multiplicity final states.  Even with these searches, we have seen  that there are corners of the parameter space where the nominal limit on $m_\go$ drops well below 1~TeV.  The robust limits are weaker still.  What causes the limits on the gluino mass to be so weak in these corners of the parameter space? We will show in this section that just eliminating \MET and tops does not capture the entire effect.  A reduced multiplicity of high $p_T$ jets also plays an important role.

\subsection{High jet multiplicity}

\begin{figure}[t]
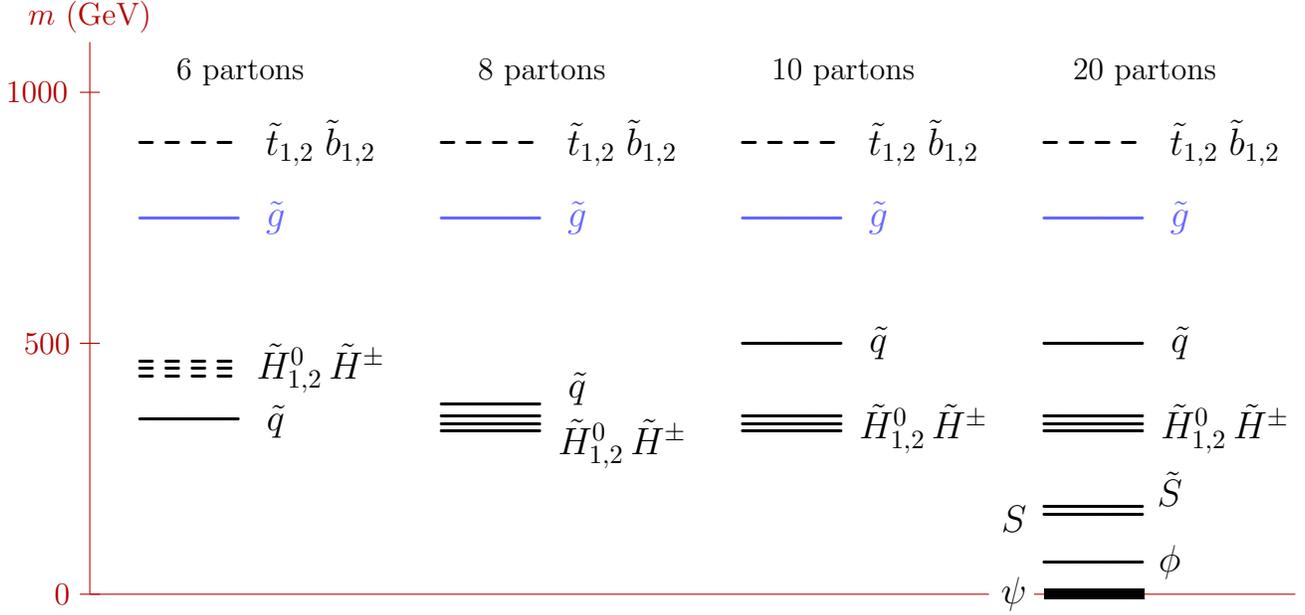

\begin{center}
$$\beginpicture
\setcoordinatesystem units <0.19\tdim,0.19\tdim>
\stpltsmbl
\makered{
\putrule from -100 0 to -100 1100
\putrule from -100 0 to 1690 0
\putrule from 1780 0 to 2300 0
\put {$m$ (GeV)} at -100 1150
\putrule from -120 0 to -80 0
\putrule from -120 500 to -80 500
\putrule from -120 1000 to -80 1000
\put {0} at -155 0
\put {500} at -185 500
\put {1000} at -205 1000 }
\put {6 partons} at 200 1040
\setsolid
\plot 0 350 200 350 /
\put {\large$\sq$} at 270 345
\setdashes
\plot 0 435 200 435 /
\plot 0 450 200 450 /
\plot 0 465 200 465 /
\put {\large$\Ho^0_{1,2}\,\Ho^\pm$} at 360 450
\setsolid
\makeblue{\plot 0 750 200 750 /
\put {\large$\go$} at 270 750 }
\setdashes
\plot 0 900 200 900 /
\put {\large$\st_{1,2}\;\sbo_{1,2}$} at 360 900
\put {8 partons} at 800 1040
\setsolid
\plot 600 325 800 325 /
\plot 600 340 800 340 /
\plot 600 355 800 355 /
\put {\large$\Ho^0_{1,2}\,\Ho^\pm$} at 960 300
\plot 600 380 800 380 /
\put {\large$\sq$} at 870 410
\setsolid
\makeblue{\plot 600 750 800 750 / \put {\large$\go$} at 870 750 }
\setdashes
\plot 600 900 800 900 /
\put {\large$\st_{1,2}\;\sbo_{1,2}$} at 960 900
\put {10 partons} at 1400 1040
\setsolid
\plot 1200 325 1400 325 /
\plot 1200 340 1400 340 /
\plot 1200 355 1400 355 /
\put {\large$\Ho^0_{1,2}\,\Ho^\pm$} at 1560 330
\plot 1200 500 1400 500 /
\put {\large$\sq$} at 1470 500
\setsolid
\makeblue{\plot 1200 750 1400 750 /
\put {\large$\go$} at 1470 750 }
\setdashes
\plot 1200 900 1400 900 /
\put {\large$\st_{1,2}\;\sbo_{1,2}$} at 1560 900
\put {20 partons} at 2000 1040
\setsolid
\plot 1800 65 2000 65 /
\put {\large$\phi$} at 2050 65
\plot 1800 160 2000 160 /
\plot 1800 175 2000 175 /
\put {\large$\tilde S$} at 2050 210
\put {\large$S$} at 1740 150
\plot 1800 325 2000 325 /
\plot 1800 340 2000 340 /
\plot 1800 355 2000 355 /
\put {\large$\Ho^0_{1,2}\,\Ho^\pm$} at 2160 330
\plot 1800 500 2000 500 /
\put {\large$\sq$} at 2070 500
\setsolid
\makeblue{\plot 1800 750 2000 750 /
\put {\large$\go$} at 2070 750 }
\setdashes
\plot 1800 900 2000 900 /
\put {\large$\st_{1,2}\;\sbo_{1,2}$} at 2160 900
\setsolid
\linethickness=4pt
\putrule from 1800 0 to 2000 0
\put {\large$\psi$} at 1740 0
\color[rgb]{0,0,0}
\endpicture$$
\caption{Mass spectra of the models studied in this section.  The label $\sq$ indicates one or more second-generation squarks.  The LSP decays to jets via RPV in the first three cases; the last shows a hidden valley model described in the text, for which a thick solid line indicates that $\psi$ is stable.  Dashed lines indicate particles that do not participate in the dominant gluino decay processes.  Blue lines indicate masses that we will vary.}
\label{fig:spectraCh7a}
\end{center}
\end{figure}

In our examples here, we will assume that second-generation squarks are relatively light, while the top squarks are sufficiently heavy that gluino decays involving tops are very rare.\footnote{In this section we allow top squarks to sometimes be slightly heavier than naturalness bounds would allow, even from theories like $\lambda$-SUSY~\cite{Hall:2011aa,Gherghetta:2012gb}.  This is mainly to allow us to illustrate some simple conceptual points without the distraction that would arise if top squarks contributed to the gluino decays.}  There is no \MET in these models because the LSP decays either via baryonic RPV or into a Hidden Valley with appropriately chosen parameters. We will start by considering four models with different multiplicities of hard partons, with the spectra shown in figure~\ref{fig:spectraCh7a}:
\begin{itemize}

\item {\it 6 partons:} The simplest case of an LSP gluino that decays to 3 jets via RPV is not interesting for us. It requires a gluino lighter than the higgsinos, which naturalness constrains to the 400~GeV range, but the ATLAS 6-7 high-$p_T$ jet search and a  CMS search for 3-jet resonances both exclude such gluinos to well above 400~GeV~\cite{ATLAS-CONF-2013-091,CMS-PAS-EXO-12-049}.  We will consider a scenario that gives a similar final state but with different kinematics.  This model has a spectrum $m_\sq < m_\go < m_\st$, where $\sq$ is a right-handed second-generation squark that decays via RPV into two quarks (using $\lambda''_{212}$), yielding a net $\go\to q\sq^* \to qqq$. For illustration, we will first assume $m_\sq = 350$~GeV.

\item {\it 8 partons:} Consider the decay chain $\go\to q\sq$, $\sq\to q\Ho$, $\Ho\to qqq$ (via RPV).  (Note this is the same scenario used in figure~\ref{fig:topdilution-grid}, but with  $m_{\tilde t}>m_\go$, eliminating decays to top quarks.) By placing $m_\sq$ just a little above $m_\Ho$, we make the jet in the $\sq\to q\Ho$ transition too soft to observe, so the final state has 8 potentially observable partons. We will take $m_\Ho = 350$~GeV and $m_\sq = 360$~GeV.

\item {\it 10 partons:} To achieve this final state, we use the same decay chain as in the previous bullet point, except that we do not require $m_\sq$ to be close to $m_\Ho$. We will choose $m_\sq = 500$~GeV, $m_\Ho=350$~GeV to illustrate our points.

\item {\it 20 partons:} Might final states with much higher partonic multiplicities be less constrained?  Since for the same gluino mass the $p_T$ of the partons would be reduced relative to a lower-multiplicity final state, one might worry that many jets would fail the $p_T$ cuts imposed, leading to a lower effective jet multiplicity and making the searches ineffective.  We explore this in a model similar to the previous one (where $m_\sq = 500$~GeV, $m_\Ho=350$~GeV) except now the higgsino decays into an all-hadronic zero-\MET Hidden Valley resulting in 10 partons per gluino.  (This Hidden Valley contains particles $S$, $\tilde S$, $\phi$, $\psi$ with masses $m_S = 160$~GeV, $m_{\tilde S} = 170$~GeV, $m_\phi = 65$~GeV, $m_\psi \approx 0$ and decay processes $\Ho^0_1 \to S\tilde S$, $\tilde S \to S\psi$, $S \to \phi\phi$, $\phi \to gg$, resulting in $\Ho^0_1 \to 8g$ with very low \MET.)
\end{itemize}

\begin{figure}[t]
\begin{center}
\includegraphics[width=0.47\textwidth]{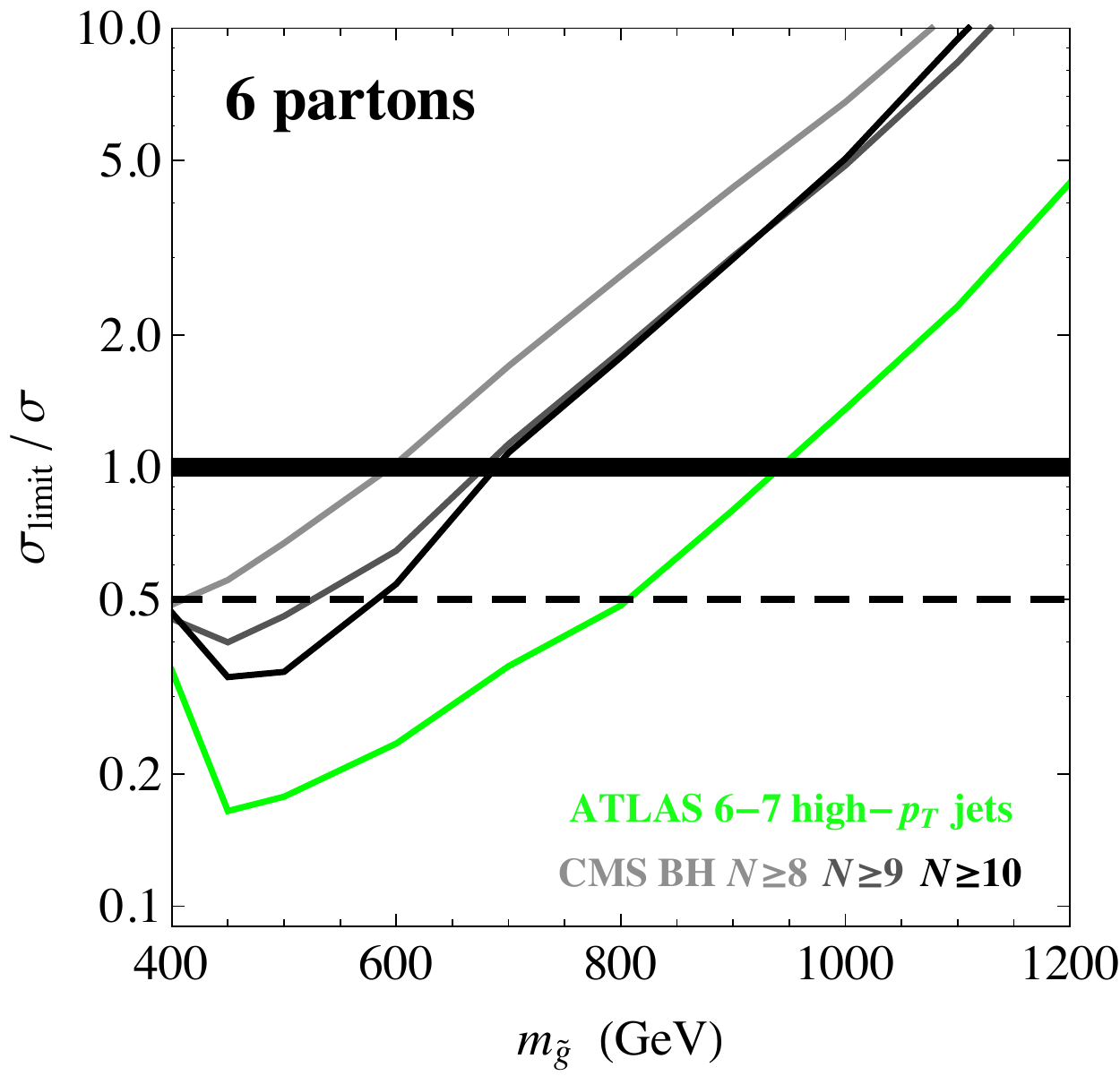}\qq
\includegraphics[width=0.47\textwidth]{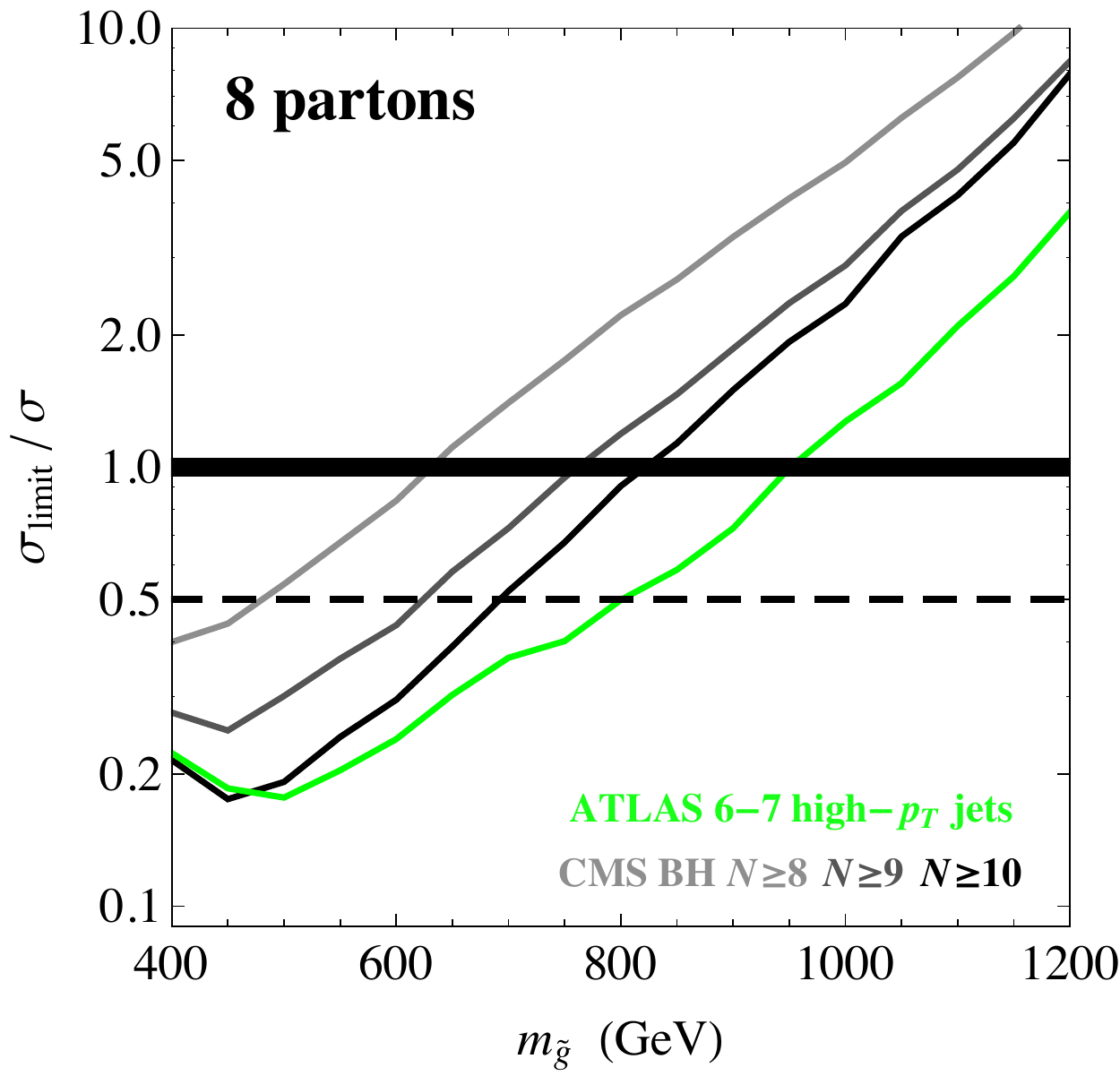}\\\vspace{5mm}
\includegraphics[width=0.47\textwidth]{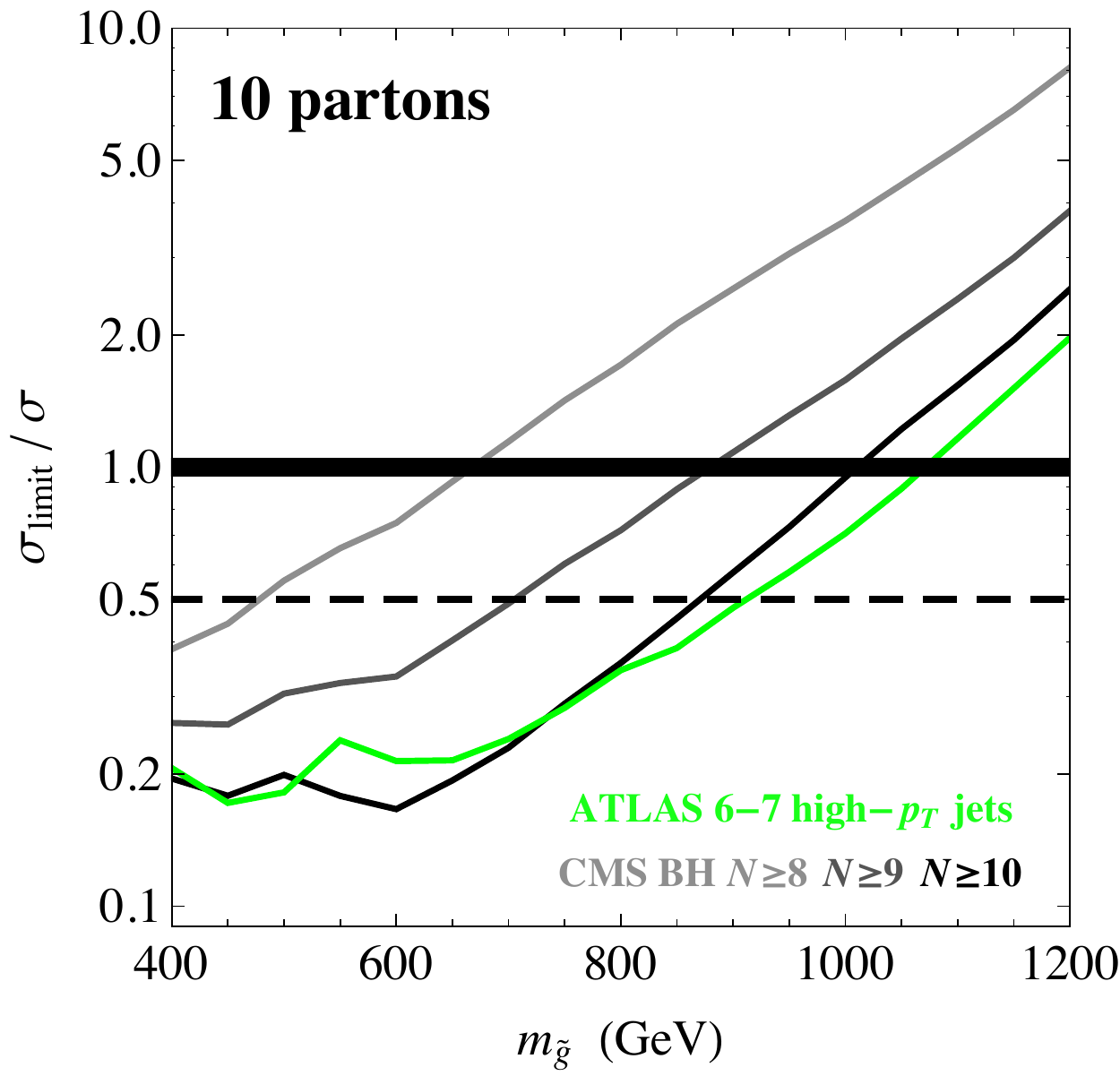}\qq
\includegraphics[width=0.47\textwidth]{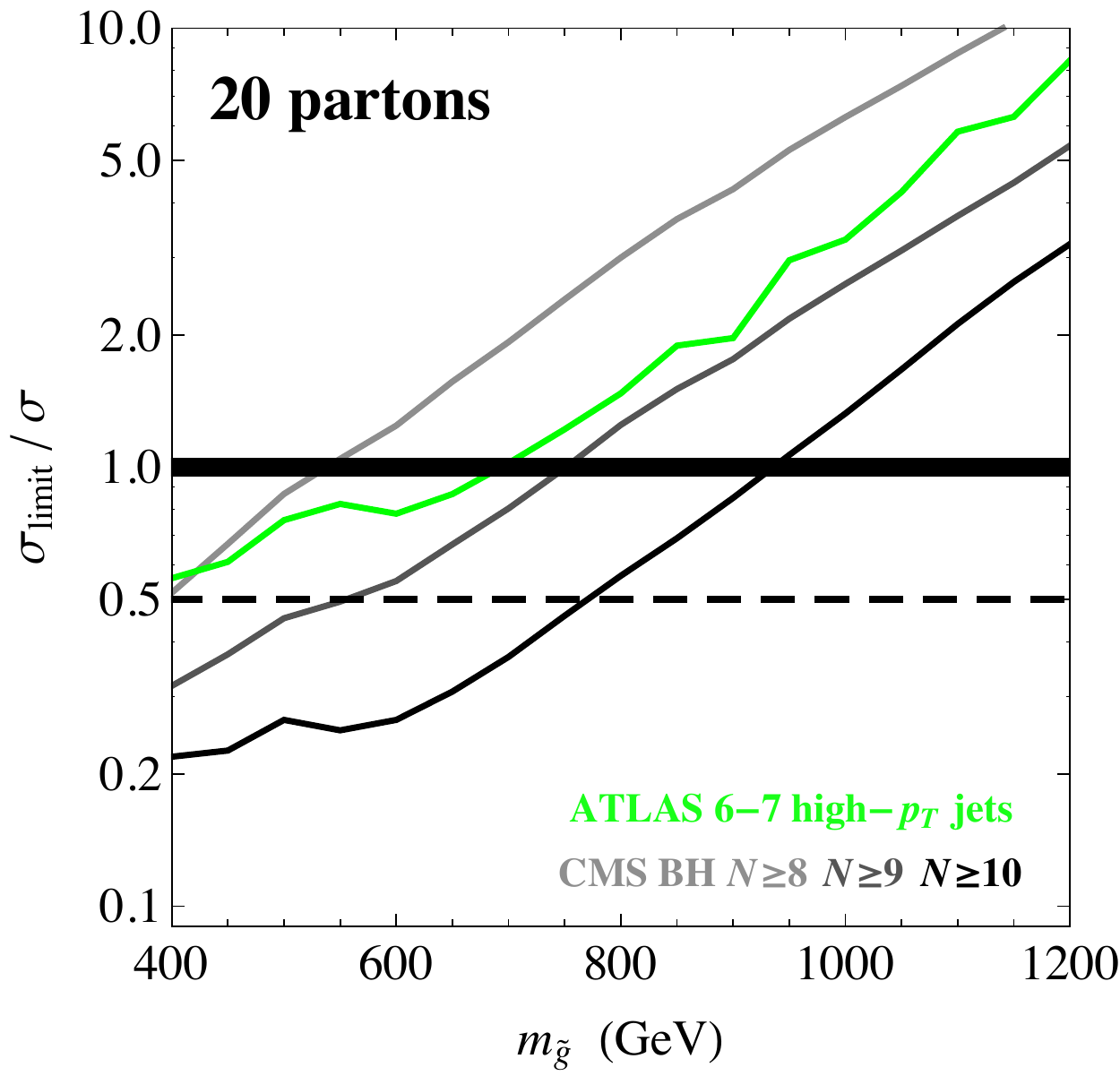}
\caption{Limits on scenarios with 6-parton (upper left), 8-parton (upper right), 10-parton (lower left), and 20-parton (lower right) final states as a function of the gluino mass. More details about the scenarios are given in the text and in figure~\ref{fig:spectraCh7a}.}
\label{fig:6-20-partons}
\end{center}
\end{figure}

Figure~\ref{fig:6-20-partons} presents the limits on these four scenarios as a function of the gluino mass.   We only show the limits from the two most constraining searches, namely those without a \MET requirement:\footnote{Formally, the \MET-based searches~\cite{CMS-PAS-SUS-13-012,Aad:2013wta} have some sensitivity here, but since these models have no intrinsic \MET, the \MET-based searches are both extremely inefficient and highly dependent on our simulation of jet energy mismeasurement.} the ATLAS 6-7 jet search~\cite{ATLAS-CONF-2013-091} and the CMS BH search~\cite{Chatrchyan:2013xva}.

These plots show that the two types of searches are nicely complementary, providing good coverage up to $m_\go\approx 950$~GeV across the different multiplicities.  At low multiplicity the ATLAS 6-7 jet search dominates, while as we move up in parton multiplicity, the CMS BH search becomes stronger and eventually dominates. This is sensible, as the ATLAS search is optimized for 6 to 10 very hard jets,  while the CMS BH search is optimized for a high multiplicity of somewhat softer objects. Indeed, the CMS BH search is effective even in a 20-parton scenario.

As an aside, let us note several important subtleties with our CMS BH search analysis. For a signal with $\leq 8$ hard partons, one might wonder why the $N \geq 9$ or $N \geq 10$ search regions dominate the limit.   This is partly due to extra jets with $p_T>50$~GeV that easily arise from initial or final state radiation. Note, however, that since we used {\sc Pythia} without any matrix-element/parton-shower matching to simulate this radiation,  these limits have larger uncertainties (which we have not accounted for) than for signals with $>8$ partons.  Additionally, the $N\geq8$ search region is significantly disadvantaged by the extreme conservatism of our analysis (as discussed in appendix~\ref{app:BH}), which can only exclude signals significantly larger than the data. If the background were measured and subtracted from the data, the $N\geq 8$ channel, which has the largest background, would improve the most.

A final subtlety with the CMS BH search is that, for all multiplicities, we have data only for $S_T>1900$~GeV, so there are large uncertainties on the signal efficiency (which we do not take into account) for low $m_\go$. In particular, a large source of uncertainty again arises from simulation of initial-state radiation (ISR).  We examine the issue of ISR further in appendix~\ref{app:simulation}, where we confirm that the biggest effects are at low gluino masses, and show that these are important for high multiplicity signals, where the CMS BH search is most relevant.  For the 20-parton signal, a softer ISR spectrum does not much change the exclusion limit at high gluino mass, but the exclusion for low gluino masses is no longer robust.  
For low-multiplicity signals, where strong limits arise from the ATLAS 6-7 jet search, this issue is less critical.

Now let us return to the question of why there are significant holes in the upper-left region of figure~\ref{fig:go_ho_200_HV1} and in the lower-left region of figure~\ref{fig:topdilution-grid}. Both holes appear where there are 8-jet final states. (In the latter case, the hole appears in the nearly-degenerate region $m_\sq \approx m_\Ho$, where, as we have just discussed, there is effectively an 8-jet final state instead of a 10-jet final state.)  However, we have just seen in our 8-parton example that the lower multiplicity {\it by itself} does not degrade sensitivity that much. So what is causing these holes?

\subsection{A hierarchy of jet $p_T$s}
\label{subsec:jetpthierarchy}

We can shed light on this question by comparing the 8-parton example considered in this section with the model of figure~\ref{fig:topdilution-grid} in the low $m_\sq$ region. The only essential difference between these two cases is that here we have taken $m_\Ho=350$~GeV while in the previous section we took $m_\Ho=200$~GeV. Evidently the lower higgsino mass has the effect of reducing the $p_T$ of three of the jets in each cascade, to the point that  kinematic cuts regularly discard the events.  Specifically, a {\it hierarchy} of jet $p_T$ values arises in this regime, which makes the event selection requirement of a certain number of jets above a fixed $p_T$ cut, as employed in the ATLAS search~\cite{ATLAS-CONF-2013-091}, ineffective. This is an interesting feature of extremely low \MET models -- by making the model {\it more} natural (lowering the higgsino mass) we are making it {\it harder} to detect.

\begin{figure}[t]
\begin{center}
$\beginpicture
\setcoordinatesystem units <0.20\tdim,0.20\tdim>
\stpltsmbl
\makered{
\putrule from -100 0 to -100 1100
\putrule from -100 0 to 530 0
\put {$m$ (GeV)} at -100 1150
\putrule from -120 0 to -80 0
\putrule from -120 500 to -80 500
\putrule from -120 1000 to -80 1000
\put {0} at -155 0
\put {500} at -185 500
\put {1000} at -205 1000 }
\setdashes
\plot 0 950 200 950 /
\put {\large$\st_{1,2}\;\sbo_{1,2}$} at 360 950
\setsolid
\plot 0 800 200 800 /
\put {\large$\go$} at 270 800
\makeblue{\plot 0 500 200 500 /
\put {\large$\sq$} at 270 500
\plot 0 265 200 265 /
\plot 0 250 200 250 /
\plot 0 235 200 235 /
\put {\large$\Ho^0_{1,2}\,\Ho^\pm$} at 360 245 }
\color[rgb]{0,0,0}
\endpicture$\qq
\includegraphics[width=0.47\textwidth]{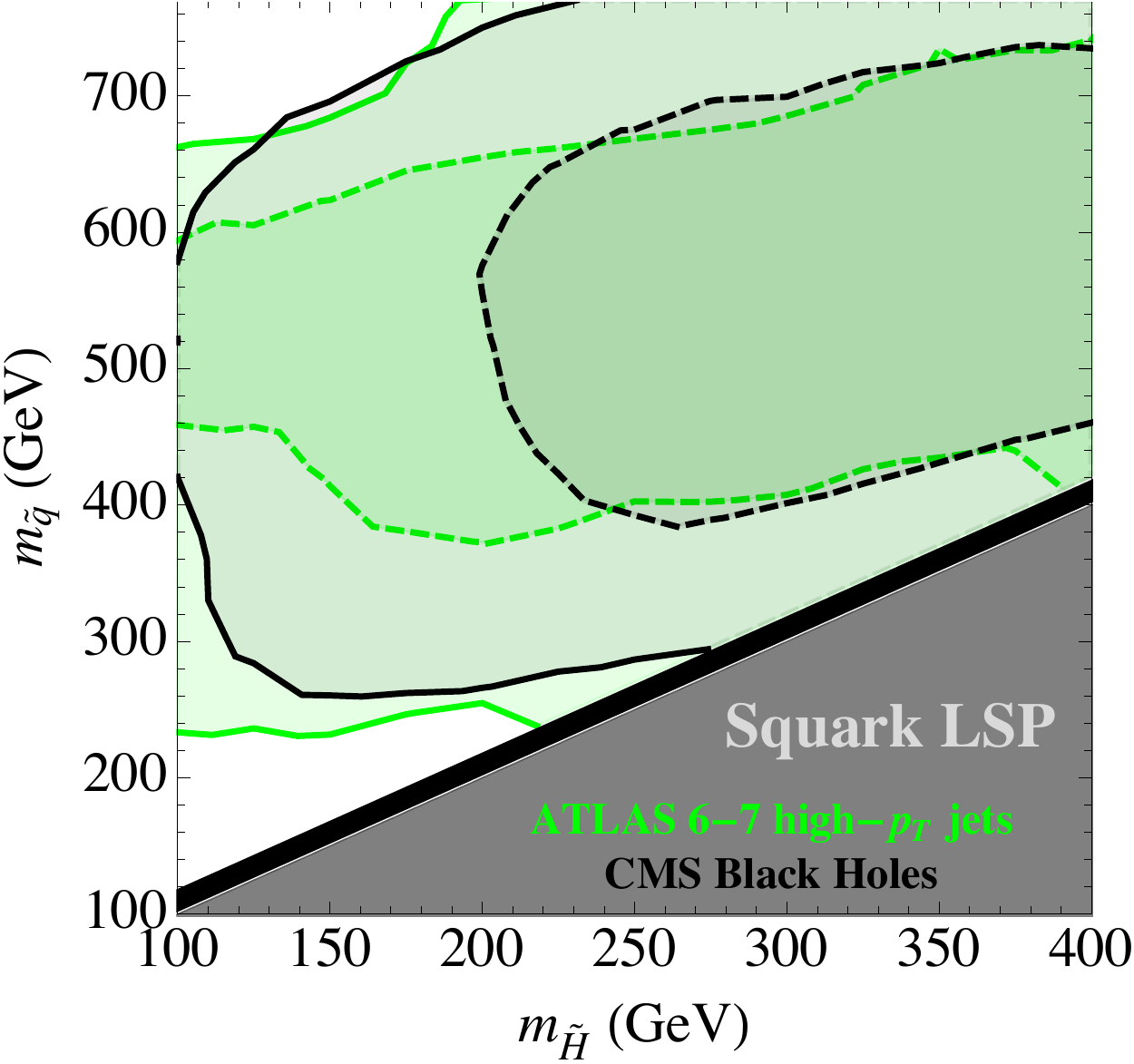}
\caption{Limits on the spectrum shown on the left, where the label $\sq$ indicates one or more second-generation squarks, the LSP decays to jets via RPV, and the dashed line indicates that the stops and sbottoms do not participate in the dominant gluino decay processes. Shown are limits, for $m_\go=800$~GeV, from the ATLAS 6-7 high-$p_T$ jets search (green) and the CMS BH search (black).  As discussed in the text, this model produces up to 10 hard partons, with this number becoming smaller in special corners of the parameter space. The dashed and solid contours correspond to robust and nominal exclusion of the model. }
\label{fig:sq2_vs_H_mgo800}
\end{center}
\end{figure}

To test this hypothesis, let us consider what happens to our 10-parton model (which again is essentially the same model as in figure~\ref{fig:topdilution-grid} but with the stops decoupled) as we vary $m_\sq$ and $m_\Ho$ for a fixed gluino mass. This model nominally has a 10-jet final state (apart from radiation jets), but squeezing $m_\sq \to m_\go$ or $m_\sq \to m_\Ho$ effectively reduces this to an 8-jet final state. Separately decreasing $m_\Ho$ also effectively reduces the multiplicity as the quarks from the $\Ho$ decay become increasingly likely to be too soft to pass the cuts; or, for $m_\Ho \ll m_\go$, some of them merge into a single jet. Taking two such limits together can even yield just a 4-jet final state. Our hypothesis therefore predicts a failure to exclude the $m_\Ho\ll m_\go$ region for both $m_\sq\to m_\go$ and $m_\sq\to m_\Ho$. The plot in figure~\ref{fig:sq2_vs_H_mgo800} supports this. Here we have fixed $m_\go=800$~GeV, and show how the limits vary as a function of $m_\sq$ and $m_\Ho$. The ATLAS 6-7 high-$p_T$ jets search is shown in green, and the CMS BH search is shown in black.  The solid contours indicate nominal exclusion (i.e., $\sigma_{\rm limit}/\sigma=1$); robust exclusion (i.e., $\sigma_{\rm limit}/\sigma=0.5$) is shown with the dashed contours.  We see precisely the expected gap in the limits, for small $m_\Ho$ together with $m_\sq\to m_\go$ or $m_\sq\to m_\Ho$.

\begin{figure}[t]
\begin{center}

\begin{picture}(0,0)(-190,0)
\put(0,0){\includegraphics[width=0.47\textwidth]{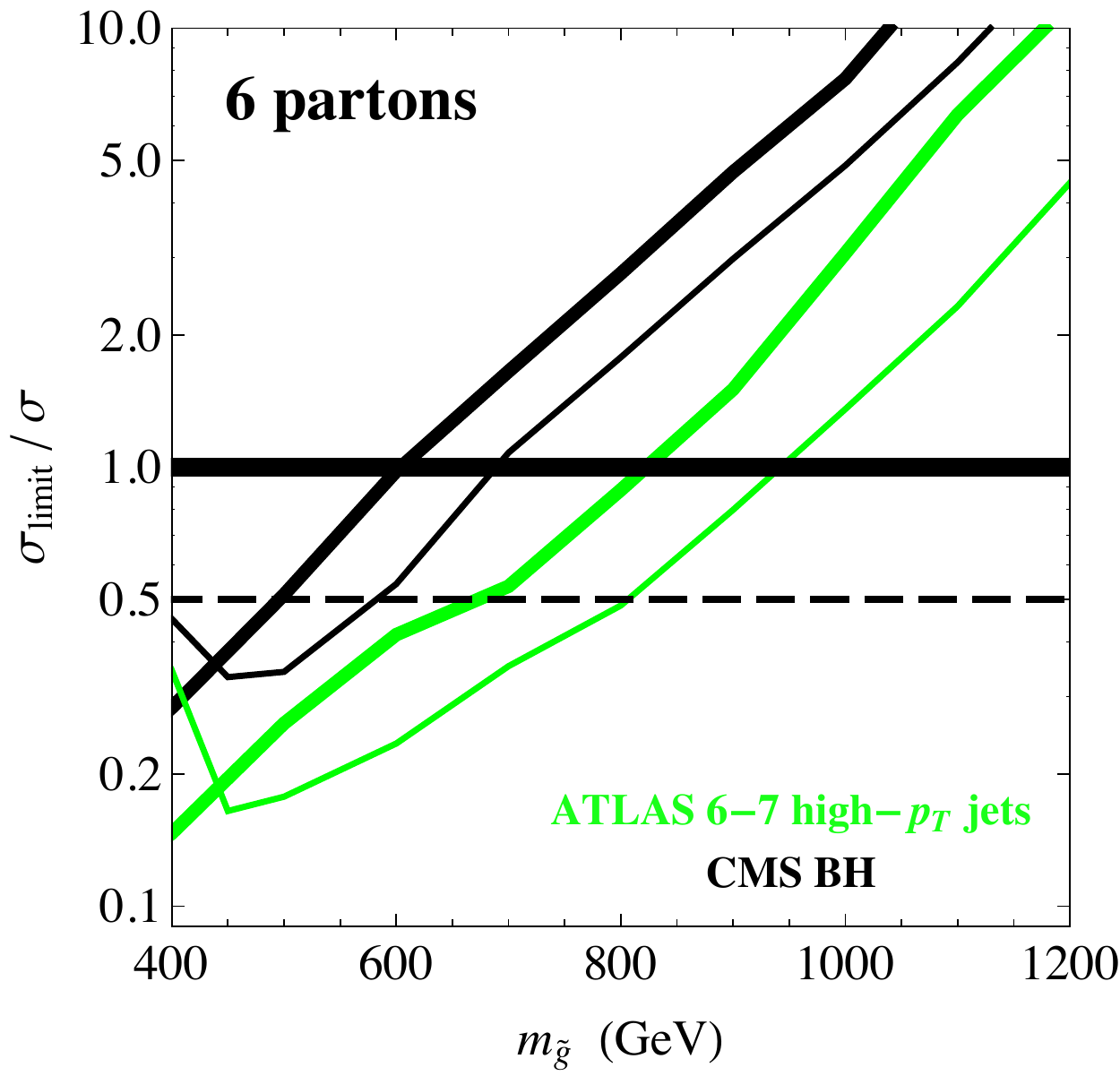}}
\end{picture}
$\beginpicture
\setcoordinatesystem units <0.20\tdim,0.20\tdim>
\stpltsmbl
\makered{
\putrule from -100 0 to -100 1100
\putrule from -100 0 to 530 0
\put {$m$ (GeV)} at -100 1150
\putrule from -120 0 to -80 0
\putrule from -120 500 to -80 500
\putrule from -120 1000 to -80 1000
\put {0} at -155 0
\put {500} at -185 500
\put {1000} at -205 1000 }
\setdashes
\plot 0 950 200 950 /
\put {\large$\st_{1,2}\;\sbo_{1,2}$} at 360 950
\setsolid
\makeblue{\plot 0 800 200 800 /
\put {\large$\go$} at 270 800 }
\setdashes
\plot 0 465 200 465 /
\plot 0 450 200 450 /
\plot 0 435 200 435 /
\put {\large$\Ho^0_{1,2}\,\Ho^\pm$} at 360 445
\setsolid
\makeblue{\plot 0 300 200 300 /
\put {\large$\sq$} at 270 300 }
\color[rgb]{0,0,0}
\endpicture$
\hspace{92mm}
\caption{Limits on the 6-parton scenario from the spectrum shown on the left, where the label $\sq$ indicates one or more second-generation squarks decaying to jets via RPV, and dashed lines indicate particles that do not participate in the dominant gluino decay processes. Limits are shown for $m_\sq = 350$~GeV (thin lines) and $m_\sq = 150$~GeV (thick lines).}
\label{fig:go_sq_150_350_UDD212-effthr}
\end{center}
\end{figure}

As a cross-check for this conclusion, we return to the 6-parton model of figure~\ref{fig:spectraCh7a}, in which the decay chain is $\go\to \bar q\sq$ and $\sq \to \bar q \bar q$. We saw in figure~\ref{fig:6-20-partons} (upper left) that limits reach $\sim$~950~GeV if $m_\sq=350$~GeV, where all of the jets are quite hard.  However, in a spectrum with a smaller $m_\sq$, we expect more of a $p_T$ hierarchy.  In this case, the two jets from the squark decay can become soft enough to drop below $p_T$ thresholds or merge into a single jet, resulting again in fewer high $p_T$ jets.  This is confirmed in figure~\ref{fig:go_sq_150_350_UDD212-effthr}, where decreasing $m_\sq$ to 150~GeV lowers the limit on $m_\go$ to approximately 820~GeV.

Although we will not explore them in detail here, we note that related gaps can be found in models with complex hidden valleys. For example, starting with the same spectrum as in figure~\ref{fig:sq2_vs_H_mgo800} ($\go-\sq-\Ho$ with stops heavier), one can replace the RPV $\Ho^0_1 \to jjj$ decay with a decay chain into a ``2-step'' Hidden Valley:
\begin{equation}
\label{eqn:hv2step}
\Ho^0_1 \to S\tilde S,\qquad \tilde S \to S \psi,\qquad S\to gg
\end{equation}
with $S$, $\tilde S$  hidden valley states, and $\psi$ the stable LSP (which could either be the gravitino or another particle within the HV sector; in any event the superpartner of $\psi$ is assumed to be irrelevant in this decay). If $m_{\tilde S}-m_S$ is very small and positive, and $\psi$ is close to massless, then this decay process yields almost no \MET; this is the kinematics typical of Stealth Supersymmetry~\cite{Fan:2011yu,Fan:2012jf}, which like RPV can serve to completely remove \MET from the event.  Without a jet $p_T$ hierarchy, the model is constrained by the ATLAS 6-7 high-$p_T$ jets search and/or the CMS BH search, but if the gluino decay chain creates a jet $p_T$ hierarchy, we again find a weakening of the limits on $m_\go$ below 1~TeV.  In other HV models, we have even found examples of isolated gaps at $m_\go\sim 500$~GeV.

We conclude, therefore, that there is a discernible gap in the search strategies here, corresponding to final states with hierarchical jet $p_T$'s, where the multiplicity of sufficiently high-$p_T$ jets drops below the thresholds of the ATLAS 6-7 jet search. It would appear that using a common $p_T$ cut for all of the jets is responsible for these holes in the coverage. However, a staggered set of $p_T$ cuts, while more efficient for signal, would admit much more QCD multi-jet background.  Presumably, if a looser set of $p_T$ cuts is used on some of the jets, then, to compensate, some use of angular information is required.  Note that in typical QCD events the hardest jets are roughly back-to-back; that will often not be the case for these signals.  This could be put to use to reduce backgrounds. It is also possible that jet substructure methods~\cite{Butterworth:2008iy,Kaplan:2008ie,Ellis:2009me,Krohn:2009th,Thaler:2010tr,Ellis:2012sn} may aid in the cases with a heavy gluino and a light and boosted unstable SM LSP.

\section{Potential Loopholes}
\label{sec:loopholes}

We have presented some evidence that the weakest limits on natural SUSY models with gluinos below 1~TeV come in signals that have no \MET, few or no tops, and only a few hard jets.  Only in special cases have we seen limits on the gluino drop far below 1~TeV.  We now must ask if this is sufficient evidence to suggest that almost all natural SUSY models with gluinos are subject to similar constraints, or whether there is a broad class of models that we did not consider for which much weaker limits could arise.

One general weakness in our study is that we have not considered models that populate a variety of final states,  so that the signal gets spread across multiple searches.  For example, if only half the events have all jets (the remainder having a photon or a lepton), and these have a variety of multiplicities, then each search may be weakened considerably. It is likely however that by combining the different searches, as in the study~\cite{Buchmueller:2013exa}, strong limits can still be obtained. On the other hand, the holes caused by a jet $p_T$ hierarchy, identified in figures~~\ref{fig:go_ho_200_HV1}, \ref{fig:topdilution-grid}, \ref{fig:sq2_vs_H_mgo800} and~\ref{fig:go_sq_150_350_UDD212-effthr}, do not seem to arise due to such an issue.

Another potential weakness is that we have only considered models where the final states contain gluons, light quarks and top quarks, and among these, only ones with the higgsinos in the cascade, neglecting the electroweak-inos and sleptons, as well as any non-MSSM states.  Cascade decays often result in leptons or taus (from emitted $W,Z,h$ bosons or sleptons) or photons (from $h$ decays or as in GMSB).  If these objects are common enough to matter, the lower backgrounds associated with them should in principle make the search program more sensitive to these models. However, at the present time, not all relevant searches have been updated for 8~TeV, and some have not been carried out at all. We should therefore consider how the presence of such objects affects {\it current} limits, rather than the potential ones.

We will also discuss our general assumption that first-generation squarks are too heavy to affect the gluino pair production rate, arguing that limits should not degrade even if the squark masses are comparable to the gluino mass. Finally, we will make some remarks about possible loopholes from other models, including non-MSSM cascade decays, which could produce many soft jets and unusual kinematics if the higgsinos are heavy.

\subsection{First-generation squarks}
\label{sec:1st-gen-sq}

First-generation squarks enter the diagrams for gluino production, making the cross section dependent on their masses. Throughout the paper, we have been using the gluino pair-production cross section that corresponds to very heavy squarks. If the squarks are lighter, the $\go\go$ cross section is decreased due to interference. To what extent can this affect our limits?

For squarks degenerate with the gluino, where the destructive interference is maximal, the {\it direct} $\go\go$ production cross section at the 8~TeV LHC gets reduced by 40\% (17\%) for a 1000~GeV (500~GeV) gluino~\cite{Beenakker:1996ch,Beenakker:1996ed}. However, production processes involving squarks largely compensate (and often even overcompensate) for this loss.

For $m_{\sq_1} > m_\go$ (where $\sq_1$ refers to the first-generation squarks), $\sq_1$ typically decays as $\sq_1 \to q\go$, so $\sq_1\go$ or $\sq_1\sq_1^{(\ast)}$ production leads in the end to gluino pairs, plus one or two jets. If we include these processes, the maximal decrease in the total $\go\go + X$ cross section as we lower $m_{\sq_1}$ towards $m_\go$ is only 24\% (5\%) for $m_\go = 1000$~GeV (500~GeV), and is obtained for $m_{\sq_1} \sim 2500$~GeV (1900~GeV). Note that for squarks of this mass or higher, their decays to the gluino are not suppressed by phase space and therefore branching fractions for their other decays, not producing gluinos, would likely be negligible. So the limits we derive cannot be weakened significantly. Moreover the increased multiplicity and $S_T$ in squark-produced gluino events may even enhance the limits in many searches, such as the CMS BH search. 

Meanwhile, if $m_\go > m_{\sq_1}$, limits depend on how the squarks decay.  However, the rates for first-generation squark pair production are enormous --- if $m_\go\sim m_{\sq_1} \sim$  1~TeV, $\sq_1\sq_1^{(\ast)}$ production is 4 times larger than gluino production would be with squarks decoupled, and the cross sections grow very rapidly as $m_{\sq_1}$ decreases.  It is very difficult to hide these particles.  If there is any significant \MET, the signal is easily excluded.  If the squarks decay resonantly to two jets, they are excluded by the CMS 2-jet pair search~\cite{Chatrchyan:2013izb}; if they decay resonantly to 3 jets, they are excluded by the ATLAS and CMS 6-7 jet searches~\cite{ATLAS-CONF-2013-091,CMS-PAS-EXO-12-049}; if they decay resonantly or non-resonantly to four or more jets, the ATLAS 6-7 jets and the CMS BH search are almost certain to be sensitive; the latter search is also sensitive if they decay to four or more objects of any type.  The situation is even more constrained if leptons or photons are regularly produced.  Thus there is very limited room for first-generation squarks lighter than a 1~TeV gluino to have escaped notice. 

\subsection{Extra $b$ jets}

To be conservative, we selected examples with $b$-jets arising only from the $b$ and $t$ quarks that are almost inevitably present in models with stops and sbottoms, but not from Higgs and $Z$ decays that in some other scenarios may be present as well. We also chose models with no $b$ quarks in RPV or HV-particle decays, whereas in fact $b$'s are quite common in explicit models.  The LSST search, the ATLAS high-multiplicity low-\MET search and the ATLAS 6-7 high-$p_T$-jets search all contain search regions which use $b$ tagging.  The presence of more $b$ jets can only improve the limits; indeed we saw strong exclusion limits from searches with $>2$ $b$ tags for the $t$- and $b$-rich signals in figures~\ref{fig:go_vs_ho_min} and~\ref{fig:go_st_350_ho_200} (left).

\subsection{Leptons, including taus}
\label{subsec:leptons}

We have discussed leptons from top quarks in section~\ref{sec:tops}, but leptons can also arise from $W$, $Z$ or $h$ emission, on- or off-shell sleptons, or lepton-number-violating RPV operators. Such leptons can be hard or soft, and need not be accompanied by a $b$ quark.  We need only consider cases where each gluino {\it commonly} produces 1 lepton, so that gluino pair events typically have 1 or 2 leptons.  Limits on models with many $n \geq 3$ lepton events at high $S_T$ and/or \MET are extremely strong, even if all leptons are taus~\cite{Chatrchyan:2013xsw,CMS-PAS-SUS-13-008,CMS-PAS-SUS-13-010,ATLAS:2012epx,ATLAS:2013rla,ATLAS:2013qla}.  Models with many same-sign dileptons also receive very strong limits~\cite{CMS-PAS-SUS-13-013,ATLAS:2013tma,CMS-PAS-B2G-12-015,ATLAS-CONF-2013-051}.

We organize the discussion by the quantity of \MET, because the largest gap we observe occurs in the low-\MET region.
\begin{itemize}
\item Large \MET: This case is powerfully constrained. In the ``minimal natural SUSY models'' studied in section~\ref{sec:minimal}, we saw the CMS search for $\ell$+$b$+$6j$+\MET~\cite{CMS:wwa} excluding gluinos almost up to the kinematic limit, except for heavy higgsinos. Similarly strong limits apply to scenarios without $b$-jets; see for example the ATLAS search~\cite{ATLAS-CONF-2013-062}.  This ATLAS search also provides sensitivity to very soft leptons. Opposite-sign dileptons are covered in~\cite{ATLAS-CONF-2013-062,TheATLAScollaboration:2013via,TheATLAScollaboration:2013gha}.  Taus are covered both by jet + \MET searches (in which hadronically decaying taus appear as jets), and by searches that specifically require hadronic taus~\cite{ATLAS:2013ama}.

\item No \MET:  Our discussion in section~\ref{sec:all-hadronic} is unaffected by the presence of soft leptons, since these do not impact the ATLAS 6-7 jets and CMS BH searches, and by hard leptons if multiplicity is high, since the CMS BH search includes hard leptons as ``objects''.  Low multiplicity models, albeit non-generic, are also covered if dileptons are common: strong constraints come from the CMS leptoquark search~\cite{CMS:zva} (which requires 2 hard leptons + $\geq 2$ hard jets) and the same-sign dilepton searches.

\item Small \MET: We saw in section~\ref{sec:MET} that, in the absence of leptons, there are searches covering low-multiplicity high-\MET models, others covering high-multiplicity low-\MET models, and still others covering models without \MET.  But with one or two detected leptons, there are no searches that fully exploit high multiplicity to compensate for low \MET. Meanwhile, although the jets + low \MET searches include hadronic taus among their jets, they veto on electrons and muons down to $\sim 10$~GeV. This means that if there are no taus among the leptons, there is currently a gap in the search strategy.  In section~\ref{sec:tops} we saw that the proposed LSST search for events with one electron or muon is powerful in the presence of $b$ jets.  To cover the full range of models, this search should be performed, and also extended (as discussed in~\cite{Lisanti:2011tm}) to include events with zero $b$ tags and events with two leptons.  In all cases $t\bar t$+jets is a dominant background, with some $W$+jets and $Z$+jets contributions.  Note that for a single lepton, it is important to use a non-leptonic trigger, such as an $H_T$ trigger, so that lepton $p_T$ cuts for the leptonic searches can be as low as possible, and complementary to the lepton veto applied in the jets + low \MET searches.
\end{itemize}

\subsection{Photons}

As in the case of leptons, if photons are rare then our results are unchanged. And models in which 3 or more photons, or a combination of 3 or more leptons and photons, are common must already be excluded. Even though no explicit limits are available, such signals in high $S_T$ events would have been very difficult to miss.

If diphoton events are common, there exist searches (although not yet with the full dataset) that set strong limits on scenarios with even a small amount of \MET~\cite{Aad:2012zza,CMS:2012kwa} and scenarios with large jet multiplicity~\cite{CMS:2012un} (in the latter case, using an LSST-type method).

If 1-photon final states are more common than 2-photon final states, the situation is less clear. While the search~\cite{CMS:2012kwa} constrains scenarios with a single hard photon and even a small amount of \MET, there are no searches for a single photon and large jet multiplicity without a \MET requirement. The CMS BH search, which counts photons as ``objects'', still applies to very high multiplicity scenarios, but scenarios with intermediate multiplicity and no \MET may or may not be constrained by the ATLAS 6-7 jet search~\cite{ATLAS-CONF-2013-091}, depending on the details of how that search (which focuses on jets) behaves in the presence of photons. Therefore, there may be a gap in the rare no-\MET models where single-photon + jets events are common.\footnote{Some limits may be obtainable by studying the photon + jets control plots for the ATLAS 2-6 jets + \MET search~\cite{ATLAS-CONF-2013-047}.  We thank A.~Barr for this suggestion.}  There are also no searches for a single soft photon + jets + \MET.\footnote{Note that these events may not be included in several of the jets + \MET searches~\cite{ATLAS-CONF-2013-047,CMS-PAS-SUS-13-012} that we rely upon in this work, as they veto or effectively veto on photons.  Some other searches, including~\cite{Aad:2013wta}, do not describe their procedures in detail.}  Clearly, since SM photons are rarer than jets, it should be possible to put limits on models with common photons that are much stronger than for models without photons, but for now, a loophole may remain.

\subsection{Other models to consider}
\label{subsec:othermods}

Though we discussed them briefly in section~\ref{sec:approach}, we have not explicitly considered the possibility of non-SM colored particles $C$, and their superpartners $\tilde C$, that might contribute to gluino decays via $\go\to C\tilde C$.   Each of these particles (given that stable $R$-hadrons of moderate mass are excluded) must decay to at least one jet, along with either \MET or additional visible objects. In most contexts, the resulting final states will not be significantly different from cases we have already discussed: typically these decays lead to high \MET, or if \MET is absent, to high object-multiplicity and often high jet-multiplicity. We do not see any simple examples that easily evade our arguments.  Except in special cases, which correspond to loopholes we have already discussed, we expect limits in the TeV range.

We have not specifically addressed compressed spectra in which the LSP does not decay (visibly) and the gluino lies close in mass to the LSP. Since the higgsino is bounded by 400~GeV, we need only consider gluinos in the 700~GeV range or below.   A tightly compressed spectrum is probably ruled out by now; see footnote~\ref{footnote:compressed}. Possibly more problematic are spectra that are substantially but not completely compressed; we saw hints of this in the large $m_\Ho$ region of figure~\ref{fig:go_vs_ho_min}, where robust limits were lost.  To tie up this loose end requires more careful simulation of ISR, using matrix-element/parton-shower matching techniques.\footnote{For heavier gluinos, a partly compressed and natural spectrum, in which the higgsinos are light but are cut out of the decay chain, is contrived, but could evade searches for some time. As an example, suppose a gluino, squark and bino have large nearly degenerate masses, $m_\go>m_\sq>m_{\tilde B}$, where the bino can decay invisibly to non-MSSM particles, bypassing the higgsinos.}

A hybrid of these two situations may also cause problems.  Suppose the gluino is not too heavy and the LSP is stable and rather heavy, for instance $m_\go=900$~GeV, $m_\Ho=400$~GeV. If the gluino decays in several steps, each of which involves emission of a jet of low to moderate $p_T$, leaving behind a massive particle, then after the cascade the LSP will be non-relativistic in the gluino rest frame.  This last is much the same as in the compressed case, except that more visible energy has been produced along the way.  As in the compressed case, back-to-back gluino pairs produce nearly back-to-back LSPs, with limited \MET.  Only recoil of the system against ISR generates much \MET, and matching techniques are required to predict the signal.  It may be that models of this class with high $m_\go$ are difficult to exclude with current searches, but more investigation is required to determine whether this is the case.

The scenarios we considered created signals with up to 20 partons produced; see figure~\ref{fig:6-20-partons}.  Slightly more complicated models with interesting dynamics could cause gluino decays to create even more partons, though some of these will merge, meaning the number of observed objects may not be quite so large.  If the signal has a very large number of soft objects, the CMS BH search might not be sensitive; jets+\MET searches might fail if a large \MET signal is accompanied only by many soft jets.  It may be that innovative approaches are needed for such models.

Models with exotic signatures --- for example, long-lived particles, non-isolated clusters of leptons and hadrons, quirks~\cite{Kang:2008ea}, or fireballs of soft pions or photons --- might not have been picked up by the existing searches.  However, to determine this requires a dedicated study of the event selection criteria, quality cuts, and analysis methods of all the relevant searches.  We have not attempted this, but searches for these exotic signatures in high $S_T$ events should be carried out.  It is common to use fully inclusive searches for these exotic object(s), but if they arise in gluino events in would be useful to require high $S_T$ and/or \MET and/or hard jets in event selection.

\section{Conclusions}
\label{sec:conclusions}

In this paper we have considered supersymmetric extensions of the Standard Model in which the lightest higgsinos are lighter than 400~GeV (for naturalness) and a gluino is around or below 1~TeV (so that it is produced abundantly at the 8~TeV LHC).  We make no other explicit assumptions concerning the model.  We have not, for instance, assumed $R$-parity conservation, or a minimal natural SUSY spectrum, or that first-generation squarks are at accessible mass scales, or that the new particles of the model are restricted to the MSSM.  In this final section, we will summarize our results regarding the status of natural SUSY models and discuss ways in which the reach for some of them could be improved.

First of all, we find that gluinos that produce significant \MET or frequently produce a top quark in their decays are excluded, usually robustly, up to and often beyond 1~TeV. Moreover, this statement is quite conservative, both because it is based solely on gluino pair production, and because we do not statistically combine search regions, either within an individual search, or between multiple searches.

For all-jet final states without \MET, the situation is more complex. Limits near 1~TeV are still common, with the ATLAS 6-7 jets search~\cite{ATLAS-CONF-2013-091} covering the low-multiplicity region and the CMS BH search~\cite{Chatrchyan:2013xva} protecting at high multiplicity. Robust exclusion limits are generally well below 1~TeV, and, especially for low gluino masses, may be sensitive to the modeling of ISR (see appendix~\ref{app:simulation} for examples).  Moreover, holes reaching to significantly lower $m_\go$ can arise when the signal has a small number of jets that have a large hierarchy in $p_T$.  In this case the ATLAS search~\cite{ATLAS-CONF-2013-091}, which requires jets above a high $p_T$ threshold, becomes insensitive. Importantly, the problems surrounding holes of this class may be exacerbated at a higher collision energy, since this will probe more massive gluinos that are further split from the LSP (which for natural models will remain below 400~GeV).  Note this issue affects not only gluinos and supersymmetry, but also colored particles with all-hadronic final states that may arise in other models.

In our exploration of the various signatures, we used simple models.  This had the effect that in large regions of parameter space, experimental sensitivity to gluino production was dominated by a single final state or a small number of final states.  In transitions between such regions, the signal gets spread across multiple searches, and it is possible that no individual search will set a strong limit.  A similar issue can arise in more complex models with a variety of possible cascade decays, as we mentioned in section~\ref{sec:loopholes}.  (We did not find many examples in which limits are weakened by this effect, but we did not explore this issue systematically.)  In such situations, it can become important to combine searches within an experiment, to obtain significantly better limits.  Since we have observed that the searches in table~\ref{tab:searches} have strong and complementary sensitivity, the experimental collaborations might want to consider, for Run II, redesigning these searches (and perhaps others) so that they could eventually be combined.

Importantly, however, this issue does {\it not} affect the holes that we have identified in section~\ref{sec:all-hadronic}, because they arise even in models with simple decay chains.   In order to fill these holes, and to deal with other loopholes discussed in section~\ref{sec:loopholes}, so that  ATLAS and CMS may cover the full territory up to $m_\go\sim 1$~TeV, we suggest that a few minor adjustments to the search program might be advisable.

\subsection{Suggestions regarding lepton + many jets searches}

The current array of leptonic searches within both ATLAS and CMS is not optimal for constraining models with high multiplicity and low \MET.  This warrants implementing something like the LSST search~\cite{Lisanti:2011tm}, which we have simulated here.  Our results suggest that a leptonic search of this type, with bins at high jet multiplicity and high $S_T$, will have reach beyond the existing leptonic searches.  Moreover, the reach for scenarios where the only source of \MET is top quarks appears comparable to and possibly greater than the no-lepton high-jet-multiplicity low-\MET searches that both experiments have carried out~\cite{CMS-PAS-SUS-13-012,Aad:2013wta}.  Since the LSST search and the non-leptonic high-multiplicity low-\MET search at each experiment would use independent event samples, the combination might have significantly increased sensitivity.  This could be even more important at a higher center-of-mass energy than 8~TeV, because in natural models it is believed there should be a top squark below 1~TeV, and once gluino searches are targeting $m_\go>1.2$~TeV it becomes almost impossible to avoid having $\go\to t\tilde t$ kinematically accessible.  Therefore, maximum sensitivity to an excess of $t$-plus-jets at high multiplicity is highly desirable at Run II of the LHC.

We have also noted that for the ATLAS and CMS high-multiplicity low-\MET searches, there appear to be risks that a gluino that produces top quarks may contaminate the one-lepton samples used as control regions, hiding the signal as part of the $t\bar t$ background.  Combining a leptonic and non-leptonic search with a complete set of control regions may help to eliminate this risk, although we have not studied this in detail.

One issue to consider in the LSST search is binning by number of $b$ tags.  In the bulk of our discussion, the motivation for the LSST search was top quarks, in which case $\geq 1$ $b$ tag should be included.  Search regions requiring larger numbers of $b$'s are also quite important in our context because many of the natural SUSY scenarios have large $b$ multiplicities. Sources of $b$'s in such scenarios can include tops from stop-mediated gluino decays, tops/bottoms from the decays of stops to neutralinos/charginos, tops and bottoms from analogous processes involving sbottoms, preferential decays of higgsinos through stops/sbottoms due to Yukawa couplings and/or RPV couplings with multiple third-generation indices.

Additionally, as we noted in section~\ref{subsec:leptons}, a bin with {\it no} $b$ tags may be important for covering low-\MET signals in which leptons come from sources other than top decays. We also mentioned that it might be important to use non-leptonic triggers, e.g., $H_T$ triggers, in order to allow the search to be extended to softer leptons. For low-\MET scenarios in which dilepton events are common, an LSST-type search looking for dileptons + many jets is motivated as well.

There may be a similar motivation for a single photon + many jets search.  In this case, the background is QCD with a real or fake photon, and measuring this background may be very difficult.  Nevertheless, the possibility that a single photon may be common in gluino pair events should be considered, though it may well be that whenever this signal is important, the sensitivity to diphoton events is better.

\subsection{Suggestions regarding all-hadronic and related searches}

For the ATLAS 6-7 jets search~\cite{Aad:2013wta}, we suggest that new bins be studied that might allow for events with 2 very high-$p_T$ and non-back-to-back jets to be accompanied by 4 to 8 jets at lower $p_T$.  The use of angles between jets as well as jet $p_T$s may allow reduction of QCD background to an acceptable level. We believe that this could help fill the holes with a jet $p_T$ hierarchy that we have identified in this paper. For instance, in figure~\ref{fig:topdilution-grid} the gap extends as low as  $m_\go\sim 800$~GeV, and can go even lower still in some models, as exhibited in figure~\ref{fig:sq2_vs_H_mgo800}.

For low multiplicity, CMS might be able to place interesting limits with their search for gluinos that decay to three jets via RPV~\cite{CMS-PAS-EXO-12-049}.  While an LSP gluino with $m_\go\gg400$~GeV is not natural, three-jet decays via a two-step cascade (e.g., $\go\to q\sq^*\to qqq$ via RPV, see figure~\ref{fig:spectraCh7a}) can occur in natural models, and we studied them in figures~\ref{fig:6-20-partons} and~\ref{fig:go_sq_150_350_UDD212-effthr}.   Unfortunately, as the CMS search~\cite{CMS-PAS-EXO-12-049} involves a fit to data that we cannot model, it is unknown whether or not cases like these (in which the jets are produced with different kinematics) are constrained.   Similarly, when the gluino decays to three jets plus a small amount of \MET (e.g., $\go\to g\Ho$, $\Ho\to S$ + \MET, $S\to gg$, as in figure~\ref{fig:go_ho_200_HV1} with $m_s\sim 150$-$190$~GeV), the resonance will be distorted. In such cases, it is even less clear what may be excluded.  A reanalysis of CMS data in~\cite{CMS-PAS-EXO-12-049} for a broader range of models would be valuable.

Lastly, our results, obtained from an extremely conservative interpretation of the data (see appendix~\ref{app:BH}), very clearly suggest repurposing the CMS black hole search~\cite{Chatrchyan:2013xva}.  Such an improved search would be sensitive not only to gluinos but to any particle that is produced with a very high rate and decays into many objects.  There are several natural modifications to consider for this search.  First, stronger limits than ours could be obtained by CMS by reinterpreting the search just as it stands now.  Second, extension of the control region down below $S_T=1900$~GeV, so that $S_T>1900$~GeV could be used as a signal region, would presumably increase the search's sensitivity to such signals.  Third, one might imagine adding bins with even more objects than $N\geq 10$; signals from gluinos with very complex cascade decays and/or decays into a complex Hidden Valley may heavily populate those bins.  Finally, one could perhaps imagine extending the search to lower $S_T$ while simultaneously reducing slightly the $p_T$ cuts on the objects, to maintain sensitivity to lower-mass particles whose decays produce softer jets.  This would be especially important for alleviating the dependence on the ISR modeling on the far tails, discussed in appendix~\ref{app:simulation}.

\subsection{Summary and outlook}

In conclusion, we have constructed an argument that can be used to understand which classes of natural SUSY models with an accessible gluino are currently excluded, or can soon be excluded, by LHC data. Our general conclusion is that the search strategies currently in use by ATLAS and CMS already provide  {\it almost} complete coverage of the natural SUSY parameter space with a gluino at or below 1~TeV. If a model is not covered by searches for the three main signatures of gluino decays (\MET, tops, and high multiplicity), it is usually ruled out by searches for even lower background objects such as multiple leptons. We believe that the few remaining gaps allowing gluinos below 1~TeV can likely be filled with more dedicated study of the Run I dataset.  We also think that the same search methods will continue to be powerful when the LHC resumes at higher energy and pushes toward $m_\go\sim 2$~TeV.

We emphasize that our strong claims about natural SUSY would not have been possible without the ATLAS 6-7 jet search~\cite{Aad:2013wta}, and our very conservative repurposing of the CMS BH search~\cite{Chatrchyan:2013xva}, which are crucial in providing powerful limits on models with very low \MET.  To avoid exclusion up to $m_\go\sim 1$~TeV, it appears a model must have a gluino which almost always decays only to jets, without large \MET or tops.  Even then, most models are ruled out nearly to $m_\go\sim 1$~TeV, unless the multiplicity is small enough to reduce the sensitivity of the BH search, and the kinematics make the jet $p_T$s so dissimilar that a fixed and large $p_T$ cut, as used in the ATLAS 6-7 jets search, has low signal efficiency. Such gaps in coverage could potentially be closed if these two searches were tailored for greater sensitivity to a wide class of models in which gluinos decay with very low \MET, either via $R$-parity violation or via a low-\MET Hidden Valley, including Stealth Supersymmetry.

We also noted possible holes that may arise when low-\MET events with a single lepton or photon are common. In this regard, and also for maximizing sensitivity to top quarks in gluino decays, we argued that the lepton-plus-many-jets search suggested in~\cite{Lisanti:2011tm}, and its variants with two leptons or a photon, still appear very well-motivated.

Note that our results have application far beyond supersymmetry.  Any particle with color and/or spin representations exceeding those of a gluino will have a comparable or larger cross section for the same mass.   So varied are gluino decays across SUSY parameter space that nearly all decays that one can think of for {\it any} colored particle (excepting decays to heavy invisible particles) have been excluded at gluino production rates up to the 1~TeV range, and for particles in higher color and spin representations, well beyond.

Our methodology involved drawing very general lessons from a small number of targeted studies.  With such an approach, loopholes are inevitable, though as yet we have found very few that we do not know how to close.  Ideally, the ``rule of thumb'' at the end of section~\ref{sec:introduction} that guided our own thinking should make it easier to find other loopholes, and to identify any additional gaps in the CMS and ATLAS search strategies.

Looking ahead, we believe this way of organizing one's thinking will continue to be useful as the LHC restarts at higher energy.  We have seen that certain types of searches are extremely powerful already, while others deserve to be further optimized.  The exercise of closing the remaining loopholes, and extending the reach as far as possible within the 7-8~TeV data set, should prove helpful as we prepare for Run II.

\section*{Acknowledgments}

We thank Nathaniel Craig, Jim Hirschauer, Xinlu Huang, Andrey Katz, Graham Kribs, Michelangelo Mangano, Maxim Perelstein, Peter Skands, Ken van Tilburg, and Brock Tweedie for useful discussions.  We thank Rob Bainbridge, Alan Barr, Erez Etzion, Claire Gwenlan, Eva Halkiadakis, Andreas Hoecker, Salvatore Rappoccio, Chris Rogan, Thomas Speer, Antonella Succurro and Frank Wuerthwein for help with specific questions regarding the searches.  J.A.E., Y.K.\ and M.J.S.\ were supported by DOE grant DE-FG02-96ER40959. The work of  M.J.S.\ was also supported by NSF grant PHY-0904069 and DOE grant DE-SC003916. The work of D.S.\ is supported by a DOE Early Career Award and a Sloan Foundation Fellowship. Y.K.\ thanks the KITP at UCSB (supported in part by NSF grant PHY11-25915) and the Theory Division at CERN for their hospitality while portions of this work were completed.

\appendix

\section{Our Methodology for Reinterpreting Searches}
\label{app:methodology}

In this appendix, we detail how results of existing searches are reinterpreted for the scenarios we study here.  Most of the ATLAS and CMS searches we use are interpreted straightforwardly, as we will shortly describe in section~\ref{app:mostsearches}.  The CMS black holes search suffers significant contamination of control regions by our signals, and we treat it in a special though very conservative fashion, as we will describe in section~\ref{app:BH}.  Finally, we simulate a lepton + many jets search proposed in~\cite{Lisanti:2011tm} but not yet carried out by either ATLAS or CMS.  This search and our simulation of it are described in section~\ref{app:LSST}.  Specifics of simulation and validation are presented in appendix~\ref{app:simulation}.

\subsection{Reinterpretation of existing searches}
\label{app:mostsearches}

\begin{table}[t]
\begin{center}
\begin{tabular}{|c|c|c|}\hline
Search                            & Data (fb$^{-1}$)  & Reference \\\hline
ATLAS 2-6 jets + large \MET       &        20.3       & \cite{ATLAS-CONF-2013-047} \\
ATLAS 7-10 jets + low \MET        &        20.3       & \cite{Aad:2013wta} \\
CMS jets + \MET                   &        19.5       & \cite{CMS-PAS-SUS-13-012} \\
CMS $b$-jets + \MET               &        19.4       & \cite{Chatrchyan:2013wxa} \\
ATLAS 3$b$-jets + \MET            &        12.8       & \cite{ATLAS:2012ksq} \\
CMS jets w/$\alpha_T$ w/$b$       &        11.7       & \cite{Chatrchyan:2013lya} \\
CMS monojet + \MET                &        19.5       & \cite{CMS:rwa} \\
ATLAS monojet + \MET              &        10.5       & \cite{ATLAS:2012zim} \\
ATLAS $\ell$+jets+\MET            &         5.8       & \cite{ATLAS:2012tna} \\
CMS $\ell$+$b$+$6j$+\MET          &        19.4       & \cite{CMS:wwa} \\
ATLAS SS dilepton w/$b$ (SUSY)    &        20.7       & \cite{ATLAS:2013tma} \\
ATLAS SS dilepton w/$b$ (Exotics) &        14.3       & \cite{ATLAS-CONF-2013-051} \\
CMS SS dilepton + jets (w/$b$)    &        19.5       & \cite{CMS-PAS-SUS-13-013} \\
CMS SS dilepton + jets w/2-3$b$   &        10.5       & \cite{Chatrchyan:2012paa} \\
CMS OS/SS dilepton ($t'$)         &        19.6       & \cite{CMS-PAS-B2G-12-015} \\
ATLAS $\ell$+$6j$ w/3-4$b$ ($t'$) &        14.3       & \cite{ATLAS:2013ima} \\
CMS 2nd gen.\ leptoquarks         &        19.6       & \cite{CMS:zva} \\
ATLAS 6-7 high-$p_T$ jets         &        20.3       & \cite{ATLAS-CONF-2013-091} \\
CMS black holes                   &        12.1       & \cite{Chatrchyan:2013xva} (reanalyzed) \\\hline
\end{tabular}
\end{center}
\caption{8~TeV LHC searches included in our study.}
\label{tab:all-searches}
\end{table}

Table~\ref{tab:all-searches} lists the potentially relevant 8~TeV LHC searches that were included in our analysis.  Several other classes of searches (e.g., multileptons) are also powerful at constraining certain specific types of natural SUSY scenarios. We did not list them here since they are not relevant to the final states of the conservative scenarios that we study in this work (even though many of these searches are actually included in our simulation framework). Many searches from the 7~TeV LHC were included as well, but none of them are competitive at constraining our scenarios. Only searches that present their results as a set of cut-and-count bins are included,  because these are the ones that we can simulate easily and reliably.  Additionally, one would expect them to be more useful for general scenarios than searches that were tuned to address very specific models.  In general we use the search regions defined by ATLAS and CMS.  For the ATLAS $t'$ search~\cite{ATLAS:2013ima}, we defined the search regions ourselves, as $H_T > 800$, 1000, 1200, 1400, 1600, 1800~GeV for bins with 3 and $\geq 4$ $b$-tags (on top of ATLAS's selection).

Our methods for limit setting are somewhat conservative. For each point in the parameter space of the model, we use the search region that sets the strongest limit, without combining search regions. We also do not combine searches (in contrast to~\cite{Buchmueller:2013exa}, for example). Some searches (notably the CMS Razor search~\cite{CMS-PAS-SUS-13-004}, the CMS search for 3-jet resonances in multijet events~\cite{CMS-PAS-EXO-12-049} and the CMS search for paired dijet resonances~\cite{Chatrchyan:2013izb}) were not included in our analysis due either to their complexity or to their computational burden. These factors may be weakening our limits in some cases, but we do not believe this significantly affects our conclusions.

Note, however, that except for the treatment of the CMS black holes search (as described in the next subsection), we have not taken into account the potential contamination of the control regions of the various searches by our signals. It would therefore be beneficial for the experimental collaborations to confirm (or perhaps modify) our conclusions by reanalyzing models of the type we consider here, especially in cases where control region contamination could be an issue.

Our method for simulating the signals for these searches and deriving limits are given in appendix~\ref{app:simulation}.

\subsection{Special reinterpretation of the CMS black holes search}
\label{app:BH}

A search along the lines of the 8~TeV (12.1 fb$^{-1}$) CMS search for microscopic black holes~\cite{Chatrchyan:2013xva} proves extremely useful in the context of many of the models we consider in this paper, especially when the gluino decays to a large number of jets.  This CMS search simply requires a large multiplicity of hard objects ($\geq N$ jets, electrons, muons, and photons with $p_T > 50$~GeV) and looks for deviations on the tail of the $S_T$ spectrum, where $S_T$ is defined as the scalar sum of the $p_T$s of these objects (and \MET, if above 50~GeV). 
However, a naive, direct application of~\cite{Chatrchyan:2013xva} cannot be employed for most of our scenarios, because the search uses the control region $1.9~{\rm TeV} < S_T < 2.2$~TeV for normalizing the background prediction for each $N$. These control regions would be badly contaminated by a gluino signal unless the gluino is significantly heavier than $\sim 1$~TeV. 

To make use of the CMS data while avoiding this contamination issue, we derive very conservative limits through a simple procedure.  We assume the observed data is potentially {\it entirely from signal, with zero background}; then, for a particular model with a particular $m_\go$, if the predicted signal significantly exceeds the observed data (based on a standard frequentist method~\cite{Conway:2000ju}), then that value of $m_\go$ is excluded. Because gluino signals for high-multiplicity decays are so large, even exceeding the QCD background, this method turns out to be quite powerful despite being so conservative.
\begin{table}[t]
\begin{center}
\begin{tabular}{|c|c|c|c|}\hline
\multicolumn{2}{|c|}{Search region} & Events  & Limit on \\\cline{1-2}
$N \geq $ &  $S_T \mbox{ (GeV)} > $ & in data & $\sigma\times\epsilon$ (fb) \\\hline
\multirow{2}{*}{8}  & 1900 & 425 & 38   \\\cline{2-4}
                    & 2200 & 122 & 11.7 \\\hline
\multirow{2}{*}{9}  & 1900 & 111 & 10.7 \\\cline{2-4}
                    & 2200 &  35 & 3.8 \\\hline
\multirow{2}{*}{10} & 1900 &  25 & 2.9 \\\cline{2-4}
                    & 2200 &  10 & 1.4 \\\hline
\end{tabular}
\end{center}
\caption{Search regions we defined for interpreting the CMS BH search data~\cite{Chatrchyan:2013xva}.  As we cannot use the CMS background estimate because our signals significantly contaminate their control regions, the limits are based on our extremely conservative assumption that none of the observed events can be definitively attributed to the background.}
\label{tab:BH}
\end{table}

We defined our search regions, for number-of-objects thresholds of $N \geq 8$, 9, 10, to be $S_T > 1.9$~TeV and $S_T > 2.2$~TeV.  CMS does not present data at lower $S_T$, which limits our sensitivity (and the reliability of our estimates) for light gluinos. The observed numbers of events in these search regions and the corresponding limits on the size of a new physics signal are shown in table~\ref{tab:BH}.

\subsection{A simulated search for lepton-plus-many-jets (with $\geq 1\ b$ tag)}
\label{app:LSST}

In mid-2011, Lisanti, Schuster, Toro and one of us (MJS)~\cite{Lisanti:2011tm} identified a large gap in the strategies then in use by ATLAS and CMS: models with low \MET but high jet multiplicity would evade the SUSY searches for \MET, low multiplicity exotica searches, and extremely high-$S_T$ black hole searches. LSST proposed a general class of searches, with one or two rare objects (leptons, photons, $Z$) plus high jet multiplicity, to close this gap; and they studied the case of a lepton + $n$ jets, perhaps with $b$ tags ($\ell$ + $n$ $j$ + $b$ for short), and with no significant \MET requirement (except a small lower $M_T$ cut to remove fake leptons).  Such a search has not been performed, and similar searches that have been carried out~\cite{Chatrchyan:2012af,CMS-PAS-B2G-12-015,ATLAS:2013ima} are not as useful for constraining gluino scenarios. To understand the maximum reach of the 8~TeV data, we include the LSST proposed search in our study, as described below.

The 7~TeV CMS search for heavy quarks decaying as $Q\to tZ, tW$~\cite{Chatrchyan:2012af} is essentially an $\ell$ + $n$ $j$ + $b$ search (with $n = 4,5,6,7$).\footnote{We are unable to repurpose a similar single-lepton channel of the 8~TeV CMS search targeting heavy quarks decaying as $T\to bW, tZ, th$~\cite{CMS-PAS-B2G-12-015} because the analysis is based on a boosted decision tree, which we cannot duplicate.} Though this search is not very effective at constraining gluino models (the interesting high-$S_T$ region is compressed into a single overflow bin, and jet multiplicity extends only up to 7) we can utilize their data as a check on our simulation of the LSST search.

\begin{table}[t]
\begin{center}
\begin{tabular}{|c|c|c|c|c|c|c|}\hline
$S_T$ cut & \multicolumn{3}{|c|}{Background events} & \multicolumn{3}{|c|}{Limit on $\sigma\times\epsilon$ (fb)} \\\cline{2-7}
(GeV)     & $n \geq 7$ & $n \geq 8$ & $n \geq 9$    & $n \geq 7$ & $n \geq 8$ & $n \geq 9$ \\\hline
 800 & 1740 & 480  & 119  &  73  &  20  & 5.1 \\
1000 &  830 & 280  &  86  &  35  & 11.8 & 3.7 \\
1200 &  370 & 141  &  52  & 15.5 &  6.0 & 2.3 \\
1400 &  164 &  64  &  27  &  7.0 &  2.8 & 1.24 \\
1600 &   74 &  30  & 13.2 &  3.2 & 1.41 & 0.71 \\
1800 &   32 & 15.5 &  7.7 & 1.49 & 0.79 & 0.46 \\
2000 & 14.4 &  6.8 &  2.8 & 0.75 & 0.42 & 0.25 \\
2200 &  8.1 &  3.7 & 1.54 & 0.50 & 0.29 & 0.2  \\
2400 &  4.7 & 1.94 & 0.70 & 0.33 & 0.2  & 0.15 \\
2600 &  2.1 & 1.06 & 0.32 & 0.25 & 0.2  & 0.15 \\
2800 & 1.20 & 0.42 & 0.13 & 0.2  & 0.15 & 0.15 \\
3000 & 0.32 &   0  &   0  & 0.15 & 0.15 & 0.15 \\\hline
\end{tabular}
\end{center}
\caption{Search regions and estimated expected background (for 20~fb$^{-1}$ at the 8~TeV LHC) for our implementation of the proposed LSST search~\cite{Lisanti:2011tm} for lepton-plus-many-jets (with $\geq 1\ b$ tag). In computing the expected limits, a systematic uncertainty of $50\%$ has been assumed for all bins.}
\label{tab:LSST}
\end{table}

To estimate the reach of an $\ell$ + $n$ $j$ + $b$ search, we simulated a modified version of~\cite{Chatrchyan:2012af}. We assumed 20~fb$^{-1}$ of data at the 8~TeV LHC, kept the object selection requirements unchanged,\footnote{The $p_T$ requirements on electrons, muons and jets are 35, 42 and 35~GeV, respectively. The $b$-tagging efficiency is assumed to be 65\%. The selection requires exactly one lepton (after acceptance, identification efficiencies and isolation requirements), at least four jets with $p_T > 100$, 60, 50, 35~GeV, at least one $b$-tagged jet, and $\slash E_T > 20$~GeV. For simplicity, and in order to be able to validate against~\cite{Chatrchyan:2012af}, we have not attempted to optimize these cuts.} and defined search regions with jet multiplicities $n \geq 7, 8$ and $9$ and $S_T$ cuts in steps of 200~GeV (namely, $S_T > 800, 1000, \ldots, 3000$~GeV), where $S_T$ is the scalar sum of the $p_T$s of the lepton, the jets and \MET. Based on 2010 ATLAS and CMS lepton + jets studies (reconfirmed more recently in the control samples for a 2012 ATLAS search~\cite{Aad:2013wta}), LSST argued~\cite{Lisanti:2011tm} that for large multiplicity and high $S_T$ the background is dominated by $t\bar t$ + jets, with $W$ + jets a small contributor, especially with a $b$-tagged jet requirement.  To estimate the expected background, we generated a $t\bar t$+jets sample, matched up to 5 extra jets (using {\sc ALPGEN}~\cite{Mangano:2002ea} with {\sc Pythia~8}~\cite{Sjostrand:2007gs}). All other backgrounds were assumed negligible, as was true for the 7~TeV CMS search~\cite{Chatrchyan:2012af}. Our $S_T$ distributions for 4, 5, 6 and 7 jets at 7~TeV agree with those of CMS, but only after multiplying by 1.6 to get the same normalization (which could be due to limitations of our detector simulation or different parameters used in the matching procedure).  We applied this same normalization factor to our 8~TeV distributions.  We anticipate the actual experiment to use data-driven methods for estimating the $t\bar t$+jets background.  It is difficult for us to predict the systematic uncertainty that such an estimate would have as a function of $n$ and $S_T$. We present our results obtained with the assumption, which we believe is conservative, that the systematic uncertainty on the background in all search regions is $50\%$; lower systematics would not significantly improve our limits, since at high $S_T$ and high multiplicity the dominant uncertainty is statistical. The expected background and sensitivity for our search regions are shown in table~\ref{tab:LSST}.

A variant of the search with 3 $b$ tags would be very powerful in $b$-enriched signals which arise quite commonly in models with light top and bottom squarks, cascades with Higgs emission, RPV decays or HV sectors.  On the other hand, a search region requiring exactly zero $b$ tags, while more difficult to model because of the now significant $W$+jets background, would be important to implement for signals without $b$s.  Such signals can arise in models where, e.g., $W$ bosons from chargino-to-neutralino or squark-to-squark transitions (rather than top quarks) are an important source of leptons.  We have not explicitly studied models of this kind in detail, but such search regions are valuable for complete coverage.

\section{Details of Simulation and Limit Computation}
\label{app:simulation}

We define the desired superpartner spectra and compute decay branching fractions using {\sc SuSpect~2.41}~\cite{Djouadi:2002ze} and {\sc SDECAY~1.3}~\cite{Muhlleitner:2003vg} (importantly, {\sc SDECAY} contains the loop decay $\go\to g\tilde\chi^0$).  Events are generated using {\sc Pythia~8.175}~\cite{Sjostrand:2007gs}, with MSSM production and decay processes imported from {\sc Pythia~6.4.26}~\cite{Sjostrand:2006za}. We also use {\sc Pythia~8} to further decay the MSSM LSP, where relevant. As discussed in the main text, the only production processes we are interested in are $pp \to \go\go$.

\begin{figure}[t]
\begin{center}
\includegraphics[width=0.47\textwidth]{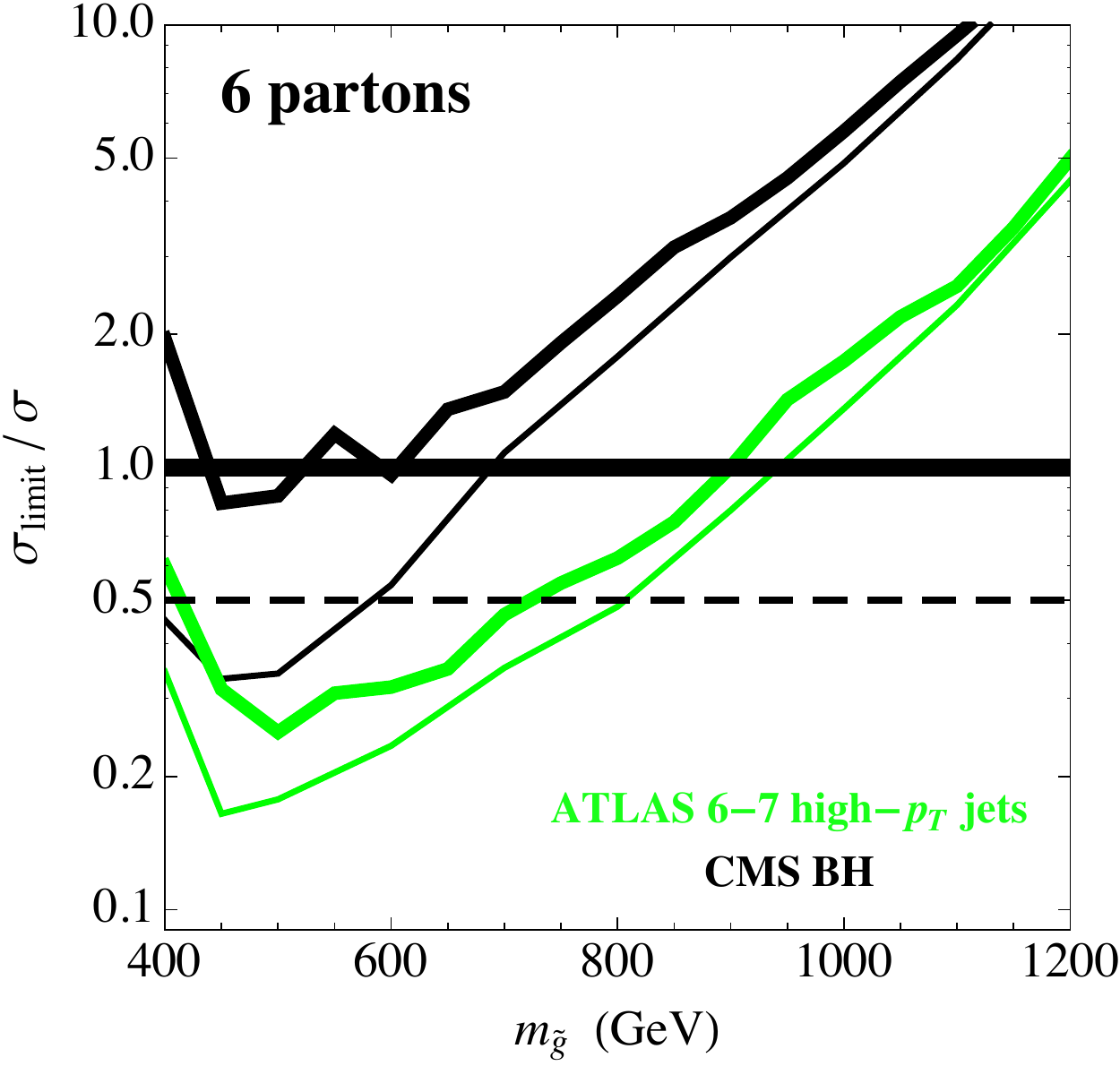}\qq
\includegraphics[width=0.47\textwidth]{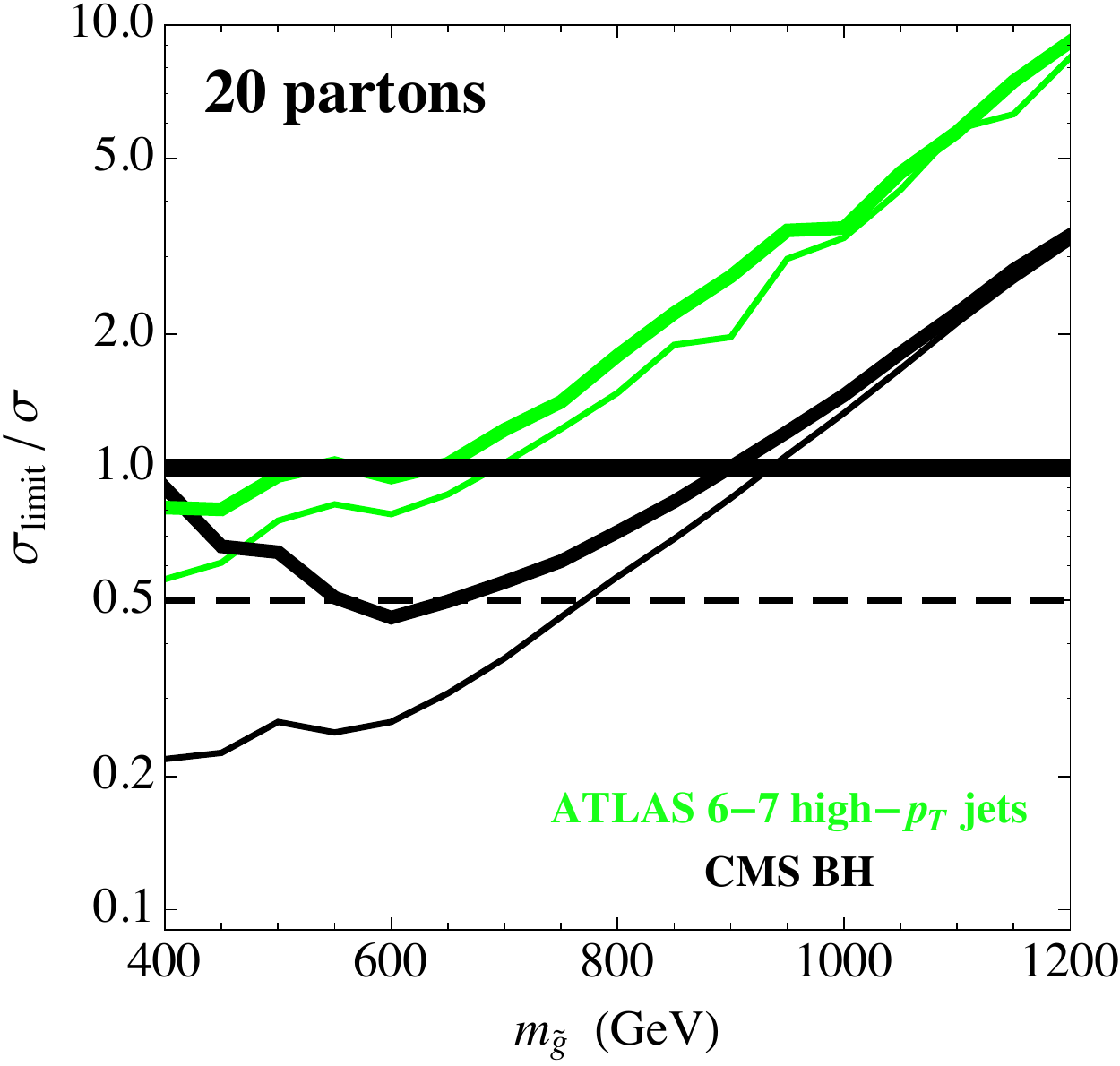}
\caption{Limits on the 6-parton and 20-parton scenarios from section~\ref{sec:all-hadronic} with the {\sc Pythia} parameter SpaceShower:pTdampMatch kept at its default value of 0 as in the main text (thin lines) and set to 1 (thick lines).}
\label{fig:ISR}
\end{center}
\end{figure}

For simplicity, the simulation of initial and final state radiation is based solely on {\sc Pythia}'s parton showering. In a study more focused on specific scenarios,  generating higher multiplicity matrix elements and matching them with the parton showers would be desirable in some of the cases, as we have indicated in the text. The potential significance of a more careful simulation is examined in figure~\ref{fig:ISR}, where we change the {\sc Pythia} parameter SpaceShower:pTdampMatch from 0 (its default value) to 1, which dampens the shower from $1/p_T^2$ to $1/p_T^4$ behavior in the region above the hard scale. It has been found in~\cite{Corke:2010zj} (see also \cite{Bai:2013xla}) that for some processes, including $\go\go$ production, this can lead to a better agreement with matrix elements for hard emissions. The results in figure~\ref{fig:ISR} demonstrate that dampening the shower can affect the robustness of limits from the CMS BH search --- for the 20-parton scenario from section~\ref{sec:all-hadronic}, the cross-section limits are significantly weaker at low gluino masses. However, the exclusion limit, which occurs at high $m_\go$, is not much affected.   Meanwhile the limits for the 6-parton scenario, which depend mainly on the ATLAS 6-7 jets search, are little changed.  More generally, among the searches that are important in our analysis, only the CMS BH search appears markedly sensitive to ISR modeling.

We pass the events through a private detector simulator (which includes the anti-$k_T$ jet algorithm from \textsc{FastJet}~\cite{Cacciari:2005hq} with jet size set to $\Delta R = 0.45$), which uses truth Monte Carlo information and includes the geometric acceptances of the various particles, jet energy resolution
\begin{equation}
\frac{\sigma_E}{E} = \frac{100\%}{\sqrt{E\mbox{ [GeV]}}} \oplus 5\%
\label{jet-smearing}
\end{equation}
(based on~\cite{CMS:2011fta}), identification of $b$-jets and hadronic $\tau$ candidates, and computation of isolation variables for leptons. We then apply trigger, lepton and $b$-jet identification efficiencies, and lepton isolation requirements, as relevant to each search, to the extent this information is provided in the experimental publications or obtained via other means. We do not simulate fake objects (e.g., jets faking leptons or $c$-jets tagged as $b$-jets) since these effects are typically small as far as the signal is concerned. We also do not simulate pile-up or its removal.

We simulate fake \MET due to jet energy mismeasurement, according to the parameterization in eq.~(\ref{jet-smearing}). One might worry that this model may not be sufficiently accurate for the purposes of \MET-based searches for scenarios that have no intrinsic \MET. Fortunately, in those cases the complementary non-\MET based searches set stronger limits anyway, as described in section~\ref{sec:MET} (e.g., figure~\ref{fig:go_750_ho_200_HV1}).

After passing the events through the analysis cuts of each particular search, we compare the NLO+NLL gluino pair production cross section (in the limit of decoupled squarks; see however section~\ref{sec:1st-gen-sq})~\cite{Kramer:2012bx}\footnote{We use the central value of the cross section, without taking the theoretical uncertainty into account. The latter varies from 16\% at $m_\go = 500$~GeV to 26\% at $m_\go  = 1000$~GeV.} with the $95\%$~CL excluded cross section (which is the limit on $\sigma\times\epsilon$ divided by the simulated efficiency) for each search region. In cases where the limits on $\sigma\times\epsilon$ are not provided in the experimental papers, we compute them with the frequentist method~\cite{Conway:2000ju} using the provided backgrounds (and their uncertainties). Due to the limited information available, we conservatively always use the single search region giving the best limit in each case, even though some of the searches envision combining multiple search regions. We do not account for a possible contamination of the control regions by our signals (except in the case of the CMS BH search, as discussed in appendix~\ref{app:BH}).

One concern is that our limits could be at times relying on extreme tails of our signal distributions, where systematic uncertainties on the efficiency (which we do not account for) could become very large.  This is especially an issue for tails in the \MET distribution, which in models with little or no \MET are dominated by our simulation of jet energy mismeasurement. As a precaution, we require that our exclusion limits employ at least $\epsilon_{\rm min} = 10^{-4}$ of the signal events.  More specifically, we have implemented a smooth threshold on the signal efficiency $\epsilon$ by modifying the excluded cross section as $\sigma \to \sigma\exp(\epsilon_{\rm min}/\epsilon)$.

\begin{table}[t]
\begin{center}
\begin{tabular}{|cc|ccc|c|} \hline
\multicolumn{2}{|c|}{Benchmark model} & \multicolumn{3}{|c|}{Search region} & \multirow{2}{*}{Ratio} \\\cline{1-5}
Decay & $m_\go$ (GeV)    & $n_j$ & $n_b$ & $p_T$ (GeV)  &  \\\hline\hline
\multirow{2}{*}{$\go \to jjj$} &  500 & 7 & 0 & 120 & 1.13 \\
                               & 1200 & 7 & 0 & 180 & 0.77 \\ \hline
\multirow{2}{*}{$\go \to bjj$} &  500 & 7 & 2 &  80 & 1.08 \\
                               & 1200 & 7 & 1 & 180 & 0.92 \\ \hline
\multirow{2}{*}{$\go \to tjj$} &  500 & 7 & 1 &  80 & 1.07 \\
                               & 1200 & 7 & 1 & 180 & 0.73 \\ \hline
\end{tabular}
\caption{Validation of the ATLAS 6-7 high-$p_T$ jets search~\cite{ATLAS-CONF-2013-091} on several benchmark scenarios. The ratio is between the number of events in our and ATLAS's simulations. 
\label{tab:Validation}}
\end{center}
\end{table} 

\begin{figure}[t]
\begin{center}
\includegraphics[width=0.465\textwidth]{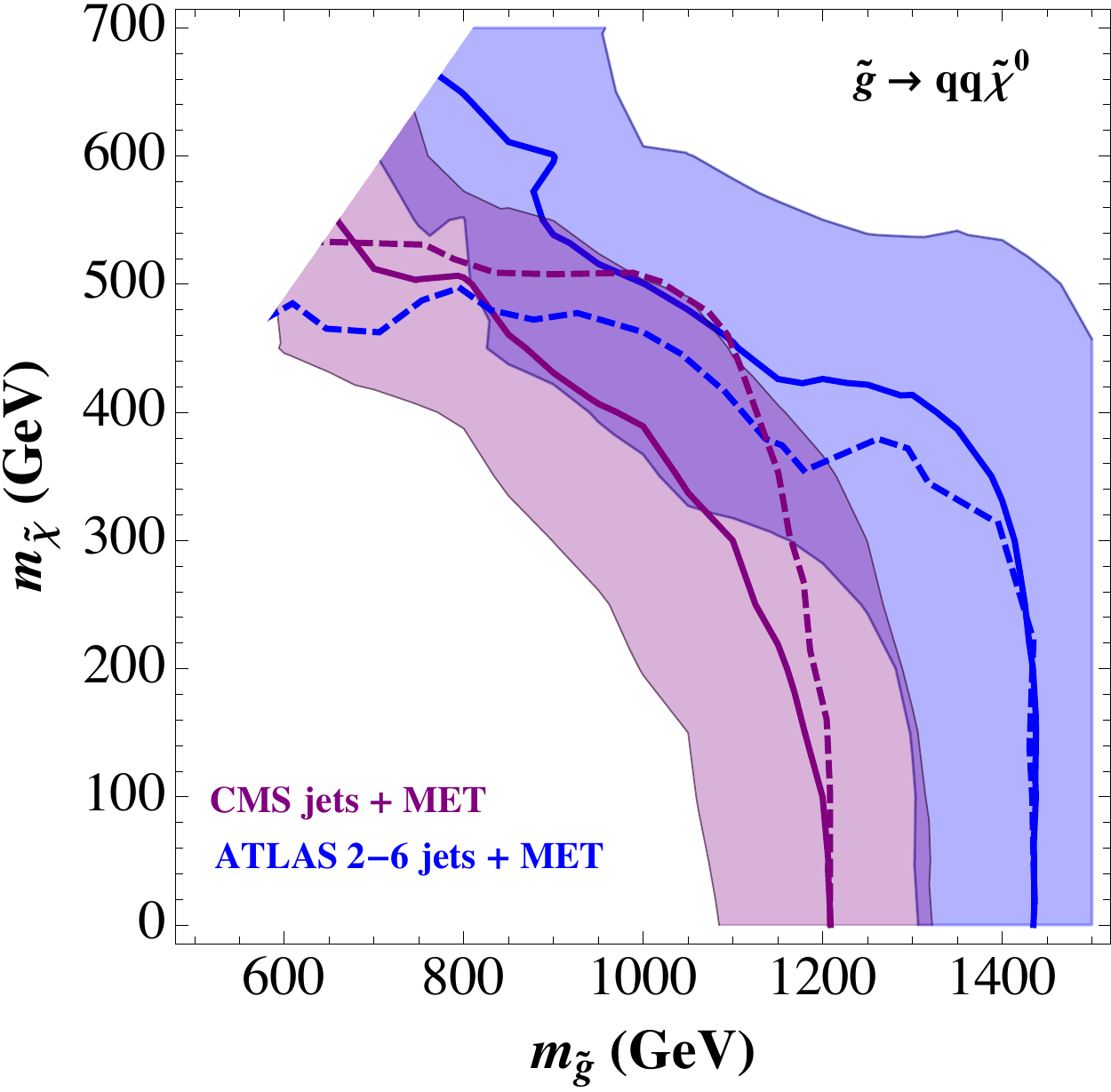}\qq
\includegraphics[width=0.48\textwidth]{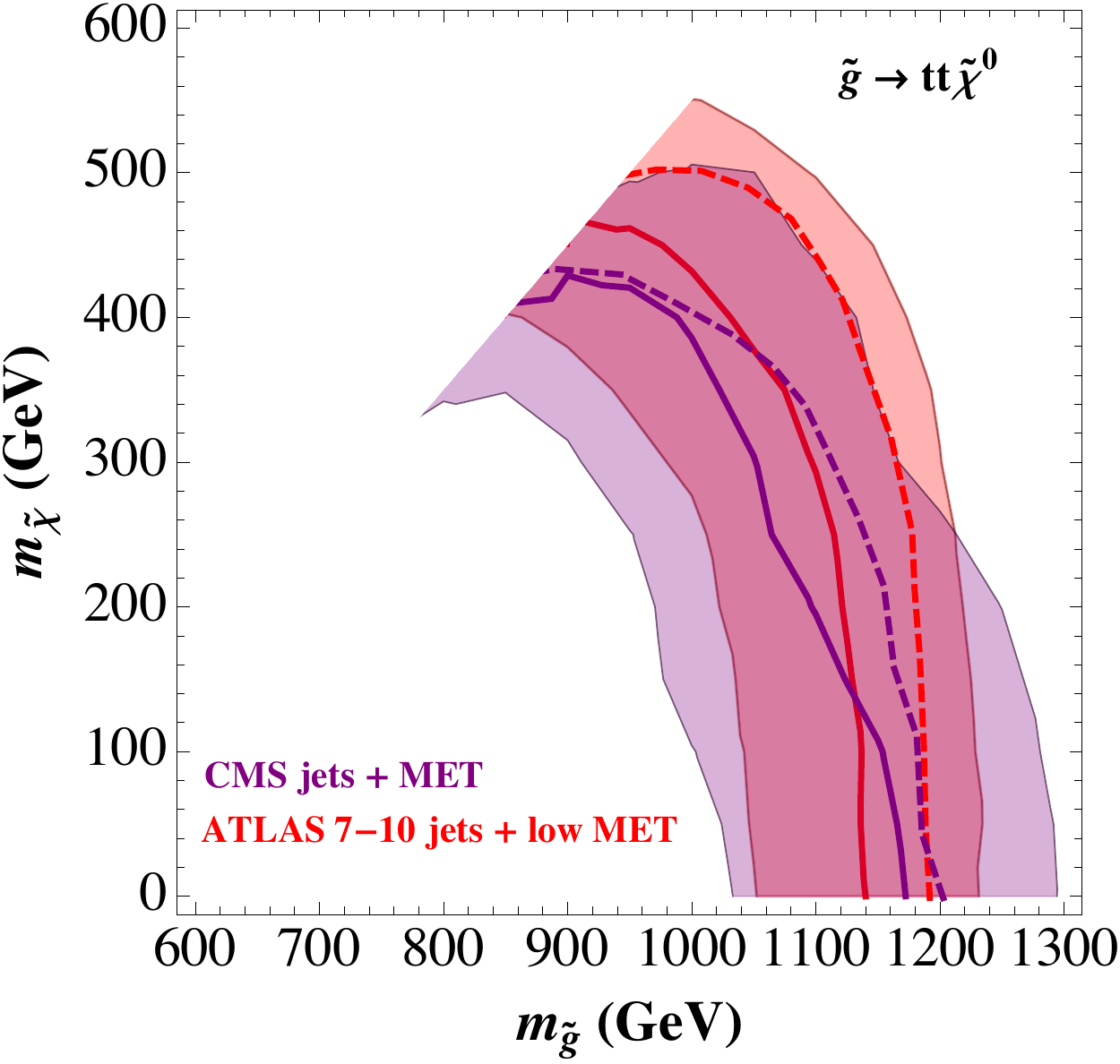}
\caption{Validation of the jets + \MET searches~\cite{ATLAS-CONF-2013-047,CMS-PAS-SUS-13-012,Aad:2013wta}.  The dashed lines correspond to the limits from the experimental studies.  The solid lines are the limits from our simulation, with the shaded bands demonstrating what would happen with a factor-of-2 error in the acceptance (see our discussion of \emph{robustness} of limits in section~\ref{sec:minimal}).   {\bf Left:} Simplified model of $\go\to q\bar q \tilde\chi^0$~\cite{CMS-PAS-SUS-13-012,ATLAS-CONF-2013-047}.  In the compressed region ($m_{\tilde \chi} \gtrsim m_\go/2$) our modeling of ISR is expected to be unreliable.  We do not study such scenarios in this paper.   {\bf Right:} Simplified model of $\go\to t\bar t \tilde\chi^0$~\cite{CMS-PAS-SUS-13-012,Aad:2013wta}. Note that both ATLAS 7--10 jets + low MET~\cite{Aad:2013wta} and CMS jets + \MET~\cite{CMS-PAS-SUS-13-012} combine search regions, which is consistent with them setting stronger limits than we do.}
\label{fig:jetsmetvalidaiton}
\end{center}
\end{figure}

Some of the experimental publications have provided their simulated events yields for easily reproducible examples of new physics signals. We used them, where available, for validating our detector simulation and analysis code. Previous versions of our detector simulation code have been also validated in~\cite{Kats:2011qh,Evans:2012bf} in a similar way. Typically, our event yields agree with those quoted in the experimental papers to within $\sim 30\%$ (although, in a few cases the discrepancy is about a factor of 2). The reader may shift our exclusion curves by these amounts to estimate by how much such uncertainties may be affecting the limit in each case.

Validation of the ATLAS 6-7 high-$p_T$ jets search~\cite{ATLAS-CONF-2013-091} is presented in table~\ref{tab:Validation}.  Validation of ATLAS and CMS jets + \MET searches~\cite{ATLAS-CONF-2013-047,CMS-PAS-SUS-13-012,Aad:2013wta} is shown in figure~\ref{fig:jetsmetvalidaiton}.  The ATLAS 2--6 jets + \MET search~\cite{ATLAS-CONF-2013-047} is modeled reliably up until the compressed region $m_{\tilde \chi} \gtrsim m_\go/2$, in which the missing energy becomes very sensitive to ISR. In part due to this modeling subtlety, we discuss such compressed spectra separately (see in particular section~\ref{subsec:othermods}).   Our limits from the CMS jets + \MET search~\cite{CMS-PAS-SUS-13-012} are a bit weaker than those computed by CMS, but this is expected, as we do not statistically combine bins.  Similarly, our limits from ATLAS 7--10 jets + low MET~\cite{Aad:2013wta} are weaker as we use only the best bin rather than the correlated fit used in the study (as discussed in section~6.1 of~\cite{Aad:2013wta}).  Overall, we find, based on these and many other tests, that despite the different simulation tools and statistical approaches, we reproduce the official experimental results quite closely, and that our claims of exclusion by a factor of 2 (in the cross section) are robust claims of exclusion.

\small{\bibliography{killNSUSY}}

\end{document}